\documentclass[prd,onecolumn]{revtex4}
\usepackage{epsfig}
\usepackage{amsmath}
\usepackage{verbatim}
\usepackage{graphicx}
\usepackage{amsfonts}
\usepackage{amssymb}
\begin{document}
\title{Gauge-invariant and infrared-improved variational analysis of the \\Yang-Mills vacuum wave functional}
\author{Hilmar Forkel}
\affiliation{Institut f\"{u}r Physik, Humboldt-Universit\"{a}t zu Berlin, D-12489 Berlin, Germany}

\begin{abstract}
We study a gauge-invariant variational framework for the Yang-Mills vacuum
wave functional. Our approach is built on gauge-averaged Gaussian trial
functionals which substantially extend previously used trial bases in the
infrared by implementing a general low-momentum expansion for the vacuum field
dispersion (which is taken to be analytic at zero momentum). When completed by
the perturbative Yang-Mills dispersion at high momenta, this results in a
significantly enlarged trial functional space which incorporates both
dynamical mass generation and asymptotic freedom. After casting the dynamics
associated with these wave functionals into an effective action for
collections of soft vacuum-field orbits, the leading infrared improvements
manifest themselves as four-gradient interactions. Those turn out to
significantly lower the minimal vacuum energy density, thus indicating a clear
overall improvement of the vacuum description. The dimensional transmutation
mechanism and the dynamically generated mass scale remain almost
quantitatively robust, however, which ensures that our prediction for the
gluon condensate is consistent with standard values. Further results include a
finite group velocity for the soft gluonic modes due to the higher-gradient
corrections and indications for a negative differential color resistance of
the Yang-Mills vacuum.

\end{abstract}
\maketitle
\preprint{HU-???}

\section{Introduction}

The variational approach was among the first theoretical methods developed to
study nonperturbative ground-state properties of quantum systems \cite{sch26}.
Soon after the inception of QCD, it was brought to bear on the physics of the
gluon vacuum as well \cite{wan88,gre79,fey81,cor88,ker89}, making increasingly
use of its capabilities to treat nonperturbative problems analytically, at
real time and at strong coupling (in contrast e.g. to semiclassical
approximations). Indeed, variational techniques remain applicable even in
non-static and curved spacetimes \cite{jac90,ebo} as well as in nonequilibrium
\footnote{Non-equilibrium processes of particular current interest in
high-energy physics are those in the early universe and in the aftermath of
ultrarelativistic nuclear collisions at the RHIC and the CERN\ LHC (for a
brief discussion see e.g. Sec. V of Ref. \cite{far09}).} situations
\cite{ker79}. Since variational problems are naturally formulated in the
Schr\"{o}dinger picture, furthermore, they often invoke quantum-mechanical
intuition to shed new light on field-theoretic states, as encoded e.g. in the
node number of their wave functionals \cite{fey81,fey88}\ or in the action of
transformation groups \cite{jac90}. On the other hand, field-theoretic
applications have to implement the UV field modes and their dynamics exactly
\cite{fey88}, they must deal with the standard renormalization issues in the
Schr\"{o}dinger representation \cite{sym81,lus85}, and they require a
manageable way of calculating matrix elements by functionally integrating over
physical intermediate states. For gauge theories, maintaining gauge invariance
becomes an additional mandatory requirement which, in the Hamiltonian
formulation, amounts to preserving Gauss' law. Traditionally, the latter is
implemented in combination with (almost) complete gauge fixing to Coulomb
gauge \cite{wan88,sze04,feu04}, although generally without full account of the
gauge-group topology and the resulting Gribov copies. In addition, this
gauge-fixed approach has to deal with a nonlocal Hamiltonian \cite{chr80} and
must take special precautions to avoid gauge-symmetry breaking approximations
\cite{kog95}.

Alternatively, the variational problem can be formulated in a (residual)
gauge-invariant manner which, at least under certain approximations, remains
analytically tractable \cite{kog95}. The mechanism of dimensional
transmutation becomes manifestly gauge-invariant and particularly transparent
from this perspective \cite{dia98}, and the dynamical mass gap \cite{mil}
emerges already from a minimal parametrization of the Gaussian
\textquotedblleft core\textquotedblright\ wave functional \cite{kog95}. The
projection onto the gauge-singlet component of this core functional may
furthermore generate an area law for spacelike Wilson loops and thereby
account for linear quark confinement \cite{kog95,zar98,dia98}, although
experience from compact electrodynamics suggests that the vacuum wave
functional will need non-minimal adjustments to describe the charged sector
\cite{kov99}. Another attractive feature of this approach is that it
reexpresses the infrared dynamics in terms of gauge-invariant matrix fields
(which subsume whole families of gauge-field orbits \cite{for06}) and that it
preserves traceable links of heuristic value between these soft
collective\ fields and the underlying Yang-Mills fields. Moreover, the
gauge-invariant reformulation preserves the global, i.e. topological
properties of the gauge fields. The contributions from all vacuum homotopy
classes as well as instanton-mediated transitions between them, for example,
are explicitly and transparently taken into account without recourse to the
semiclassical approximation \cite{jac90,kog95,bro299,for06}. Besides such sets
of (multi-) instanton and meron orbits, the dynamics contained in the vacuum
wave functional sustains a rich variety of additional gauge-invariant infrared
modes, including Faddeev-Niemi knots and other topological as well as
nontopological solitons \cite{for06,for07}.

Our main objective in the present article will be to improve the vacuum
description provided by the gauge-invariant variational approach. To this end,
we develop a trial functional family which accommodates a rather general
momentum distribution for the infrared vacuum modes while remaining
analytically tractable. Due to the explicit representation of the ground
state, structural aspects of the Yang-Mills dynamics and their impact on the
vacuum can then be understood in terms of the resulting gauge-field
population. Moreover, our generalized trial basis will aid in the full
exploration of the gauge-projected Gaussian functional space, e.g. by
evaluating and analyzing relevant amplitudes, and thereby in the assessment of
its limitations and further improvement potential. In the longer run, this
type of analysis should reveal the extent to which Yang-Mills vacuum physics
can be captured by the most general Gaussian core\ functionals, their
subsequent gauge projection and the mean-field treatment of the resulting
soft-mode dynamics. One may hope that these insights will eventually help to
find a systematic and manageable approximation to the Yang-Mills vacuum wave
functional beyond the Gaussian ansatz.

After having set up our extended trial functional space, the next major task
will be to establish the framework for calculating the associated vacuum
amplitudes. This will enable us, in particular, to evaluate the vacuum energy
density and its reduction in the generalized trial space, thus providing a
quantitative measure for the improvement of the vacuum description. The
analysis of the infrared mode distribution encoded in the energetically
favored wave functional will further shed new light on the resulting vacuum
physics. It will, in particular, reveal the physical significance of the
variational parameters and clarify the impact of their optimized values on the
soft-mode dynamics. We will furthermore examine the influence of the enlarged
trial space on the phase structure and the order-disorder phase transition of
the underlying infrared dynamics (which has, as the Yang-Mills dynamics, its
origin in the imposition of gauge invariance). In fact, the robustness of this
phase transition ensures a reliable variational determination of the
ground-state energy. Our analysis of the resulting vacuum properties will be
complemented by evaluating the lowest-dimensional gluon condensate, which also
provides a quantitative test of the dynamically generated mass scale and its
relation to the underlying trace anomaly. Beyond\ the evaluation and analysis
of specific amplitudes, finally, the extended variational framework will
furnish a transparent theoretical laboratory well-suited to build up
nonperturbative real-time intuition about the Yang-Mills vacuum \footnote{It
may, for example, shed new light on the interplay between gauge symmetry and
topological vacuum properties, which emerge transparently in the Schroedinger
picture \cite{jac90,for06}, and on microscopic aspects of the confinement
mechanism \cite{fey81,fey88}.}.

The paper is organized as follows: in Sec. \ref{vwf} we introduce our
infrared-generalized, gauge-invariant trial functional family for the
Yang-Mills ground state. In Sec. \ref{voa} we rewrite the associated vacuum
overlap amplitude in terms of an effective bare action, summarize our strategy
for calculating the variational bound on the Yang-Mills vacuum energy, and
review the perturbative integration over the hard modes which results in a
renormalized soft-mode action. In Sec. \ref{ircf} we develop the theoretical
framework for calculating relevant amplitudes by integrating over the soft
modes, and in Sec. \ref{excor} we derive explicit expressions for the 2- and
4-point functions. Their further evaluation is prepared in Sec. \ref{gap}
where we set up the saddle-point expansion for the integral over the soft
modes. We further derive the corresponding gap equation and find its
solutions, which provide quantitative information on the phase structure of
the soft-mode dynamics. In Sec. \ref{vend} we calculate the various
contributions to the vacuum energy density (both in the ordered and disordered
phase) which we minimize in Sec. \ref{vmin}. We further compute the gluon
condensate, discuss the physical implications of our results and suggest a few
directions for future work. In Sec. \ref{suc}, finally, we collect our
principal results and our main conclusions.

\section{Vacuum wave functional}

\label{vwf}

The success of variational estimates depends mostly on the choice of the
underlying trial function(al) space. Ideally, it should be both comprehensive
enough to accommodate the relevant physics and concise enough to remain
analytically tractable and reasonably transparent. In the following sections
we establish the basis of our subsequent\ analysis by introducing a
gauge-invariant trial-functional family for the Yang-Mills ground state which
is designed to compromise between these conflicting goals. Although our trial
space expands the minimal one used in Ref. \cite{kog95} substantially, the
ensuing variational analysis will turn out to require only a moderate increase
in computational effort.

Our basis implements Feynman's three requirements for trial wave functionals
of gauge theories, i.e. the at least approximate\ calculability of matrix
elements, the correct UV asymptotics and gauge invariance \cite{fey88}. The
major methodological advance of Ref. \cite{kog95} was to show that a trial
space of manifestly gauge-invariant wave functionals, namely gauge-projected
Gaussians, still permits an (approximately) analytically manageable
variational analysis. In the following sections we first discuss the essential
features of the minimal ansatz of Ref. \cite{kog95}, and then generalize its
momentum distribution for the infrared vacuum modes.

\subsection{Gauge invariance by projection}

\label{ginv}

Variational analyses are naturally set up in the Hamiltonian formulation of
Yang-Mills theory and in the \textquotedblleft coordinate\textquotedblright%
\ Schr\"{o}dinger picture. This restricts gauge transformations to a fixed
reference time, thereby effectively decoupling them from the dynamical time
evolution. In the temporal (or Weyl) gauge $A_{0}^{a}=0$ which we adopt in the
following, the residual gauge transformations are static, furthermore, and
problems with negative-norm states (as e.g. in covariant gauges) are avoided
from the outset \footnote{Perturbation theory, on the other hand, becomes more
complex in temporal gauge (or other axial gauges) \cite{ros80}.}. The gauge
invariance\ of the vacuum wave functional is imposed by Gauss' law which acts
as a subsidiary condition in Fock space. This is particularly crucial for
variational treatments because the Yang-Mills Hamiltonian becomes unique only
when acting on physical states. Hence, as pointed out in Ref. \cite{kog95},
trying to minimize the vacuum expectation value of the Hamiltonian in the
whole space of normalizable functionals could favor contributions from
unphysical parts of the Hilbert space which may be almost arbitrarily enhanced
by adding functionals of gauge-group generators to a given Hamiltonian.

Starting from an approximate and hence typically
gauge-dependent\ \textquotedblleft core\textquotedblright\ functional
$\psi_{0}$ to be specified in Sec. \ref{gcf} ($\psi_{0}$ depends on the static
gauge fields $\vec{A}\left(  \vec{x}\right)  $, i.e. on half of the canonical
variables), we impose gauge invariance by projecting on its gauge-singlet
component, which amounts to averaging over the gauge group \cite{pol78}. The
result is a trial vacuum wave functional of the form%
\begin{equation}
\Psi_{0}\left[  \vec{A}\right]  =\sum_{Q\in Z}e^{iQ\theta}\int D\mu\left[
U^{\left(  Q\right)  }\right]  \psi_{0}\left[  \vec{A}^{U^{\left(  Q\right)
}}\right]  =:\int DU\psi_{0}\left[  \vec{A}^{U}\right]  \label{ginvvwf}%
\end{equation}
where $d\mu$ is the invariant Haar measure of the SU$\left(  N_{c}\right)  $
gauge group \footnote{The gauge transformations of Yang-Mills theory without
matter span the coset SU$\left(  N_{c}\right)  /Z_{N_{c}}$ since center
elements of SU$\left(  N_{c}\right)  $ act trivially on the gauge fields.
Following Ref. \cite{kog95} we will refrain from making the deviation of the
gauge group from SU$\left(  N_{c}\right)  $ manifest, although it could be
implemented by gauging the action (\ref{effact}) and may become relevant,
e.g., for the discussion of center vortices.}, $Q$ is the homotopy degree or
winding number of the group element $U^{\left(  Q\right)  }$, and $\theta$ is
the vacuum angle. The expression (\ref{ginvvwf}) acquires the obligatory
$\theta$ phase under large (i.e. topologically nontrivial) gauge
transformations and renders Gauss' law manifest.

Instead of dealing with the gauge-invariant trial functional (\ref{ginvvwf}),
one may alternatively evaluate the energy density of gauge-\emph{dependent}
Gaussian wave functionals and then correct for the lack of gauge invariance
before minimiziation. This approach was pursued in Ref. \cite{hei00} where the
spurious kinetic energy due to gauge rotations was subtracted by
Thouless-Valatin projection (as originally developed for the treatment of
deformed nuclei). The main advantages of the approach based on wave
functionals of the type (\ref{ginvvwf}) are that gauge invariance of
subsequent calculations and approximations is maintained exactly, and that it
applies universally to \emph{all} matrix elements.

\subsection{Gaussian core wave\ functional}

\label{gcf}

In order to determine the gauge-invariant trial functionals (\ref{ginvvwf})
and the related amplitudes completely, it remains to adopt a core functional
family $\psi_{0}$. As in any variational calculation, this choice generally
remains an uncontrolled approximation without systematic improvement strategy.
In order to motivate our choice for $\psi_{0}$ and to discuss the underlying
physics, we first recall that the vacuum wave functional cannot have nodes,
i.e. that $\psi_{0}\left[  A\right]  \geq0$ for all $A$ \cite{fey81}. (This
provides an example for how quantum-mechanical insights remain applicable to
field theory in the Schr\"{o}dinger representation as long as its infinitely
many degrees of freedom do not generate qualitatively new effects.) Hence one
may write without loss of generality
\begin{equation}
\psi_{0}\left[  \vec{A}\right]  =\frac{1}{\mathcal{N}}e^{-\Phi\left[  \vec
{A}\right]  }%
\end{equation}
where $\mathcal{N}$\ is\ a generally infinite normalization constant and
$\Phi$ is a real and typically nonlocal functional of the gauge field. In
order to find a viable approximation for $\Phi$, we represent it as a
functional power series in $A$ and determine physically reasonable and
analytically tractable truncations. The first two terms of this series are
generally discarded: constant terms can be absorbed into the normalization
constant $\mathcal{N}$, while the term linear in the gauge field (by itself)
corresponds to a coherent vacuum state which is known to be unstable
\cite{leu81}.

The next term is quadratic in $A$ and plays several crucial roles. The first
originates from an ambiguity in $\Phi$ which is due to the invariance of the
Haar measure in Eq. (\ref{ginvvwf}) under right (or left) multiplication by
any given group element. As a consequence, infinitely many choices for
$\Phi\left[  A\right]  $ lead up to an unphysical redefinition of the
normalization constant to the same $\Psi_{0}$ \cite{zar98}. This ambiguity can
be removed, however, by prescribing the quadratic term of $\Phi$. Furthermore,
this term leads to a product of Gaussian vacuum wave functionals of the
Abelian U$\left(  1\right)  $ gauge theory \footnote{The dynamical mass gap
and other nonperturbative features of 2+1 dimensional compact photodynamics
are described exactly by a Gaussian vacuum wave functional as well
\cite{nol04}.} which asymptotic freedom renders exact in the ultraviolet.
Finally, and from the practical perspective most importantly, the Gaussian
functional $\psi_{0}$ resulting from a quadratic term can be integrated over
$A$ analytically, while higher-order contributions may at best be treated
perturbatively \cite{con85}. Hence one generally truncates the series for
$\Phi$ after the quadratic term, which leads to the \textquotedblleft
squeezed\textquotedblright\ approximation for the core\ functional, i.e. to
the Gaussian
\begin{equation}
\psi_{0}^{\left(  G\right)  }\left[  \vec{A}\right]  =\frac{1}{\mathcal{N}%
_{G}}\exp\left[  -\frac{1}{2}\text{ }\int d^{3}x\int d^{3}yA_{i}^{a}\left(
\vec{x}\right)  G_{ij}^{-1ab}\left(  \vec{x}-\vec{y}\right)  A_{j}^{b}\left(
\vec{y}\right)  \right]  \label{ga}%
\end{equation}
with the normalization factor $\mathcal{N}_{G}^{-1}=\left[  \det\left(
G/2\right)  \right]  ^{-1/4}$ and a real kernel or \textquotedblleft
covariance\textquotedblright\ $G^{-1}$ which satisfies a normalizability
condition (cf. Eq. (\ref{nbility})). Eq. (\ref{ga}) appears to be the
\textquotedblleft richest\textquotedblright\ core functional\ family whose
matrix elements can be dealt with by the currently available analytical
methods of field theory. (Adding $c$ number sources to the variable $A$ would
still allow for an analytical treatment and generate finite vacuum expectation
values for $A$ in gauge-fixed approaches, while such local \textquotedblleft
condensates\textquotedblright\ (except of the time component) are not
sustained in our residually gauge-invariant vacuum and would be rendered mute
by the gauge projection in Eq. (\ref{ginvvwf}).)

As already mentioned, the core functional (\ref{ga}) represents an infinite
product of Gaussians, one for each Fourier mode of the gauge field with
momentum $\vec{k}$. Hence the components of the Fourier-transformed
covariance
\begin{equation}
\omega_{ij}^{ab}\left(  \vec{k}\right)  :=G_{ij}^{-1,ab}\left(  \vec
{k}\right)  =g_{1}^{ab}\left(  k\right)  \delta_{ij}+g_{2}^{ab}\left(
k\right)  \hat{k}_{i}\hat{k}_{j} \label{cov}%
\end{equation}
(where $\hat{k}_{i}:=k_{i}/k$ with $k:=\sqrt{k_{i}k_{i}}$) turn after
diagonalization into mode frequencies or energies. The squeezed\ state thus
generalizes the ground state of the quantum-mechanical harmonic oscillator. In
our context, it corresponds to a vacuum consisting of\ color-singlet
gauge-field pairs which may be regarded as the protostate of a glueball
condensate \cite{han82}. (This should be contrasted to the \textquotedblleft
condensation\textquotedblright\ of single gluons in a coherent state.)

Since Gaussian functionals transform nontrivially under non-Abelian gauge
groups, the gauge averaging in Eq. (\ref{ginvvwf}) is an integral part of our
approximation to the physical vacuum state. Gauge-fixed variational analyses
in Coulomb gauge (for references see e.g. \cite{wan88,cor88,ker89,sze04,feu04}%
) are generally based on Gaussian functionals as well, however. They were
found to generate a mass gap \cite{wan88,sze04} and, when multiplied by the
inverse square root of the Faddeev-Popov determinant, an approximately
linearly rising confinement potential \cite{feu04}.

\subsection{General properties and UV asymptotics of the covariance}

\label{guv}

We are now going to specify the properties of the covariance (\ref{cov}) which
characterizes the members of our core trial functional family (\ref{ga}) and
which contains the variational parameters whose values will be adapted below
to optimally approximate the Yang-Mills vacuum wave functional. Translational
invariance implies $G_{ij}^{-1,ab}\left(  \vec{x},\vec{y}\right)
=G_{ij}^{-1,ab}\left(  \vec{x}-\vec{y}\right)  $ and was already anticipated
in Eq. (\ref{ga}). Without loss of generality at the perturbative level (and
beyond, since the integration over the gauge group in Eq. (\ref{ginvvwf})
averages out longitudinal contributions), we will further specialize to a
purely transverse covariance with $g_{2}\equiv0$ \cite{kog95}, i.e.
\begin{equation}
G_{ij}^{-1,ab}\left(  k\right)  =\delta_{ij}G^{-1,ab}\left(  k\right)  ,
\label{gspat}%
\end{equation}
which allows for a direct comparison with the results of Ref. \cite{kog95} and
should be sufficient for our explorative purposes. (The impact of longitudinal
contributions $\propto g_{2}^{ab}$ was discussed in Refs. \cite{dia98,bro98},
and a specific prescription for $g_{2}^{ab}$ was shown to reproduce the 1-loop
Yang-Mills $\beta$-function. Due to the ambiguity of the core functional
mentioned in Sec. \ref{gcf}, a longitudinal term in the covariance can always
be removed from $\Phi$ without changing the physical part of the vacuum wave
functional, although generally in exchange for terms containing higher powers
of the gauge field \cite{zar98}.)

In order to motivate the color structure of our covariance, we note that only
the homogeneous part of the gauge transformations keeps the exponent $\Phi$ of
the core functional bilinear in $A$. Gauge transformations which vary little
over distances for which $G^{-1}\left(  \vec{x}-\vec{y}\right)  $ has
appreciable support therefore leave the Gaussian part of the exponent with a
covariance of the form \cite{kog95}
\begin{equation}
G^{-1,ab}\left(  k\right)  =\delta^{ab}G^{-1}\left(  k\right)  \label{gcol}%
\end{equation}
approximately invariant. This will hold, in particular, for those gauge group
elements which determine our soft-mode dynamics (see below). Since the same
color structure is appropriate for the hard gauge-field modes, which we will
treat perturbatively in the small bare coupling $g_{\text{b}}$, we adopt Eq.
(\ref{gcol}) for the remainder of this paper. As a consequence, the core
functionals $\psi_{0}$ become invariant both under global U$\left(
N_{c}\right)  $ transformations and under $N_{c}^{2}-1$ copies of the
U$\left(  1\right)  $ gauge group.

Combining the spacial (\ref{gspat}) and color (\ref{gcol}) structures, our
covariance assumes the form
\begin{equation}
G_{ij}^{-1ab}\left(  \vec{x},\vec{y}\right)  =\delta^{ab}\delta_{ij}%
G^{-1}\left(  \vec{x}-\vec{y}\right)  =\delta^{ab}\delta_{ij}\int\frac{d^{3}%
k}{\left(  2\pi\right)  ^{3}}e^{i\vec{k}\left(  \vec{x}-\vec{y}\right)
}G^{-1}\left(  k\right)
\end{equation}
which is symmetric in each of the 3 \textquotedblleft index\textquotedblright%
\ pairs. The covariance is further restricted by the requirement that wave
functionals of physical states have to be normalizable. Since the norm of the
functional (\ref{ga}) involves an integral over the Fourier modes $A\left(
k\right)  $, this implies that the integrand $\psi_{0}^{\ast}\left(  k\right)
\psi_{0}\left(  k\right)  $ has to damp large gauge fields $A\left(  k\right)
$ for all $k$. The corresponding localization in field space is implemented by
demanding
\begin{equation}
G^{-1}\left(  k\right)  >0 \label{nbility}%
\end{equation}
which ensures vacuum stability and a positive energy spectrum of the
associated quantum field theory. The condition (\ref{nbility}) will limit the
domain of the variational parameters to be introduced below.

As noted by Feynman, a further mandatory requirement is that all trial
functionals reproduce the asymptotically free gauge dynamics for
$k\rightarrow\infty$ exactly \cite{fey88}. This prevents the infinitely many
ultraviolet modes (which are irrelevant for the vacuum physics) from
artificially dominating the soft-mode energy density through their IR-mode
couplings. In order to implement the correct UV behavior, we factorize the
unprojected core functional (\ref{ga}) as
\begin{equation}
\psi_{0}^{\left(  G\right)  }\left[  \vec{A}\right]  =\psi_{0}^{\left(
G_{<}\right)  }\left[  \vec{A}_{<}\right]  \psi_{0}^{\left(  G_{>}\right)
}\left[  \vec{A}_{>}\right]
\end{equation}
by splitting the $\vec{k}$ integration domain in the exponentials into
soft/hard regions with momenta $k\gtrless\mu$ relative to a separation scale
$\mu$, i.e. \textbf{ }%
\begin{equation}
\psi_{0}^{\left(  G_{\lessgtr}\right)  }\left[  \vec{A}_{\lessgtr}\right]
=\exp\left\{  -\frac{1}{2}\int\frac{d^{3}k}{\left(  2\pi\right)  ^{3}}%
\theta\left(  \pm\mu^{2}\mp\vec{k}^{2}\right)  A_{\lessgtr,i}^{a}\left(
k\right)  G_{\lessgtr}^{-1}\left(  k\right)  A_{\lessgtr,i}^{a}\left(
k\right)  \right\}  . \label{gir}%
\end{equation}
This allows to incorporate the asymptotic freedom of Yang-Mills theory (i.e.
the Gaussian UV fixed point) by requiring that $G$ approaches the
non-interacting, massless static vector field propagator $G_{0}\left(
k\right)  =1/k$ when $k\rightarrow\infty$. As long as $\mu\gg\Lambda
_{\text{YM}}$ where $\Lambda_{\text{YM}}$ is the Yang-Mills scale, a natural
approximation is therefore \footnote{As pointed out in Ref. \cite{kog95},
minimizing the mode energy by using the more general power ansatz
$G_{>}\left(  k\right)  =\alpha k^{\beta}$ for the UV covariance results in
$\alpha=1,\beta=-1$, i.e. precisely the covariance (\ref{gm1uv}) required by
asymptotic freedom.}
\begin{equation}
G_{>}^{-1}\left(  k\right)  =k \label{gm1uv}%
\end{equation}
which we adopt in the following \footnote{Adopting Eq. (\ref{gm1uv}) for
\emph{all} momenta $0\leq k\leq\Lambda_{\text{UV}}$, on the other hand, would
turn the functional (\ref{ginvvwf}) into the exact ground state of $N_{c}%
^{2}-1$ copies of U$\left(  1\right)  $ photodynamics (since averaging over
the gauge group $\left[  U\left(  1\right)  \right]  ^{N_{c}}$ removes the
longitudinal gauge-field modes).}. More specifically, the value of $\mu$ must
be chosen large enough for perturbative corrections from the hard modes, or
equivalently the renormalization-group (RG) improved running coupling
$\alpha\left(  \mu\right)  =g^{2}\left(  \mu\right)  /\left(  4\pi\right)  $,
to remain small. (RG-improved perturbative corrections to the leading
$k\rightarrow\infty$ behavior will enter when integrating out the hard modes
perturbatively, cf. Sec. \ref{huen} and App. \ref{puint}.) Since $\mu$ will be
treated as a variational parameter, this has to be checked \emph{a
posteriori}, i.e. for the value $\mu^{\ast}$ which turns out to minimize the
vacuum energy. For the infrared momenta $k<\mu$, finally, the nonperturbative
Yang-Mills dynamics is expected to induce a qualitatively different covariance
$G_{<}^{-1}\left(  k\right)  $ which will be the subject of the following section.

\subsection{Generalized IR\ (soft-mode) covariance}

\label{gengir}

Due to the logarithmically slow running of the Yang-Mills coupling in the
ultraviolet, the high-momentum behavior of our trial functional family is
rather accurately reproduced by the hard-mode covariance (\ref{gm1uv}). Hence
only the IR covariance $G_{<}^{-1}\left(  k\right)  $, which encodes the more
complex and less understood nonperturbative vacuum physics, remains to be
determined variationally. To this end, we have to implement a parametrization
for $G_{<}^{-1}$ which is sufficiently \textquotedblleft
rich\textquotedblright\ to accommodate the relevant physics without impeding
an analytically tractable variational analysis. The minimal choice
\begin{equation}
G_{<,\text{KK}}^{-1}\left(  k\right)  =\mu\label{gm1kk}%
\end{equation}
was adopted in the pioneering study \cite{kog95} and shown to generate a mass
gap when the separation scale $\mu$ is simultaneously utilized as a
variational parameter. While it therefore provides an efficient starting point
for describing the Yang-Mills vacuum, one may also worry about too much bias
since it leaves only one characteristic vacuum mass scale to be determined by
energy minimization. This involves the risk of essentially \textquotedblleft
building in\textquotedblright\ the mass gap without gaining much further
insight into the underlying vacuum structure.

In the following, we will therefore rely on a more comprehensive
parametrization of $G_{<}^{-1}$ which better accommodates the $k<\mu$ mode
dynamics while still allowing for a variational analysis without the need for
solving a functional differential equation. It is based on the under
reasonable analyticity assumptions general gradient expansion \cite{for06}%
\begin{equation}
G_{<}^{-1}\left(  \vec{x}-\vec{y}\right)  =m_{\text{g}}\left[  1+c_{1}%
\frac{\partial_{x}^{2}}{\mu^{2}}+c_{2}\left(  \frac{\partial_{x}^{2}}{\mu^{2}%
}\right)  ^{2}+c_{3}\left(  \frac{\partial_{x}^{2}}{\mu^{2}}\right)
^{3}+...\right]  \delta_{<}^{3}\left(  \vec{x}-\vec{y}\right)  \label{gm1}%
\end{equation}
which can be efficiently truncated to yield a manageable trial basis for the
$k^{2}\ll\mu^{2}$ soft-mode physics. Besides $\mu$, the variational parameter
space then contains the IR gluon mass\ $m_{\text{g}}>0$ and a few of the
low-momentum constants $c_{n}$ which characterize dispersive properties of the
vacuum (cf. Sec. \ref{phimpl}). (The parameters $c_{n}$ should be considered
as renormalized at $\mu$ since they do not receive UV contributions from
integrating over the $k>\mu$ modes.) The derivatives in Eq. (\ref{gm1}) act on
the regularized delta function%
\begin{equation}
\delta_{<}^{3}\left(  \vec{x}-\vec{y}\right)  :=\int\frac{d^{3}k}{\left(
2\pi\right)  ^{3}}\theta\left(  \mu^{2}-\vec{k}^{2}\right)  e^{i\vec{k}\left(
\vec{x}-\vec{y}\right)  } \label{regdel}%
\end{equation}
which encodes the slow variation $\left\Vert \partial A_{<}\right\Vert
/\left\Vert A_{<}\right\Vert \leq\mu$ of the soft modes and ensures that the
higher-order terms of the above expansion are parametrically suppressed. It
further restricts the support of $G_{<}^{-1}$:
\begin{equation}
G_{<}^{-1}\left(  \vec{x}\right)  \sim0\text{ \ \ \ for }\left\vert \vec
{x}\right\vert >\frac{1}{\mu}.
\end{equation}
Together with the hard-mode covariance (\ref{gm1uv}), Eq. (\ref{gm1}) provides
a rather general parametrization of the core functional (\ref{ga}) which
includes the minimal one-parameter ansatz (\ref{gm1kk}) for $c_{n}=0$ and
$m_{\text{g}}=\mu$.

(Note, incidentally, that the perturbative covariance (\ref{gm1uv}) has a
branch point at $k^{2}=0$ and thus cannot be directly expanded as in Eq.
(\ref{gm1}). Note further that the expression (\ref{gm1}) can be extended
beyond the limited\ spacial tensor structure (\ref{gspat}) by writing
\begin{equation}
G_{<,ij}^{-1}\left(  \vec{x}-\vec{y}\right)  =G_{<,ij}^{\left(  L\right)
-1}\left(  \vec{x}-\vec{y}\right)  +G_{<,ij}^{\left(  T\right)  -1}\left(
\vec{x}-\vec{y}\right)  \label{gm1gen}%
\end{equation}
with ($X\equiv L,T$)%
\begin{equation}
G_{<,ij}^{\left(  X\right)  -1}\left(  \vec{x}-\vec{y}\right)  =m_{\text{g}%
}^{\left(  X\right)  }\left[  1+c_{1}^{\left(  X\right)  }\frac{\partial
_{x}^{2}}{\mu^{2}}+c_{2}^{\left(  X\right)  }\left(  \frac{\partial_{x}^{2}%
}{\mu^{2}}\right)  ^{2}+c_{3}^{\left(  X\right)  }\left(  \frac{\partial
_{x}^{2}}{\mu^{2}}\right)  ^{3}...\right]  \delta_{X,ij}^{3}\left(  \vec
{x}-\vec{y}\right)
\end{equation}
where%
\begin{equation}
\delta_{X,ij}^{3}\left(  \vec{x}\right)  =\int\frac{d^{3}k}{\left(
2\pi\right)  ^{3}}\theta\left(  \mu^{2}-\vec{k}^{2}\right)  \delta
_{X,ij}\left(  \hat{k}\right)  e^{i\vec{k}\vec{x}}%
\end{equation}
with $\delta_{T,ij}\left(  \hat{k}\right)  \equiv\delta_{ij}-k_{i}k_{j}%
/\vec{k}^{2}$ and $\delta_{L,ij}\left(  \hat{k}\right)  \equiv k_{i}k_{j}%
/\vec{k}^{2}$. This decouples the longitudinal and transverse contributions
and implies, in particular,
\begin{equation}
\partial_{x,i}G_{ij}^{\left(  T\right)  -1}\left(  \vec{x}-\vec{y}\right)  =0.
\end{equation}
For $m_{\text{g}}^{\left(  T\right)  }=m_{\text{g}}^{\left(  L\right)  }$ and
$c_{n}^{\left(  T\right)  }=c_{n}^{\left(  L\right)  }$ the more general
expression (\ref{gm1gen}) simplifies to Eq. (\ref{gm1}). In the following
analysis we adopt the latter, in order not to compromise transparency by
additional parameters and to allow for a direct comparison with the results of
Ref. \cite{kog95}.)

In momentum space Eq. (\ref{gm1}) becomes%
\begin{equation}
G_{<}^{-1}\left(  k\right)  =m_{\text{g}}\left[  1-c_{1}\frac{k^{2}}{\mu^{2}%
}+c_{2}\left(  \frac{k^{2}}{\mu^{2}}\right)  ^{2}-c_{3}\left(  \frac{k^{2}%
}{\mu^{2}}\right)  ^{3}+...\right]  \theta\left(  \mu^{2}-k^{2}\right)  ,
\label{gm}%
\end{equation}
which renders the relation between the parameters $c_{n}$ and the dispersion
properties of the IR quasigluons (cf. Sec. \ref{phimpl}) more explicit. It
further implies that the effective IR gluon propagator%
\begin{equation}
G_{<}\left(  k\right)  =\frac{1}{m_{\text{g}}}\left[  1+c_{1}\frac{k^{2}}%
{\mu^{2}}+\left(  c_{1}^{2}-c_{2}\right)  \frac{k^{4}}{\mu^{4}}+\left(
c_{1}^{3}-2c_{1}c_{2}+c_{3}\right)  \frac{k^{6}}{\mu^{6}}+...\right]
\theta\left(  \mu^{2}-k^{2}\right)  , \label{gk}%
\end{equation}
resulting from $G_{<}^{-1}\left(  k\right)  G_{<}\left(  k\right)  =1$, is
analytic at $k^{2}=0$. (The finiteness of $G_{<}$ at zero momentum,
$G_{<}\left(  0\right)  =m_{\text{g}}^{-1}$, is reminiscent of lattice results
for the gluon propagator in Landau gauge \cite{latland}.) The adjustable
parameters $m_{\text{g}},$ $\mu$ and $c_{n}$ are to be determined
variationally (cf. Sec. \ref{vmin}). Their physically sensible domain is
restricted by several constraints, however. Indeed, as a consequence of the
normalizability condition (\ref{nbility}) the low-momentum constants must
satisfy the bounds%
\begin{equation}
c_{1}<1,\text{ \ \ \ \ }c_{2}>-1,... \label{cbnds}%
\end{equation}
(for $m_{\text{g}}>0$). The value of the IR gluon mass $m_{\text{g}}>0$ is
further constrained by requiring continuity of $G^{-1}\left(  k\right)  $ at
the matching point $k=\mu$ between soft and hard momenta. This fixes
$m_{\text{g}}$ as a function of the other variational parameters. When
approximating the series (\ref{gm}) by the truncation $c_{n\geq2}=0$, for
example, one has%
\begin{equation}
m_{\text{g}}\left(  \mu,c_{1}\right)  =\frac{\mu}{1-c_{1}}. \label{mgc}%
\end{equation}
Note that the requirement of a non-negative IR gluon mass then restricts the
$c_{1}$ domain to $c_{1}<1$, in agreement with the bound (\ref{cbnds}) from
normalizability. The singularity of $m_{\text{g}}\left(  \mu,c_{1}\right)  $
for $c_{1}\rightarrow1$ reflects the onset of the vacuum instability and is
inherited by $G_{<}^{-1}\left(  k\right)  $, cf. Eq. (\ref{gm}).

In addition to the lowest-order approximation (\ref{gm1kk}), our expansion
(\ref{gm1}) encompasses another previously used ansatz for the IR covariance,
namely the inverse of the non-interacting massive vector propagator,
\begin{equation}
G_{0}^{-1}\left(  k;\mu\right)  =\sqrt{k^{2}+\mu^{2}}\overset{k<\mu}{=}%
\mu\allowbreak\left[  1+\frac{1}{2}\frac{k^{2}}{\mu^{2}}-\frac{1}{8}\left(
\frac{k^{2}}{\mu^{2}}\right)  ^{2}+\frac{1}{16}\left(  \frac{k^{2}}{\mu^{2}%
}\right)  ^{3}-...\right]  , \label{g0}%
\end{equation}
which obviously corresponds to $m_{\text{g}}=\mu$, $c_{1}^{\left(  0\right)
}=-1/2,$ $c_{2}^{\left(  0\right)  }=-1/8$ etc.. Its truncation to $c_{n\geq
2}=0$ was adopted as the basis of a gauge-invariant saddle-point expansion for
the IR amplitudes and found to reproduce the soft-mode action (cf. Sec.
\ref{voa}) at its saddle points within a few percent accuracy \cite{for06}.
(Lattice simulations similarly found the Landau-gauge gluon propagator to
behave like a massive vector propagator at intermediate momenta \cite{sil04}.)

In Fig. \ref{epdel} we compare the expressions (\ref{gm1kk}) and (\ref{g0})
for the IR covariance to Eq. (\ref{gm}) with $c_{1}=\pm0.15$ (the positive
value will turn out to minimize the vacuum energy, cf. Sec. \ref{res}) and
$c_{n\geq2}=0$. A more detailed discussion of the physics encoded in the
covariance (\ref{gm}) will be postponed until the variationally optimized
values of the low-momentum constants are found (cf. Sec. \ref{phimpl}).

In summary, the Gaussian ansatz (\ref{ga}) for the core functional encodes the
correct ultraviolet asymptotics, can be partly motivated in the infrared and
is amenable to existing analytical methods of field theory. As such, it
provides the best currently available, analytically manageable and
gauge-invariant approximation to the Yang-Mills vacuum wave functional.

\section{Energy and soft-mode dynamics of the trial functional family}

\label{voa}

\subsection{Calculational strategy}

\label{strat}

Variational analyses \`{a} la Rayleigh-Ritz amount to minimizing the
expectation value of a given Hamiltonian in a suitable space of trial states.
In our context, the energy density of the trial vacuum is%
\begin{equation}
\left\langle \mathcal{H}\left(  A,E\right)  \right\rangle =\frac{\int D\vec
{A}\Psi_{0}^{\ast}\left[  \vec{A}\right]  \mathcal{H}\left(  \vec{A}^{a}%
,\frac{i\delta}{\delta\vec{A}^{a}\left(  \vec{x}\right)  }\right)  \Psi
_{0}\left[  \vec{A}\right]  }{\int D\vec{A}\Psi_{0}^{\ast}\left[  \vec
{A}\right]  \Psi_{0}\left[  \vec{A}\right]  } \label{hexp}%
\end{equation}
where $\mathcal{H}$ is the Yang-Mills Hamiltonian (cf. Eq. (\ref{hamYM})) in
temporal gauge and $\Psi_{0}$ the generalized vacuum wave functional family
introduced in Sec. \ref{vwf}. This expectation value is most efficiently
evaluated by means of a generating functional, as summarized in App.
\ref{stacor}. In the present section we prepare for the calculation of Eq.
(\ref{hexp}) by sketching our basic strategy (which straightforwardly
generalizes to more general matrix elements, cf. App. \ref{stacor}).

After inserting the gauge-projected vacuum wave functional (\ref{ginvvwf})
into Eq. (\ref{hexp}) and interchanging the order of the integration over
fields and group elements, the gauge invariance of the $\vec{A}$ integral
allows to factor out one gauge group volume. Eq. (\ref{hexp}) can be then be
rewritten as
\begin{equation}
\left\langle \mathcal{H}\left(  A,E\right)  \right\rangle =\frac{\int DU\int
D\vec{A}\psi_{0}\left[  \vec{A}^{U}\right]  \mathcal{H}\left(  \vec{A}%
^{a},\frac{i\delta}{\delta\vec{A}^{a}\left(  \vec{x}\right)  }\right)
\psi_{0}\left[  \vec{A}\right]  }{\int DU\int D\vec{A}\psi_{0}\left[  \vec
{A}^{U}\right]  \psi_{0}\left[  \vec{A}\right]  }%
\end{equation}
(where $DU$ is the functional SU$\left(  N_{c}\right)  $ measure as defined in
Eq. (\ref{ginvvwf})). After evaluating the functional derivatives contained in
$\mathcal{H}$, the Gaussian integration over $A$ can be performed exactly (cf.
App. \ref{stacor} for more details), resulting in%
\begin{equation}
\left\langle \mathcal{H}\right\rangle =\frac{\int DU\left\langle \left\langle
\left\langle \mathcal{H}\right\rangle \right\rangle \right\rangle \exp\left\{
-\Gamma_{\text{b}}\left[  U\right]  \right\}  }{\int DU\exp\left\{
-\Gamma_{\text{b}}\left[  U\right]  \right\}  }%
\end{equation}
where we introduced the abbreviation
\begin{equation}
\left\langle \left\langle \left\langle \vec{A}...\vec{A}...\vec{E}...\vec
{E}\right\rangle \right\rangle \right\rangle \exp\left\{  -\Gamma_{\text{b}%
}\left[  U\right]  \right\}  \equiv\int D\vec{A}\psi_{0}\left[  \vec{A}%
^{U}\right]  \vec{A}...\vec{A}...\frac{i\delta}{\delta\vec{A}}...\frac
{i\delta}{\delta\vec{A}}\psi_{0}\left[  \vec{A}\right]  .
\end{equation}
The above expression defines an effective bare action $\Gamma_{\text{b}%
}\left[  U\right]  $ which describes dynamical correlations originating from
the gauge projection of the functional $\psi_{0}$. More specifically, from the
normalization $\left\langle \left\langle \left\langle 1\right\rangle
\right\rangle \right\rangle =1$ it follows that
\begin{equation}
\Gamma_{\text{b}}\left[  U\right]  =-\ln\int D\vec{A}\,\ \psi_{0}^{\ast
}\left[  \vec{A}^{U}\right]  \psi_{0}\left[  \vec{A}\right]  . \label{gammab}%
\end{equation}
This gauge-invariant, 3-dimensional Euclidean action would become $U$
independent if $\psi_{0}$ were gauge invariant by itself. Hence $\Gamma
_{\text{b}}\left[  U\right]  $ gathers all those gauge-field contributions to
the generating functional whose approximate vacua $\psi_{0}$ at $t=\pm\infty$
differ by a relative gauge orientation $U$. The variable $U$ thus represents
the contributions from a specifically weighted ensemble of all gluon field
orbits to the vacuum overlap and is gauge invariant by construction.

After splitting $U\left(  \vec{x}\right)  =U_{<}\left(  \vec{x}\right)
U_{>}\left(  \vec{x}\right)  $ with $U_{>}\left(  \vec{x}\right)  =\exp\left(
-ig\phi^{a}\left(  \vec{x}\right)  \lambda^{a}/2\right)  $ into hard- and
soft-mode contributions and integrating over the hard UV modes $\phi$
perturbatively (as done in Ref. \cite{kog95} which used the same UV covariance
(\ref{gm1uv}), cf. App. \ref{stacor}), one arrives at%
\begin{equation}
\left\langle \mathcal{H}\right\rangle =\frac{\int DU_{<}\int D\phi\left\langle
\left\langle \left\langle \mathcal{H}\right\rangle \right\rangle \right\rangle
\exp\left\{  -\Gamma_{\text{b}}\left[  \phi,U_{<}\right]  \right\}  }{\int
DU_{<}\int D\phi\exp\left\{  -\Gamma_{\text{b}}\left[  \phi,U_{<}\right]
\right\}  }.
\end{equation}
After further defining%
\begin{equation}
\left\langle \left\langle \mathcal{O}\right\rangle \right\rangle \exp\left\{
-\Gamma_{<}\left[  U_{<}\right]  \right\}  :=\int D\phi\left\langle
\left\langle \left\langle \mathcal{O}\right\rangle \right\rangle \right\rangle
\exp\left\{  -\Gamma_{\text{b}}\left[  \phi,U_{<}\right]  \right\}  \label{o2}%
\end{equation}
and in particular the effective soft-mode action
\begin{equation}
\Gamma_{<}\left[  U_{<}\right]  :=-\ln\int D\phi\exp\left\{  -\Gamma
_{\text{b}}\left[  \phi,U_{<}\right]  \right\}  , \label{gsm0}%
\end{equation}
we end up with a reformulation of the matrix element (\ref{hexp}) in terms of
the dynamics for the $U_{<}$ field, i.e. for the low-momentum components of
the integration variable originating from the gauge projection of the vacuum
functional (\ref{ginvvwf}):%
\begin{equation}
\left\langle \mathcal{H}\right\rangle =\frac{\int DU_{<}\left\langle
\left\langle \mathcal{H}\right\rangle \right\rangle \exp\left\{  -\Gamma
_{<}\left[  U_{<}\right]  \right\}  }{\int DU_{<}\exp\left\{  -\Gamma
_{<}\left[  U_{<}\right]  \right\}  }. \label{hsmvev}%
\end{equation}
Hence the calculation of $\left\langle \mathcal{H}\right\rangle $ boils down
to an integral over $U_{<}$ in which the \textquotedblleft
reduced\textquotedblright\ (i.e., fixed $U_{<}$) matrix element $\left\langle
\left\langle \mathcal{H}\right\rangle \right\rangle $ is weighted by a
Boltzmann factor containing just the soft-mode action $\Gamma_{<}\left[
U_{<}\right]  $. This action is governed by our IR expansion (\ref{gm1}) of
the covariance $G_{<}^{-1}$ which produces interactions to be determined in
Sec. \ref{sma}. Since $\left\langle \left\langle \mathcal{H}\right\rangle
\right\rangle $ consists of nonlocal functionals of the soft modes $U_{<}$,
furthermore, the evaluation of Eq. (\ref{hsmvev}) amounts to calculating
(equal-time) soft-mode correlation functions. The necessary framework for such
calculations will be set up in Sec. \ref{ircf}.

\subsection{Soft-mode action}

\label{sma}

In this section we derive explicit expressions for the low-momentum dynamics
(\ref{gsm0}) induced by the wave functional (\ref{ginvvwf}) with the Gaussian
core (\ref{ga}) and the generalized covariance (\ref{gm1uv}), (\ref{gm}).
After specializing the expression (\ref{gammab}) for the bare action to the
core functional (\ref{ga}), the integral over the gauge fields can be carried
out exactly (cf. App. \ref{stacor}). The result is the action of a
3-dimensional, bilocal nonlinear sigma model:
\begin{equation}
\Gamma_{\text{b}}\left[  U\right]  =\frac{1}{2g_{\text{b}}^{2}}\int d^{3}x\int
d^{3}yL_{i}^{a}\left(  \vec{x}\right)  \mathcal{D}^{ab}\left(  \vec{x}-\vec
{y}\right)  L_{i}^{b}\left(  \vec{y}\right)  . \label{effact}%
\end{equation}
(Above we have omitted a term of higher order in the small bare coupling
$g_{\text{b}}$, cf. App. \ref{stacor}.) The $U$ fields enter $\Gamma
_{\text{b}}$ both in terms of the left-invariant SU$\left(  N_{c}\right)  $
Maurer-Cartan forms%
\begin{equation}
L_{i}\left(  \vec{x}\right)  =U^{\dagger}\left(  \vec{x}\right)  \partial
_{i}U\left(  \vec{x}\right)  =:L_{i}^{a}\left(  \vec{x}\right)  \frac
{\lambda^{a}}{2i} \label{L}%
\end{equation}
(with real components $L_{i}^{a}$ and the SU$\left(  N_{c}\right)  $ Gell-Mann
matrices $\lambda^{a}$) and through higher-order\ corrections to the bilocal
operator
\begin{equation}
\mathcal{D}^{ab}=\left[  \left(  G+G^{U}\right)  ^{-1}\right]  ^{ab}%
\simeq\frac{1}{2}G^{-1}\delta^{ab}+...
\end{equation}
where $G^{U}=G^{ab}\left(  \vec{x}-\vec{y}\right)  U^{\dagger}\left(  \vec
{x}\right)  t^{a}U\left(  \vec{x}\right)  \otimes U\left(  \vec{y}\right)
t^{b}U^{\dagger}\left(  \vec{y}\right)  $ and $t^{a}=\lambda^{a}/2$. The above
reformulation of the vacuum functional overlap can alternatively be obtained
by a saddle-point evaluation \cite{dia98} of the Gaussian integral in Eq.
(\ref{gammab}).

According to the strategy outlined in Sec. \ref{strat}, the calculation of
infrared amplitudes involves an effective action (\ref{gsm0}) which only
retains soft field modes explicitly. After factorizing $U=U_{<}U_{>}$, the
hard modes $U_{>}$ with $k^{2}>\mu^{2}$ can be integrated out of Eq.
(\ref{gsm0}) perturbatively (cf. App. \ref{stacor}) as long as the
renormalized coupling \cite{bro99}
\begin{equation}
g\left(  \mu\right)  =g_{\text{b}}+\frac{g_{\text{b}}^{3}N_{c}}{\left(
2\pi\right)  ^{2}}\ln\frac{\Lambda_{\text{UV}}}{\mu}+O\left(  g_{\text{b}}%
^{5}\right)  \label{g(mu)}%
\end{equation}
(for $G_{>}\left(  k\right)  =k^{-1}$) stays sufficiently small. Note that Eq.
(\ref{g(mu)}) exhibits asymptotic freedom and differs from the one-loop
Yang-Mills\ coupling only by a small correction factor $1/11$ which could e.g.
be generated by an anisotropic component for $G_{>}^{-1}$ \cite{dia98,bro98}.
The one-loop integration over the high-momentum modes was found to be reliable
down to $\mu\simeq1.3$ GeV\ \cite{bro99}. In this range, the resulting
renormalized soft-mode action
\begin{equation}
\Gamma_{<}\left[  U_{<}\right]  =\frac{1}{4g^{2}\left(  \mu\right)  }\int
d^{3}x\int d^{3}yL_{<,i}^{a}\left(  \vec{x}\right)  G_{<}^{-1}\left(  \vec
{x}-\vec{y}\right)  L_{<,i}^{a}\left(  \vec{y}\right)  \label{gamless}%
\end{equation}
(cf. Eq. (\ref{gsm})) is obtained from Eq. (\ref{effact}) by replacing the
bare coupling $g_{b}$ with the running coupling $g\left(  \mu\right)  $. One
may further simplify the action (\ref{gamless}) by exploiting the strongly
local support of the IR covariance (\ref{gm1}) (compared to the minimal
soft-mode wavelength $\mu^{-1}$) which manifests itself in the regularized
delta function (\ref{regdel}) and its derivatives. Indeed, this allows to
apply the unitarity constraint for the $U_{<}$ approximately at neighboring
points, i.e.
\begin{equation}
G^{-1}\left(  \vec{x}-\vec{y}\right)  U_{<}^{\dagger}\left(  \vec{x}\right)
U_{<}\left(  \vec{y}\right)  \simeq G^{-1}\left(  \vec{x}-\vec{y}\right)  ,
\label{unapr}%
\end{equation}
which could be systematically improved by Taylor expansion of the slowly
varying soft modes. To leading order, one may therefore replace Eq.
(\ref{gamless}) by the $2U$ contribution%
\begin{equation}
\Gamma_{<}^{\left(  2U\right)  }\left[  U_{<}\right]  =\frac{1}{2g^{2}\left(
\mu\right)  }\int d^{3}x\int d^{3}ytr\left\{  \partial_{i}U_{<}\left(  \vec
{x}\right)  G^{-1}\left(  \vec{x}-\vec{y}\right)  \partial_{i}U_{<}^{\dagger
}\left(  \vec{y}\right)  \right\}  . \label{gapr}%
\end{equation}
Inserting our full expansion (\ref{gm1}) of the generalized IR\ covariance
into the low-momentum dynamics (\ref{gamless}), on the other hand, yields the
complete soft-mode action in the form
\begin{equation}
\Gamma_{<}\left[  U_{<}\right]  =\int d^{3}z\left[  \mathcal{L}_{<,0}\left(
\vec{z}\right)  +\mathcal{L}_{<,c_{1}}\left(  \vec{z}\right)  +\mathcal{L}%
_{<,c_{2}}\left(  \vec{z}\right)  +...\right]  \label{smd}%
\end{equation}
where the $\mathcal{L}_{<,c_{n}}\left(  \vec{z}\right)  $ are (quasi-) local
Lagrangians. (A saddle-point expansion of this dynamics, with the $c_{n}$
determined by the massive vector covariance (\ref{g0}), was employed in Ref.
\cite{for06} to identify gluonic IR\ degrees of freedom.) The leading-order,
two-derivative Lagrangian%
\begin{equation}
\mathcal{L}_{<,0}\left(  \vec{z}\right)  =-\frac{m_{g}}{2g^{2}}tr\left\{
U_{<}^{\dagger}\left(  \vec{z}\right)  \partial_{i}U_{<}\left(  \vec
{z}\right)  U_{<}^{\dagger}\left(  \vec{z}\right)  \partial_{i}U_{<}\left(
\vec{z}\right)  \right\}  =\frac{m_{g}}{2g^{2}}tr\left\{  \partial_{i}%
U_{<}^{\dagger}\partial_{i}U_{<}\right\}  \label{L0}%
\end{equation}
(where we used $U_{<}^{\dagger}U_{<}=1$ and neglected a surface term) is just
the standard nonlinear $\sigma$ model. The $c_{n}\neq0$ corrections (with the
low-momentum constants $c_{n}$ restricted by the requirements of a positive
static action and a bounded vacuum energy, cf. Sec. \ref{gengir}) generate
$2\left(  n+1\right)  $-derivative interactions which are, relative to the
leading term (\ref{L0}), suppressed by $n$ powers of $k^{2}/\mu^{2}$:%
\begin{equation}
\mathcal{L}_{<,c_{n}}\left(  \vec{z}\right)  =\frac{c_{n}m_{g}}{2g^{2}\mu
^{2n}}tr\left\{  U_{<}^{\dagger}\left(  \vec{z}\right)  \partial_{i}%
U_{<}\left(  \vec{z}\right)  \partial^{2n}\left[  \partial_{i}U_{<}^{\dagger
}\left(  \vec{z}\right)  U_{<}\left(  \vec{z}\right)  \right]  \right\}  .
\label{lc}%
\end{equation}
The action (\ref{smd}), based on the Lagrangians (\ref{L0}), (\ref{lc}),
exhibits the dynamics generated both by the generalized IR covariance
(\ref{gm1}) and by gauge projection. It encodes, in particular, information on
the vacuum topology which shows up in the form of instanton \cite{sch98}
(through the Atiyah-Manton holonomy \cite{ati89}), meron \cite{dea76},
monopole \cite{tho76}, Fadeev-Niemi-Cho knot \cite{fad70,fad97} etc.
contributions and can be made explicit \cite{for06}. The approximation
(\ref{unapr}) can be used to reduce the $c_{n}\neq0$ interaction terms to
their bilinear (i.e. $2U$) parts%
\begin{equation}
\mathcal{L}_{<,c_{n}}^{\left(  2U\right)  }\left(  \vec{z}\right)
=\frac{c_{n}m_{g}}{2g^{2}\mu^{2n}}tr\left\{  \partial_{i}U_{<}^{\dagger
}\left(  \vec{z}\right)  \partial^{2n}\left[  \partial_{i}U_{<}\left(  \vec
{z}\right)  \right]  \right\}  , \label{l2}%
\end{equation}
furthermore, which will turn out to generate the dominant contributions to the
vacuum energy density and other matrix elements.

We close this section by recalling that the soft-mode dynamics (\ref{smd})
follows uniquely from the adopted vacuum wave functional and preserves
traceable links between the $U_{<}$\ fields and the underlying gauge fields
\cite{for06}. Nevertheless, it remains reasonably transparent and allows for
efficient and controlled truncations resulting e.g. in the analytical
expressions for the vacuum energy to be derived below. These benefits
originate in large part from reexpressing the dynamics in terms of the
gauge-invariant fields $U$\ which gather collective contributions from whole
gauge-field orbits instead of dealing with each gauge field individually
\cite{for06}.

\section{Soft-mode correlation functions}

\label{ircf}

As outlined in Sec. \ref{strat}, the calculation of the vacuum energy density
requires the evaluation of the reduced matrix element $\left\langle
\left\langle \mathcal{H}\right\rangle \right\rangle $ which contains
correlations functions of the soft-mode fields $U_{<}$. In the following
section we set up our framework for evaluating these correlators.

\subsection{Generating functional}

\label{gfun}

Soft-mode correlation functions, as they appear in the integrand $\left\langle
\left\langle \mathcal{H}\right\rangle \right\rangle $ of the vacuum
expectation value (\ref{hsmvev}) and in the gap equation to be derived below,
are efficiently calculated by means of the generating functional%
\begin{equation}
Z\left[  j,j^{\dagger}\right]  =\int DU_{<}\exp\left[  -\Gamma_{<}\left[
U_{<}\right]  -\int d^{3}ztr\left\{  jU_{<}^{\dagger}+j^{\dagger}%
U_{<}\right\}  \right]  \label{gf}%
\end{equation}
where $j,j^{\dagger}$ are matrix sources. In order to prepare for the
(approximate) integration over the soft modes in Eq. (\ref{gf}), we first
release the unitarity constraint on the $U_{<}$ fields in the usual manner by
inserting a delta functional, i.e.%
\begin{align}
Z\left[  j,j^{\dagger}\right]   &  =\int DV\delta\left[  V^{\dagger
}V-1\right]  \exp\left[  -\Gamma_{<}\left[  V\right]  -\int d^{3}ztr\left\{
jV^{\dagger}+j^{\dagger}V\right\}  \right] \\
&  =\int D\Sigma\int DV\exp\left[  -\Gamma_{<}\left[  V\right]  -\Gamma
_{\Sigma}\left[  V,\Sigma\right]  -\int d^{3}ztr\left\{  jV^{\dagger
}+j^{\dagger}V\right\}  \right]  , \label{zsigv}%
\end{align}
where the Hermitian matrix fields $\Sigma\left(  \vec{x}\right)  $ act as
Lagrange multipliers (with a normalization chosen for later convenience) and
where $\Gamma_{\Sigma}\left[  V,\Sigma\right]  $ contains the interactions
between the $\Sigma$ and $V$ fields,
\begin{equation}
\Gamma_{\Sigma}\left[  V,\Sigma\right]  =\int d^{3}x\mathcal{L}_{\Sigma
}\left(  \vec{z}\right)  =\frac{m_{g}}{2g^{2}}\int d^{3}xtr\left[
\Sigma\left(  V^{\dagger}V-1\right)  \right]  . \label{gamsigm}%
\end{equation}
The integral over the unconstrained complex matrices $V$ has a linear
Euclidean measure $DV$. Hence, after pulling out the subleading (cf. Sec.
\ref{sma}), non-Gaussian $4U$ interactions as functional derivatives with
respect to the sources, it can be performed analytically. This results in%
\begin{align}
Z\left[  j,j^{\dagger}\right]   &  =\int D\Sigma DV\exp\left\{  -\int
d^{3}z\left[  \mathcal{L}_{<,0}+\mathcal{L}_{\Sigma}+\mathcal{L}_{<,c_{1}%
}+\mathcal{L}_{<,c_{2}}+...+tr\left\{  jV^{\dagger}+j^{\dagger}V\right\}
\right]  \right\} \\
&  =\int D\Sigma DV\exp\left[  -\int d^{3}z\left(  \mathcal{L}_{<,c_{1}%
}^{\left(  4U\right)  }+\mathcal{L}_{<,c_{2}}^{\left(  4U\right)
}+...\right)  \right]  \exp\left[  -\int d^{3}z\left(  \mathcal{L}^{\left(
2U\right)  }+tr\left\{  jV^{\dagger}+j^{\dagger}V\right\}  \right)  \right] \\
&  =\mathcal{V}^{\left(  4U\right)  }\left[  \frac{\delta}{\delta j}%
,\frac{\delta}{\delta j^{\dagger}}\right]  Z^{\left(  2U\right)  }\left[
j,j^{\dagger}\right]  , \label{z3}%
\end{align}
where we have gathered the part of the generating functional which originates
from the bilinear Lagrangian
\begin{equation}
\mathcal{L}^{\left(  2U\right)  }:=\mathcal{L}_{<,0}+\mathcal{L}_{<,c_{1}%
}^{\left(  2U\right)  }+\mathcal{L}_{<,c_{2}}^{\left(  2U\right)
}+...+\mathcal{L}_{\Sigma}=:tr\left\{  V^{\dagger}\Delta V\right\}
-\frac{m_{g}}{2g^{2}}tr\left\{  \Sigma\right\}
\end{equation}
with the kernel
\begin{equation}
\Delta\left(  \vec{x}-\vec{y};\left\{  c_{n}\right\}  ,\Sigma\right)
:=-\frac{m_{g}}{2g^{2}}\left(  \partial_{x}^{2}+\frac{c_{1}}{\mu^{2}}%
\partial_{x}^{4}+\frac{c_{2}}{\mu^{4}}\partial_{x}^{6}+...-\Sigma\right)
\delta_{<}^{3}\left(  \vec{x}-\vec{y}\right)  \label{smpr}%
\end{equation}
(which defines the inverse soft-mode propagator) as%
\begin{equation}
Z^{\left(  2U\right)  }\left[  j,j^{\dagger}\right]  =\int D\Sigma
DV\exp\left[  -\int d^{3}z\left(  \mathcal{L}^{\left(  2U\right)  }+tr\left\{
jV^{\dagger}+j^{\dagger}V\right\}  \right)  \right]  . \label{z2u}%
\end{equation}
All contributions from the $4U$ vertices, on the other hand, are collected in
the functional potential%
\begin{equation}
\mathcal{V}^{\left(  4U\right)  }\left[  U^{\dagger},U\right]  =\exp\left[
-\int d^{3}z\left(  \mathcal{L}_{<,c_{1}}^{\left(  4U\right)  }+\mathcal{L}%
_{<,c_{2}}^{\left(  4U\right)  }+...\right)  \right]  .
\end{equation}
The Gaussian integral over the $V$ fields in Eq. (\ref{z2u}) can then be
performed analytically, with the result
\begin{equation}
\frac{Z^{\left(  2U\right)  }\left[  j,j^{\dagger}\right]  }{Z^{\left(
2U\right)  }\left[  0,0\right]  }=\int D\Sigma\exp\left[  \int d^{3}x\int
d^{3}ytr\left\{  j^{\dagger}\left(  \vec{x}\right)  \Delta^{-1}\left(  \vec
{x}-\vec{y};\left\{  c_{n}\right\}  ,\Sigma\right)  j\left(  \vec{y}\right)
\right\}  \right]  . \label{z2}%
\end{equation}
The whole $\Sigma$ dependence of the correlators, as well as the $c_{n}$
dependence originating from the $2U$ interactions, is now concentrated in the
gauge-invariant soft-mode propagator $\Delta^{-1}$ which we will analyze in
Sec. \ref{smp}. The perturbative treatment of the $4U$ vertices for
$\left\vert c_{n}\right\vert \ll1$ starts from the expansion
\begin{align}
\mathcal{V}^{\left(  4U\right)  }\left[  \frac{\delta}{\delta j},\frac{\delta
}{\delta j^{\dagger}}\right]   &  \equiv\exp\left[  -\int d^{3}z\mathcal{L}%
_{<,c_{1}}^{\left(  4U\right)  }\left(  \frac{\delta}{\delta j},\frac{\delta
}{\delta j^{\dagger}}\right)  -...\right] \\
&  =1-\int d^{3}z\mathcal{L}_{<,c_{1}}^{\left(  4U\right)  }\left(
\frac{\delta}{\delta j},\frac{\delta}{\delta j^{\dagger}}\right)  +O\left(
c_{1}^{2},c_{n\geq2}\right)  \label{vapr}%
\end{align}
of the functional potential which contains the $c_{n}$ dependence induced by
the $4U$ interactions.

\subsection{Mean-field approximation and vacuum phases}

\label{mfa}

After having rewritten the generating functional as an integral over the
auxiliary field $\Sigma$, we evaluate the latter in the saddle-point or
mean-field approximation (cf. Appendix \ref{spint}). When the integrand in Eq.
(\ref{zsigv}) is expressed as a Boltzmann factor%
\begin{equation}
\exp\left\{  -\tilde{\Gamma}\left[  \Sigma\right]  \right\}  :=\int
DV\exp\left\{  -\Gamma_{<}\left[  V\right]  -\Gamma_{\Sigma}\left[
V,\Sigma\right]  \right\}
\end{equation}
(in the present section we are interested only in the vacuum overlap amplitude
and therefore set all sources to zero), the saddle-point equation takes the
form
\begin{align}
\frac{\delta\tilde{\Gamma}\left[  \Sigma\right]  }{\delta\Sigma\left(  \vec
{x}\right)  }  &  =\exp\tilde{\Gamma}\left[  \Sigma\right]  \int
DV\frac{\delta\Gamma_{\Sigma}\left[  V,\Sigma\right]  }{\delta\Sigma\left(
\vec{x}\right)  }\exp\left\{  -\Gamma_{<}\left[  V\right]  -\Gamma_{\Sigma
}\left[  V,\Sigma\right]  \right\} \\
&  =\frac{m_{g}}{2g^{2}}\exp\tilde{\Gamma}\left[  \Sigma\right]  \int
DV\left[  V^{\dagger}\left(  \vec{x}\right)  V\left(  \vec{x}\right)
-1\right]  \exp\left\{  -\Gamma_{<}\left[  V\right]  -\Gamma_{\Sigma}\left[
V,\Sigma\right]  \right\}  =0 \label{ge}%
\end{align}
which ensures that its solutions extremize the effective action $\tilde
{\Gamma}$. To leading order in the saddle-point expansion, integrals over
$\Sigma$ are then approximated by their integrands where $\Sigma$ is replaced
by a solution $\Sigma^{\left(  \text{mf}\right)  }$ of Eq. (\ref{ge}). This
turns Eq. (\ref{ge}) into%
\begin{equation}
\left\langle U^{\dagger}\left(  \vec{x}\right)  U\left(  \vec{x}\right)
\right\rangle =1, \label{gapeq}%
\end{equation}
in particular, and thereby restores the unitarity constraint at the mean-field
level. In a homogeneous vacuum one expects the solution of Eq. (\ref{gapeq})
to be a constant field.\ The interactions (\ref{gamsigm}) then turn into a
mass term for $V$ which triggers dimensional transmutation (as in $O\left(
n\right)  $ models) and generates a mass gap. Hence Eq.\ (\ref{gapeq}) plays
the role of a gap equation. Since for $\Sigma>0$ the\ SU$_{L}(N_{c})\times$
SU$_{R}(N_{c})$\ symmetry of the soft-mode action (\ref{smd}) is unbroken, the
mean-field solution should further be proportional to the unit matrix, i.e.
\begin{equation}
\Sigma^{\left(  \text{mf}\right)  }=\bar{\Sigma}\times1, \label{mfdef}%
\end{equation}
where $\bar{\Sigma}$ is a real constant.

In order to exhibit the vacuum phase structure encoded in the wave functional
(\ref{ginvvwf}) more fully, an analogous mean-field treatment of the $V$
integral in Eq. (\ref{zsigv}) (which reproduces its Gaussian part exactly)
turns out to be useful as well. Defining the associated effective action
(again for vanishing sources) by%
\begin{equation}
\exp\left\{  -\tilde{\Gamma}\left[  V\right]  \right\}  :=\int D\Sigma
\exp\left\{  -\Gamma_{<}\left[  V\right]  -\Gamma_{\Sigma}\left[
V,\Sigma\right]  \right\}
\end{equation}
one obtains the corresponding saddle-point equation%
\begin{align}
\frac{\delta\tilde{\Gamma}\left[  V\right]  }{\delta V^{\dagger}\left(
\vec{x}\right)  }  &  =\exp\tilde{\Gamma}\left[  V\right]  \int D\Sigma
\frac{\delta\left\{  \Gamma_{<}\left[  V\right]  +\Gamma_{\Sigma}\left[
V,\Sigma\right]  \right\}  }{\delta V^{\dagger}\left(  \vec{x}\right)  }%
\exp\left\{  -\Gamma_{<}\left[  V\right]  -\Gamma_{\Sigma}\left[
V,\Sigma\right]  \right\} \\
&  \rightarrow\frac{m_{g}}{2g^{2}}\exp\tilde{\Gamma}\left[  \bar{V}\right]
\left[  \bar{\Sigma}\bar{V}\right]  \exp\left\{  -\Gamma_{\Sigma}\left[
\bar{V},\bar{\Sigma}\right]  \right\}  =0. \label{m2}%
\end{align}
In the second line we have saturated the $\Sigma$ integral with the constant
(i.e. vacuum) mean fields $\bar{V}$ and $\bar{\Sigma}$. This implies
$\Gamma_{<}\left[  \bar{V}\right]  =0$, in particular, since $\Gamma_{<}$ only
contains derivative interactions. Hence Eq. (\ref{m2}) reduces to%
\begin{equation}
\left\langle \Sigma U\left(  \vec{x}\right)  \right\rangle =0. \label{2mfe}%
\end{equation}
(The SU$_{L}(N_{c})\times$ SU$_{R}(N_{c})$\ symmetry of $\Gamma$ (under which
$U,V$ transform as $V\rightarrow RVL^{\dagger}$) turns a nonzero mean field
$\bar{V}$ into a continuous family of degenerate saddle points. The associated
Goldstone zero-mode contributions are not suppressed by the Gaussian weight
and have to be integrated exactly.)

The solutions of the saddle-point equations\ (\ref{gapeq}) and (\ref{2mfe})
characterize the vacuum phases of the effective $\sigma$ model (\ref{smd}).
The gap equation\ (\ref{gapeq}) determines the auxiliary mean field
$\bar{\Sigma}$ and will be solved in Sec. \ref{gap}. The saddle-point equation
(\ref{2mfe}) shows that the vacuum can exist in two phases, as expected on
general grounds and confirmed by lattice simulations \cite{kog82,bro90} and
the $\varepsilon$-expansion \cite{pis84}. More specifically, with increasing
\textquotedblleft analog temperature\textquotedblright\ $g^{2}(\mu)$ or
decreasing $\mu$ one expects the vacuum to pass through an order-disorder
phase transition, as in the analogous statistical spin model. In the ordered
low-temperature (i.e. weakly coupled) phase one has
\begin{equation}
\left\langle U\right\rangle \neq0,\text{ \ \ \ \ }\left\langle \Sigma
\right\rangle =0,
\end{equation}
i.e. the SU$_{L}(N_{c})\times$ SU$_{R}(N_{c})$\ symmetry is broken to its
diagonal subgroup and $N_{c}^{2}-1$ massless Goldstone bosons are generated.
In the disordered high-temperature (strong-coupling) phase with
\begin{equation}
\left\langle U\right\rangle =0,\text{ \ \ \ \ }\left\langle \Sigma
\right\rangle \neq0,
\end{equation}
on the other hand, the\ symmetry of the action is restored \footnote{In the
weakly-coupled phase (RG-improved) perturbation theory should work, while the
mean-field approximation becomes increasingly reliable with growing coupling
in the strongly coupled phase (at least as long as the mass gap is large
enough to effectively suppress fluctuations). It is well known, however,\ that
the above saddle-point approximation does not become exact at large $N_{c}$
\cite{pol87}. An improved saddle-point expansion could be developed if a
practicable analog of the SU$\left(  2\right)  $ quaternion representation
could be found for $N_{c}\geq3$ \cite{dia98}. However, for the value $N_{c}=3$
of interest in the context of QCD the ensuing large-$N_{c}$ suppression of
deviations from the mean field might still be rather weak.}. The
disorder-order transition occurs when $\left\langle \Sigma\right\rangle $
reaches zero in the disordered phase. As usual, it is the result of a
competition between the ordering tendency of the energy and the disordering
propensity of the entropy. Our above arguments imply, furthermore, that the
\emph{qualitative} phase structure is independent of the detailed interactions
(\ref{lc}) as long as the soft-mode dynamics contains only derivative
interactions. (This is guaranteed by the unitarity of the $U_{<}$ fields.) As
a consequence of $\Gamma_{<}\left[  \bar{V}\right]  =0$, in particular, the
mean-field equations and hence the qualitative phase structure have no
\emph{explicit} dependence on the low-momentum constants $c_{n}$.

\subsection{Soft-mode propagator}

\label{smp}

The static soft-mode propagator $\Delta^{-1}$ determines the Gaussian part
(\ref{z2}) of the generating functional and thereby encapsulates most of the
impact of the higher-derivative interactions on vacuum structure and
amplitudes. From the Fourier transform
\begin{equation}
\Delta\left(  \vec{x}-\vec{y}\right)  =\frac{m_{g}}{2g^{2}}\int\frac{d^{3}%
k}{\left(  2\pi\right)  ^{3}}\theta\left(  \mu^{2}-k^{2}\right)  \left(
k^{2}-\frac{c_{1}}{\mu^{2}}k^{4}+\frac{c_{2}}{\mu^{4}}k^{6}-...+\bar{\Sigma
}\right)  e^{i\vec{k}\left(  \vec{x}-\vec{y}\right)  }%
\end{equation}
of its inverse (\ref{smpr}), which we have specialized to the constant mean
field (\ref{mfdef}), the propagator is obtained as the solution of%
\begin{equation}
\int d^{3}z\Delta^{-1}\left(  \vec{x}-\vec{z}\right)  \Delta\left(  \vec
{z}-\vec{y}\right)  =\delta_{<}^{3}\left(  \vec{x}-\vec{y}\right)
\end{equation}
where $\delta_{<}^{3}$ is the regularized delta function (\ref{regdel}). In
terms of the dimensionless variables%
\begin{equation}
\vec{\kappa}\equiv\frac{\vec{k}}{\mu},\text{ \ \ \ \ \ }\xi\equiv\frac
{\sqrt{\bar{\Sigma}}}{\mu}%
\end{equation}
one therefore has
\begin{equation}
\Delta^{-1}\left(  \vec{x}-\vec{y}\right)  =\int\frac{d^{3}\kappa}{\left(
2\pi\right)  ^{3}}e^{i\mu\vec{\kappa}\left(  \vec{x}-\vec{y}\right)  }%
\Delta^{-1}\left(  \kappa\right)  \label{d}%
\end{equation}
with%
\begin{equation}
\Delta^{-1}\left(  \kappa\right)  =\frac{2g^{2}\mu}{m_{g}}\frac{\theta\left(
1-\kappa^{2}\right)  }{\kappa^{2}+\xi^{2}+\mathcal{M}^{2}\left(  \kappa
^{2}\right)  }. \label{dm1(k)}%
\end{equation}
The Fourier representation of the soft-mode propagator can be simplified by
performing the angular integrals analytically,
\begin{equation}
\Delta^{-1}\left(  \vec{x}-\vec{y}\right)  =\frac{g^{2}}{\pi^{2}m_{g}}\frac
{1}{\left\vert x-y\right\vert }\int_{0}^{1}d\kappa\frac{\kappa\sin\left(
\kappa\mu\left\vert x-y\right\vert \right)  }{\kappa^{2}+\xi^{2}%
+\mathcal{M}^{2}\left(  \kappa^{2}\right)  },
\end{equation}
which becomes useful for numerical implementations and demonstrates that the
isotropic regularization (i.e. $\kappa\leq1$) preserves
\begin{equation}
\Delta^{-1}\left(  \vec{x}-\vec{y}\right)  =\Delta^{-1}\left(  \vec{y}-\vec
{x}\right)  . \label{sym}%
\end{equation}

Equation (\ref{dm1(k)}) reveals that the higher-derivative interactions
generate a momentum-dependent selfenergy%
\begin{equation}
\mathcal{M}^{2}\left(  \kappa^{2}\right)  =-c_{1}\kappa^{4}+c_{2}\kappa
^{6}-c_{3}\kappa^{8}+...
\end{equation}
for the soft modes. Hence the propagator (\ref{dm1(k)}) may be regarded as the
resummation of a static self-energy (i.e. $\mathcal{M}^{2}$) insertion. This
geometric series converges for $\left\vert \mathcal{M}^{2}\left(  \kappa
^{2}\right)  /\left(  \kappa^{2}+\xi^{2}\right)  \right\vert <1$. Expecting
the $c_{1}$ contribution to dominate at small momenta $k\ll\mu$ and therefore
truncating to $c_{n\geq2}=0$ (see below), this implies%
\begin{equation}
\left\vert c_{1}\right\vert <\frac{\kappa^{2}+\xi^{2}}{\kappa^{4}}.
\end{equation}
For $\kappa\in\left[  0,1\right]  $ and $\xi\geq0$, the above inequality is
satisfied as long as $\left\vert c_{1}\right\vert <1$. This does not
additionally constrain the parameter space, however, since $c_{1}<1$
guarantees the normalizability of the vacuum wave functional (cf. Eq.
(\ref{cbnds})) and $\left\vert c_{1}\right\vert \ll1$ ensures the validity of
our perturbative $O\left(  c_{1}\right)  $ treatment (\ref{vapr}) of the
residual\ $4U$ interactions.

In order to analyze the singularity structure of the soft-mode propagator, we
rewrite Eq. (\ref{dm1(k)}) for $c_{n\geq2}=0$ as%
\begin{equation}
\Delta^{-1}\left(  \kappa\right)  =\frac{2g^{2}\mu}{m_{g}}\frac{\theta\left(
1-\kappa^{2}\right)  }{\kappa^{2}\left(  1-c_{1}\kappa^{2}\right)  +\xi^{2}%
}=\frac{2g^{2}\mu}{m_{g}}\frac{\theta\left(  1-\kappa^{2}\right)  }%
{\sqrt{1+4c_{1}\xi^{2}}}\left(  \frac{1}{\kappa^{2}-\kappa_{2}^{2}}-\frac
{1}{\kappa^{2}-\kappa_{1}^{2}}\right)
\end{equation}
which reveals two poles at the momenta%
\begin{equation}
\kappa_{1,2}^{2}\left(  \xi,c_{1}\right)  =\frac{1}{2c_{1}}\left(  1\pm
\sqrt{1+4c_{1}\xi^{2}}\right)  . \label{poles}%
\end{equation}
The pole$\allowbreak$ position $\kappa_{1}^{2}$ (with the positive sign in Eq.
(\ref{poles})) diverges for $c_{1}\rightarrow0$ but decreases monotonically
for $c\rightarrow1$ to reach $\kappa_{1}^{2}\left(  \xi,c_{1}=1\right)  \geq
1$. Hence the pole at $\kappa_{1}^{2}$ lies outside of the integration range
$\kappa\in\left[  0,1\right]  $ for all $c_{1}\in\left[  -\infty,1\right]  $
(and even for $\xi\rightarrow0$). The second pole at $\kappa_{2}^{2}$ lies
inside the integration range for $c_{1}>-\infty$ and $\xi=0$. For $c_{1}=0$,
in particular, it turns into the standard infrared pole of the $c_{n}=0$
propagator, showing that the momentum-dependent selfenergy does not create
singularities beyond the $\xi\rightarrow0$ pole of the uncorrected soft-mode propagator.

\section{Evaluation of the 2- and 4-point correlators for $c_{n\geq2}=0$}

\label{excor}

Having established our calculational framework for the generating functional,
we now derive explicit expressions for the soft modes' two- and four-point
functions. The results will underlie our subsequent evaluation of the matrix
elements $\left\langle U_{<}^{\dagger}\left(  \vec{x}\right)  U_{<}\left(
\vec{x}\right)  \right\rangle $ in the gap equation (\ref{gapeq}) and
$\left\langle L_{<,i}^{a}\left(  \vec{x}\right)  L_{<,i}^{a}\left(  \vec
{y}\right)  \right\rangle $ in the chromo-electric contribution (\ref{t}) to
the vacuum energy density.

As stated previously, among the contributions from the higher-derivative
interactions (\ref{lc}) those associated with $c_{1}$ are expected to dominate
at small momenta $k^{2}\ll\mu^{2}$. This expectation is further supported by
evidence from Ref. \cite{for06} where relevant infrared physics was found to
be captured by the truncation of the covariance expansion (\ref{gm1}) to
$c_{n\geq2}=0$. (More specifically, the first correction term $\mathcal{L}%
_{<,c_{1}}$ turned out to reproduce the full (nonlocal) IR action
(\ref{gamless}) at its saddle points with better than 10\% accuracy.)
Furthermore, the $c_{n\geq2}$ corrections (although straightforward to
implement) obscure the (e.g. graphical) analysis of the vacuum energy density
and the interpretation of other amplitudes. Hence we will restrict the
remainder of our investigation to the $c_{1}$ corrections, associated with the
four-gradient interactions%
\begin{align}
\mathcal{L}_{<,c_{1}}\left(  \vec{z}\right)   &  =\frac{c_{1}m_{g}}{2g^{2}%
\mu^{2}}tr\left\{  U_{<}^{\dagger}\left(  \vec{z}\right)  \partial_{i}%
U_{<}\left(  \vec{z}\right)  \partial^{2}\left[  \partial_{i}U_{<}^{\dagger
}\left(  \vec{z}\right)  U_{<}\left(  \vec{z}\right)  \right]  \right\}
\label{lc1}\\
&  =\mathcal{L}_{<,c_{1}}^{\left(  2U\right)  }\left(  \vec{z}\right)
+\mathcal{L}_{<,c_{1}}^{\left(  4U\right)  }\left(  \vec{z}\right)
\end{align}
(where derivatives are implied to act on the nearest field only) which we have
split as in Sec. \ref{sma} into the dominant $2U$ interaction term
$\mathcal{L}_{<,c_{1}}^{\left(  2U\right)  }$, given by Eq. (\ref{l2}) for
$n=1$, and the $4U$ contribution%
\begin{equation}
\mathcal{L}_{<,c_{1}}^{\left(  4U\right)  }\left(  \vec{z}\right)
=\frac{c_{1}m_{g}}{2g^{2}\mu^{2}}tr\left\{  U_{<}^{\dagger}\partial_{i}%
U_{<}\left[  2\partial_{i}\partial_{k}U_{<}^{\dagger}\partial_{k}%
U_{<}+\partial_{i}U_{<}^{\dagger}\partial^{2}U_{<}\right]  \right\}  .
\label{l4u}%
\end{equation}
As discussed in Sec. \ref{gfun}, we will evaluate the $2U$ contributions to
the relevant $n$-point functions exactly and treat the residual contributions
from the $4U$ vertices perturbatively \footnote{Alternatively, one could trade
the $4U$ interactions for another composite field of the form $\chi
_{ik}=\left(  \partial_{i}\partial_{k}U^{\dagger}\right)  U$ and treat its
contributions at the mean-field level.} to $O\left(  c_{1}\right)  $. The
latter has to be justified \textit{a posteriori}, by showing that the
variational results favor small enough coupling values $\left\vert c_{1}%
^{\ast}\right\vert \ll1$ (cf. Sec. \ref{res}).

\subsection{The 2-point function}

\label{2pf}

In this section we evaluate the 2-point soft-mode correlator (for $c_{i\geq
2}=0$) which simultaneously provides the basis for the calculation of the
4-point correlator in the next section. Since we are working to $O\left(
c_{1}\right)  $ in the $4U$ interactions, we will only keep one insertion of
the $\mathcal{L}_{<,c_{1}}^{\left(  4U\right)  }$ vertex in the functional
potential operator (\ref{vapr}), which thus reduces to%
\begin{equation}
\mathcal{V}^{\left(  4U,c_{1}\right)  }\left[  \frac{\delta}{\delta j}%
,\frac{\delta}{\delta j^{\dagger}}\right]  =1-\frac{c_{1}m_{g}}{2g^{2}\mu^{2}%
}\int d^{3}z\partial^{2}\left[  \frac{\partial_{i}\delta}{\delta j_{MN}\left(
\vec{z}\right)  }\frac{\delta}{\delta j_{PM}^{\dagger}\left(  \vec{z}\right)
}\right]  \frac{\delta}{\delta j_{QP}\left(  \vec{z}\right)  }\frac
{\partial_{i}\delta}{\delta j_{NQ}^{\dagger}\left(  \vec{z}\right)  }%
\end{equation}
(where the notation $\left(  \partial_{i}\delta\right)  /\delta j\left(
\vec{z}\right)  $ implies that the partial derivative is acting \emph{only} on
the result of the associated functional derivative). The 2-point function is
then obtained by taking derivatives of the generating functional (\ref{z3})
with respect to the matrix sources $j$ and $j^{\dagger},$ i.e.
\begin{equation}
\left\langle U_{<,AB}^{\dagger}\left(  \vec{x}\right)  U_{<,CD}\left(  \vec
{y}\right)  \right\rangle =\left.  \frac{-\delta}{\delta j_{BA}\left(  \vec
{x}\right)  }\frac{-\delta}{\delta j_{DC}^{\dagger}\left(  \vec{y}\right)
}\mathcal{V}^{\left(  4U,c_{1}\right)  }\left[  \frac{\delta}{\delta j}%
,\frac{\delta}{\delta j^{\dagger}}\right]  \frac{Z^{\left(  2U\right)
}\left[  j,j^{\dagger}\right]  }{Z^{\left(  2U\right)  }\left[  0,0\right]
}\right\vert _{j,j^{\dagger}=0}, \label{u2}%
\end{equation}
where the superscript $\left(  2U\right)  $ indicates as in Eq. (\ref{z2u})
that the corresponding quantity is evaluated by using only the $2U$
contribution to the full soft-mode action. Eq. (\ref{u2}) can then be
rewritten as%
\begin{align}
\left\langle U_{<,AB}^{\dagger}\left(  \vec{x}\right)  U_{<,CD}\left(  \vec
{y}\right)  \right\rangle  &  =\left\langle U_{<,AB}^{\dagger}\left(  \vec
{x}\right)  U_{<,CD}\left(  \vec{y}\right)  \right\rangle ^{\left(  2U\right)
}\nonumber\\
&  -\left\langle U_{<,AB}^{\dagger}\left(  \vec{x}\right)  \int d^{3}%
z\mathcal{L}_{<,c_{1}}^{\left(  4U\right)  }\left[  U_{<,}^{\dagger}\left(
\vec{z}\right)  ,U_{<,}\left(  \vec{z}\right)  \right]  U_{<,CD}\left(
\vec{y}\right)  \right\rangle ^{\left(  2U\right)  }%
\end{align}
where%
\begin{equation}
\left\langle U_{<,AB}^{\dagger}\left(  \vec{x}\right)  U_{<,CD}\left(  \vec
{y}\right)  \right\rangle ^{\left(  2U\right)  }=\left.  \frac{\delta^{2}%
}{\delta j_{BA}\left(  \vec{x}\right)  \delta j_{DC}^{\dagger}\left(  \vec
{y}\right)  }\frac{Z^{\left(  2U\right)  }\left[  j,j^{\dagger}\right]
}{Z^{\left(  2U\right)  }\left[  0,0\right]  }\right\vert _{j,j^{\dagger}%
=0}=\delta_{AD}\delta_{BC}\Delta^{-1}\left(  \vec{y},\vec{x}\right)
\end{equation}
is determined by the $2U$ part of the higher-derivative interactions only,
while the $4U$ dynamics generates the perturbative $O\left(  c_{1}\right)  $
correction
\begin{equation}
\left\langle U_{<,AB}^{\dagger}\left(  \vec{x}\right)  \int d^{3}%
z\mathcal{L}_{<,c_{1}}^{\left(  4U\right)  }\left(  \vec{z}\right)
U_{<,CD}\left(  \vec{y}\right)  \right\rangle ^{\left(  2U\right)  }%
=\frac{c_{1}m_{g}}{2g^{2}\mu^{2}}\left.  \frac{\delta^{2}}{\delta
j_{BA}\left(  \vec{x}\right)  \delta j_{DC}^{\dagger}\left(  \vec{y}\right)
}v\left[  j,j^{\dagger}\right]  \right\vert _{j,j^{\dagger}=0} \label{uc}%
\end{equation}
which is due to the insertion of the vertex\
\begin{align}
v\left[  j,j^{\dagger}\right]   &  :=\frac{2g^{2}\mu^{2}}{c_{1}m_{g}}\int
d^{3}z\mathcal{L}_{<,c_{1}}^{\left(  4U\right)  }\left(  \frac{\delta}{\delta
j},\frac{\delta}{\delta j^{\dagger}}\right)  \frac{Z^{\left(  2U\right)
}\left[  j,j^{\dagger}\right]  }{Z^{\left(  2U\right)  }\left[  0,0\right]
}\\
&  =\int d^{3}z\partial^{2}\left[  \frac{\partial_{i}\delta}{\delta
j_{MN}\left(  \vec{z}\right)  }\frac{\delta}{\delta j_{PM}^{\dagger}\left(
\vec{z}\right)  }\right]  \frac{\delta}{\delta j_{QP}\left(  \vec{z}\right)
}\frac{\partial_{i}\delta}{\delta j_{NQ}^{\dagger}\left(  \vec{z}\right)
}\frac{Z^{\left(  2U\right)  }\left[  j,j^{\dagger}\right]  }{Z^{\left(
2U\right)  }\left[  0,0\right]  } \label{v}%
\end{align}
(where $\partial^{2}$ acts only on the square bracket).

After evaluating the functional derivatives in Eq. (\ref{v}), the
contributions to the vertex $v$ can be grouped according to the number of
included source fields as
\begin{equation}
v\left[  j,j^{\dagger}\right]  =\frac{Z^{\left(  2U\right)  }\left[
j,j^{\dagger}\right]  }{Z^{\left(  2U\right)  }\left[  0,0\right]  }\left(
v_{d}+v_{jj^{\dagger}}\left[  j,j^{\dagger}\right]  +v_{jj^{\dagger
}jj^{\dagger}}\left[  j,j^{\dagger}\right]  \right)  .
\end{equation}
The constant $v_{d}=v\left[  0,0\right]  $ is the IR-divergent, disconnected
part which originates from the \textquotedblleft8\textquotedblright%
\ vacuum-bubble diagram, with all four legs of the vertex $v\left[
j,j^{\dagger}\right]  $ closed among themselves, and reads
\begin{equation}
v_{d}=-N^{3}\mu^{6}\left(  \frac{g^{2}}{\pi^{2}m_{g}}\right)  ^{2}%
\tilde{\imath}_{2}^{2}\left(  \xi,c_{1}\right)  \left(  2\pi\right)
^{3}\delta^{3}\left(  0\right)  \label{vdisc}%
\end{equation}
(the integral $\tilde{\imath}_{2}\left(  \xi,c_{1}\right)  $ is defined in
App. \ref{int}). More generally, disconnected contributions to the $n$-point
functions arise when all functional derivatives associated with external lines
hit the $Z^{\left(  2U\right)  }\left[  j,j^{\dagger}\right]  v_{d}$ part of
$v\left[  j,j^{\dagger}\right]  $. The resulting products of free Green
functions with $v_{d}$ will play no role in our following discussion. All
remaining terms require one (two) pair(s) $\left(  \delta/\delta j\right)
\left(  \delta/\delta j^{\dagger}\right)  $ to hit $v_{jj^{\dagger}}$
($v_{jj^{\dagger}jj^{\dagger}}$) and thus contribute exclusively to the\emph{
connected} 2-(4-)point function. The 2-line connected part of the vertex thus
becomes
\begin{equation}
v_{jj^{\dagger}}\left[  j,j^{\dagger}\right]  =-\frac{g^{2}N\mu^{3}}{\pi
^{2}m_{g}}\tilde{\imath}_{2}\left(  \xi,c_{1}\right)  \int d^{3}z\left[
-j_{MN}^{\dagger}\Delta^{-1}\partial^{2}\Delta^{-1}j_{NM}+\left(  \partial
_{i}\Delta^{-1}j_{MN}\right)  j_{NM}^{\dagger}\Delta^{-1}\overleftarrow
{\partial}_{i}\right]
\end{equation}
(where the $\partial_{i}$ ($\overleftarrow{\partial}_{i}$) act on the first
(second) argument of $\Delta^{-1}$ and integrals over the arguments of
$\Delta^{-1}$ folded with a source are implied but not written explicitly).
The 4-line connected part, finally, is
\begin{align}
v_{jj^{\dagger}jj^{\dagger}}\left[  j,j^{\dagger}\right]   &  =\int
d^{3}z\left(  \partial_{i}\Delta^{-1}j_{QN}\right)  j_{PQ}^{\dagger}%
\Delta^{-1}\times\nonumber\\
&  \times\left[  2\left(  \partial_{j}\Delta^{-1}j_{MP}\right)  \left(
j_{NM}^{\dagger}\Delta^{-1}\overleftarrow{\partial}_{i}\overleftarrow
{\partial}_{j}\right)  +\left(  \partial^{2}\Delta^{-1}j_{MP}\right)  \left(
j_{NM}^{\dagger}\Delta^{-1}\overleftarrow{\partial}_{i}\right)  \right]
\end{align}
and contributes only to $n\geq4$ point functions. (In both of the above
expressions the \textquotedblleft open\textquotedblright\ arguments\ (i.e.
those not integrated over) are always the vertex coordinates $\vec{z}$ with
$\partial_{i}\equiv\partial/\partial z_{i}$.)

We now proceed with the calculation of the $2$-point (and $2n$-point)
functions by evaluating the $4U$-vertex-induced functional
\begin{equation}
\Pi_{ABCD}^{\left(  2\right)  }\left[  j,j^{\dagger}\right]  \left(  \vec
{x},\vec{y}\right)  :=\frac{-\delta}{\delta j_{BA}\left(  \vec{x}\right)
}\frac{-\delta}{\delta j_{DC}^{\dagger}\left(  \vec{y}\right)  }v\left[
j,j^{\dagger}\right]  =:\frac{Z^{\left(  2U\right)  }\left[  j,j^{\dagger
}\right]  }{Z^{\left(  2U\right)  }\left[  0,0\right]  }\bar{\Pi}%
_{ABCD}^{\left(  2\right)  }\left[  j,j^{\dagger}\right]  \left(  \vec{x}%
,\vec{y}\right)  \label{p2}%
\end{equation}
which yields%
\begin{align}
\bar{\Pi}_{ABCD}^{\left(  2\right)  }\left[  j,j^{\dagger}\right]  \left(
\vec{x},\vec{y}\right)   &  =\delta_{AD}\Delta_{CB}^{-1}\left(  \vec{y}%
-\vec{x}\right)  \left(  v_{jj^{\dagger}}+v_{jj^{\dagger}jj^{\dagger}}\right)
\nonumber\\
&  +\int d^{3}z^{^{\prime\prime}}j_{AB}^{\dagger}\left(  \vec{z}^{\prime
\prime}\right)  \Delta^{-1}\left(  \vec{z}^{^{\prime\prime}}-\vec{x}\right)
\int d^{3}z^{\prime}\Delta^{-1}\left(  \vec{y}-\vec{z}^{\prime}\right)
j_{CD}\left(  \vec{z}^{\prime}\right)  \left(  v_{jj^{\dagger}}+v_{jj^{\dagger
}jj^{\dagger}}\right) \nonumber\\
&  +\int d^{3}z^{\prime}\Delta^{-1}\left(  \vec{y}-\vec{z}^{\prime}\right)
j_{CD}\left(  \vec{z}^{\prime}\right)  \frac{\delta\left(  v_{jj^{\dagger}%
}\left[  j,j^{\dagger}\right]  +v_{jj^{\dagger}jj^{\dagger}}\left[
j,j^{\dagger}\right]  \right)  }{\delta j_{BA}\left(  \vec{x}\right)
}\nonumber\\
&  +\int d^{3}z^{^{\prime\prime}}j_{AB}^{\dagger}\left(  \vec{z}^{\prime
\prime}\right)  \Delta^{-1}\left(  \vec{z}^{^{\prime\prime}}-\vec{x}\right)
\frac{\delta\left(  v_{jj^{\dagger}}\left[  j,j^{\dagger}\right]
+v_{jj^{\dagger}jj^{\dagger}}\left[  j,j^{\dagger}\right]  \right)  }{\delta
j_{DC}^{\dagger}\left(  \vec{y}\right)  }\nonumber\\
&  +\frac{\delta^{2}\left(  v_{jj^{\dagger}}\left[  j,j^{\dagger}\right]
+v_{jj^{\dagger}jj^{\dagger}}\left[  j,j^{\dagger}\right]  \right)  }{\delta
j_{BA}\left(  \vec{x}\right)  \delta j_{DC}^{\dagger}\left(  \vec{y}\right)
}.
\end{align}
The $2$-point function is obtained from $\Pi^{\left(  2\right)  }$ by setting
the sources $j,j^{\dagger}$ to zero. Since then $v_{jj^{\dagger}}\left[
0,0\right]  =v_{jj^{\dagger}jj^{\dagger}}\left[  0,0\right]  =0$ and also the
one-derivative terms as well as the term with two derivatives on
$v_{jj^{\dagger}jj^{\dagger}}$ vanish, one is left with%
\begin{equation}
\Pi_{ABCD}^{\left(  2\right)  }\left[  0,0\right]  \left(  \vec{x},\vec
{y}\right)  =\frac{\delta^{2}v_{jj^{\dagger}}\left[  j,j^{\dagger}\right]
}{\delta j_{BA}\left(  \vec{x}\right)  \delta j_{DC}^{\dagger}\left(  \vec
{y}\right)  } \label{t02p}%
\end{equation}
(the right-hand side is independent of $j,j^{\dagger}$ -- hence one does not
have to impose $j,j^{\dagger}=0$ explicitly). It thus remains to calculate
\begin{align}
\frac{\delta^{2}v_{jj^{\dagger}}\left[  j,j^{\dagger}\right]  }{\delta
j_{BA}\left(  \vec{x}\right)  \delta j_{DC}^{\dagger}\left(  \vec{y}\right)
}  &  =\delta_{AD}\delta_{CB}\frac{4Ng^{4}}{\pi^{2}}\frac{\mu}{m_{g}^{2}%
}\tilde{\imath}_{2}\left(  \xi,c_{1}\right)  \partial_{x}^{2}\frac{-\partial
}{\partial\xi^{2}}\Delta^{-1}\left(  \vec{y}-\vec{x}\right) \\
&  =-\delta_{AD}\delta_{BC}\frac{2g^{2}}{m_{g}}\frac{\gamma}{\zeta\mu}%
\bar{\Delta}^{-1}\left(  \vec{x}-\vec{y}\right)  ,
\end{align}
where we used the symmetry property (\ref{sym}) and defined the
\textquotedblleft$4U$-vertex-inserted\textquotedblright\ soft-mode propagator%
\begin{equation}
\bar{\Delta}^{-1}\left(  \vec{x}\right)  :=2\tilde{\imath}_{2}\mu\partial
_{x}^{2}\frac{\partial}{\partial\xi^{2}}\Delta^{-1}\left(  \vec{x}\right)
=\frac{4g^{2}\mu^{4}}{m_{g}}\tilde{\imath}_{2}\left(  \xi,c_{1}\right)
\int\frac{d^{3}\kappa}{\left(  2\pi\right)  ^{3}}\frac{\kappa^{2}\theta\left(
1-\kappa^{2}\right)  e^{i\vec{\kappa}\mu\vec{x}}}{\left[  \kappa^{2}\left(
1-c_{1}\kappa^{2}\right)  +\xi^{2}\right]  ^{2}} \label{m}%
\end{equation}
as well as the dimensionless parameters%
\begin{equation}
\text{\ }\xi=\frac{\sqrt{\bar{\Sigma}}}{\mu},\text{ \ \ \ \ }\gamma\equiv
\frac{g^{2}N_{c}}{\pi^{2}}=4N_{c}\frac{\alpha}{\pi},\text{ \ \ \ \ }%
\zeta\equiv\frac{m_{g}}{\mu}. \label{dimlp}%
\end{equation}
Hence the $c_{1}$ correction (\ref{uc}) to the connected 2-point function,
which originates from the perturbative insertion of the $4U$ part (\ref{l4u})
of the four-derivative interactions, becomes
\begin{equation}
\left\langle U_{<,AB}^{\dagger}\left(  \vec{x}\right)  \int d^{3}%
z\mathcal{L}_{<,c_{1}}^{\left(  4U\right)  }\left(  \vec{z}\right)
U_{<,CD}\left(  \vec{y}\right)  \right\rangle ^{\left(  2U\right)  }%
=\frac{c_{1}m_{g}}{2g^{2}\mu^{2}}\Pi_{ABCD}^{\left(  2\right)  }\left[
0,0\right]  \left(  \vec{x},\vec{y}\right)  =-\delta_{AD}\delta_{BC}%
\frac{c_{1}\gamma}{\zeta\mu^{3}}\bar{\Delta}^{-1}\left(  \vec{x}-\vec
{y}\right)  .
\end{equation}
Our final result for the 2-point function is therefore%
\begin{equation}
\left\langle U_{<,AB}^{\dagger}\left(  \vec{x}\right)  U_{<,CD}\left(  \vec
{y}\right)  \right\rangle =\delta_{AD}\delta_{BC}\left[  \Delta^{-1}\left(
\vec{x}-\vec{y}\right)  +\frac{c_{1}\gamma}{\zeta\mu^{3}}\bar{\Delta}%
^{-1}\left(  \vec{x}-\vec{y}\right)  \right]  . \label{uu}%
\end{equation}
Its diagonal dependence on the group indices reflects the SU$_{L}(N_{c}%
)\times$ SU$_{R}(N_{c})$\ symmetry of the soft-mode dynamics (\ref{smd}).

\subsection{The 4-point function}

In the following section we sketch the analogous evaluation of the 4-point
function. Although our main goal of computing $c_{n}$ corrections to the
vacuum energy density involves (in our approximation) only the 2-point
function, the explicit expression for the 4-point function%

\begin{align}
\delta_{4}  &  \equiv\left\langle U_{<,AB}^{\dagger}\left(  \vec{z}%
_{1}\right)  U_{<,CD}\left(  \vec{z}_{2}\right)  U_{<,EF}^{\dagger}\left(
\vec{z}_{3}\right)  U_{<,GH}\left(  \vec{z}_{4}\right)  \right\rangle \\
&  =\left.  \frac{-\delta}{\delta j_{BA}\left(  \vec{z}_{1}\right)  }%
\frac{-\delta}{\delta j_{DC}^{\dagger}\left(  \vec{z}_{2}\right)  }%
\frac{-\delta}{\delta j_{FE}\left(  \vec{z}_{3}\right)  }\frac{-\delta}{\delta
j_{HG}^{\dagger}\left(  \vec{z}_{4}\right)  }\mathcal{V}^{\left(
4U,c_{1}\right)  }\left[  \frac{\delta}{\delta j},\frac{\delta}{\delta
j^{\dagger}}\right]  \frac{Z^{\left(  2U\right)  }\left[  j,j^{\dagger
}\right]  }{Z^{\left(  2U\right)  }\left[  0,0\right]  }\right\vert
_{j,j^{\dagger}=0}\\
&  =:\delta_{4}^{\left(  2U\right)  }-\delta_{4}^{\left(  4U,O\left(
c_{1}\right)  \right)  }+...
\end{align}
will be useful, too, because it allows for consistency checks and an
alternative estimate of the nonperturbative energy density. Above, we have
defined%
\begin{align}
\delta_{4}^{\left(  2U\right)  }  &  \equiv\left\langle U_{<,AB}^{\dagger
}\left(  \vec{z}_{1}\right)  U_{<,CD}\left(  \vec{z}_{2}\right)
U_{<,EF}^{\dagger}\left(  \vec{z}_{3}\right)  U_{<,GH}\left(  \vec{z}%
_{4}\right)  \right\rangle ^{\left(  2U\right)  }\\
&  =\left.  \frac{\delta^{4}}{\delta j_{FE}\left(  \vec{z}_{3}\right)  \delta
j_{HG}^{\dagger}\left(  \vec{z}_{4}\right)  \delta j_{BA}\left(  \vec{z}%
_{1}\right)  \delta j_{DC}^{\dagger}\left(  \vec{z}_{2}\right)  }%
\frac{Z^{\left(  2U\right)  }\left[  j,j^{\dagger}\right]  }{Z^{\left(
2U\right)  }\left[  0,0\right]  }\right\vert _{j,j^{\dagger}=0}\\
&  =\delta_{AD}\delta_{CB}\Delta^{-1}\left(  \vec{z}_{2},\vec{z}_{1}\right)
\delta_{EH}\delta_{GF}\Delta^{-1}\left(  \vec{z}_{4},\vec{z}_{3}\right)
+\delta_{AH}\delta_{GB}\Delta^{-1}\left(  \vec{z}_{4},\vec{z}_{1}\right)
\delta_{ED}\delta_{CF}\Delta^{-1}\left(  \vec{z}_{2},\vec{z}_{3}\right)
\label{b}%
\end{align}
which shows that in the absence of $4U$ interactions the 4-point function
factorizes as
\begin{align}
\left\langle U_{LAB}^{\dagger}\left(  \vec{z}_{1}\right)  U_{LCD}\left(
\vec{z}_{2}\right)  U_{LEF}^{\dagger}\left(  \vec{z}_{3}\right)
U_{LGH}\left(  \vec{z}_{4}\right)  \right\rangle  &  =\left\langle
U_{LAB}^{\dagger}\left(  \vec{z}_{1}\right)  U_{LCD}\left(  \vec{z}%
_{2}\right)  \right\rangle \left\langle U_{LEF}^{\dagger}\left(  \vec{z}%
_{3}\right)  U_{LGH}\left(  \vec{z}_{4}\right)  \right\rangle \nonumber\\
&  +\left\langle U_{LAB}^{\dagger}\left(  \vec{z}_{1}\right)  U_{LGH}\left(
\vec{z}_{4}\right)  \right\rangle \left\langle U_{LEF}^{\dagger}\left(
\vec{z}_{3}\right)  U_{LCD}\left(  \vec{z}_{2}\right)  \right\rangle ,
\label{4ufac}%
\end{align}
and%
\begin{align}
\delta_{4}^{\left(  4U,O\left(  c_{1}\right)  \right)  }  &  \equiv
\left\langle U_{<,AB}^{\dagger}\left(  \vec{z}_{1}\right)  U_{<,CD}\left(
\vec{z}_{2}\right)  \int d^{3}z\mathcal{L}_{<,c_{1}}^{\left(  4U\right)
}\left(  \vec{z}\right)  U_{<,EF}^{\dagger}\left(  \vec{z}_{3}\right)
U_{<,GH}\left(  \vec{z}_{4}\right)  \right\rangle ^{\left(  2U\right)  }\\
&  =\frac{c_{1}m_{g}}{2g^{2}\mu^{2}}\left.  \frac{\delta^{4}v\left[
j,j^{\dagger}\right]  }{\delta j_{FE}\left(  \vec{z}_{3}\right)  \delta
j_{HG}^{\dagger}\left(  \vec{z}_{4}\right)  \delta j_{BA}\left(  \vec{z}%
_{1}\right)  \delta j_{DC}^{\dagger}\left(  \vec{z}_{2}\right)  }\right\vert
_{j,j^{\dagger}=0}\\
&  =:-\frac{c_{1}m_{g}}{2g^{2}\mu^{2}}\Pi_{ABCDEFGH}^{\left(  4\right)
}\left(  \vec{z}_{1},\vec{z}_{2},\vec{z}_{3},\vec{z}_{4}\right)  .
\end{align}
The above reordering of the functional derivatives reveals that Eq. (\ref{p2})
from the calculation of the 2-point function can be used as an intermediate
step for the further evaluation. Indeed, to $O\left(  c_{1}\right)  $ the
correction due to the $4U$ interaction becomes%
\begin{equation}
\Pi_{ABCDEFGH}^{\left(  4\right)  }\left(  \vec{z}_{1},\vec{z}_{2},\vec{z}%
_{3},\vec{z}_{4}\right)  :=\frac{\delta}{\delta j_{FE}\left(  \vec{z}%
_{3}\right)  }\frac{\delta}{\delta j_{HG}^{\dagger}\left(  \vec{z}_{4}\right)
}\left.  \Pi_{ABCD}^{\left(  2\right)  }\left[  j,j^{\dagger}\right]  \left(
\vec{z}_{1},\vec{z}_{2}\right)  \right\vert _{j,j^{\dagger}=0}%
\end{equation}
which simplifies in the zero-distance limit of some of its arguments (as it
occurs in our context) and then renders, together with Eq. (\ref{b}), the
coordinate dependence of the 4-point function explicit.

\section{Solution of the gap equation and phase diagram}

\label{gap}

As explained in Sec. \ref{mfa}, the above expressions for the 2- and 4-point
functions are to be evaluated at the saddle point $\bar{\Sigma}=:\left(
\mu\bar{\xi}\right)  ^{2}$ of the integral over the auxiliary $\Sigma$ field,
i.e. at the minimal-action solution of the gap equation
\begin{equation}
\left\langle U_{<,AB}^{\dagger}\left(  \vec{x}\right)  U_{<,BC}\left(  \vec
{x}\right)  \right\rangle =\delta_{AC}. \label{geq2}%
\end{equation}
After inserting the result (\ref{uu}) for the connected 2-point function (and
recalling $\delta_{AA}=N_{c}$), Eq. (\ref{geq2}) turns into%
\begin{equation}
N_{c}\left[  \Delta^{-1}\left(  0\right)  +\frac{c_{1}\gamma}{\zeta\mu^{3}%
}\bar{\Delta}^{-1}\left(  0\right)  \right]  =1.
\end{equation}
The zero-distance limits of the propagator (\ref{d}), (\ref{dm1(k)}) and of
the $4U$ correction (\ref{m}) are then expressed in terms of the integrals
$\tilde{\imath}_{n},$ $\tilde{j}_{n}$ (defined and evaluated in App.
\ref{int}) as%
\begin{align}
\Delta^{-1}\left(  0\right)   &  =\frac{\gamma}{\zeta N_{c}}\tilde{\imath}%
_{1}\left(  \xi,c_{1}\right)  ,\\
\bar{\Delta}^{-1}\left(  0\right)   &  =\frac{2\gamma}{\zeta N_{c}}\mu
^{3}\tilde{\imath}_{2}\left(  \xi,c_{1}\right)  \tilde{j}_{2}\left(  \xi
,c_{1}\right)
\end{align}
(both are of $O\left(  g^{2}/\zeta\right)  $), so that the gap equation
assumes its final form%
\begin{equation}
\frac{\gamma}{\zeta}\left[  \tilde{\imath}_{1}\left(  \xi,c_{1}\right)
+2c_{1}\frac{\gamma}{\zeta}\tilde{\imath}_{2}\left(  \xi,c_{1}\right)
\tilde{j}_{2}\left(  \xi,c_{1}\right)  \right]  =1. \label{geq}%
\end{equation}
(For $c_{1}=0$ and $\zeta=1$, Eq. (\ref{geq}) reduces as expected \cite{kog95}
to $\gamma\tilde{\imath}_{1}\left(  \xi\right)  =1$ (except when $\gamma$
diverges, see below).) This equation and the $\xi$ dependence of the integrals
(\ref{in}), (\ref{jn}) render the nonperturbative character of the solutions
$\bar{\xi}\left(  \mu,c_{1},\zeta\right)  $ explicit. As already mentioned, it
reflects the infinite subset of diagrams required to generate a finite mass gap.

Before finding the solutions of Eq. (\ref{geq}) as functions of the
variational parameters, we specialize $\zeta$ to the form required by
continuity of $G^{-1}\left(  k\right)  $ at $k=\mu$ (cf. Eq. (\ref{mgc}),%
\begin{equation}
\zeta_{\text{ct}}\left(  c_{1}\right)  =\frac{m_{\text{g}}}{\mu}=\frac
{1}{1-c_{1}}, \label{zct}%
\end{equation}
and adopt the one-loop Yang-Mills coupling%
\begin{equation}
\gamma\left(  \mu\right)  =\frac{g_{\text{YM}}^{2}\left(  \mu\right)  N_{c}%
}{\pi^{2}}\overset{N_{c}=3}{=}\frac{24}{11\ln\frac{\mu}{\Lambda_{\text{YM}}}}
\label{g}%
\end{equation}
to render the $\mu$ dependence of $\gamma$ explicit. The coupling
$g_{\text{YM}}\left(  \mu\right)  $ is just $11/10$ times larger
\cite{bro98,bro99} than the coupling (\ref{g(mu)}) obtained from integrating
out the high-momentum modes \cite{bro99} governed by the UV covariance
(\ref{gm1uv}). (This discrepancy can be mended by admitting non-transverse
modes to the UV covariance \cite{dia98}.) We prefer to use the Yang-Mills
coupling because it facilitates direct comparison with the numerical results
of Ref. \cite{kog95}.

\begin{figure}[ptb]
\begin{center}
\includegraphics[height = 9cm]{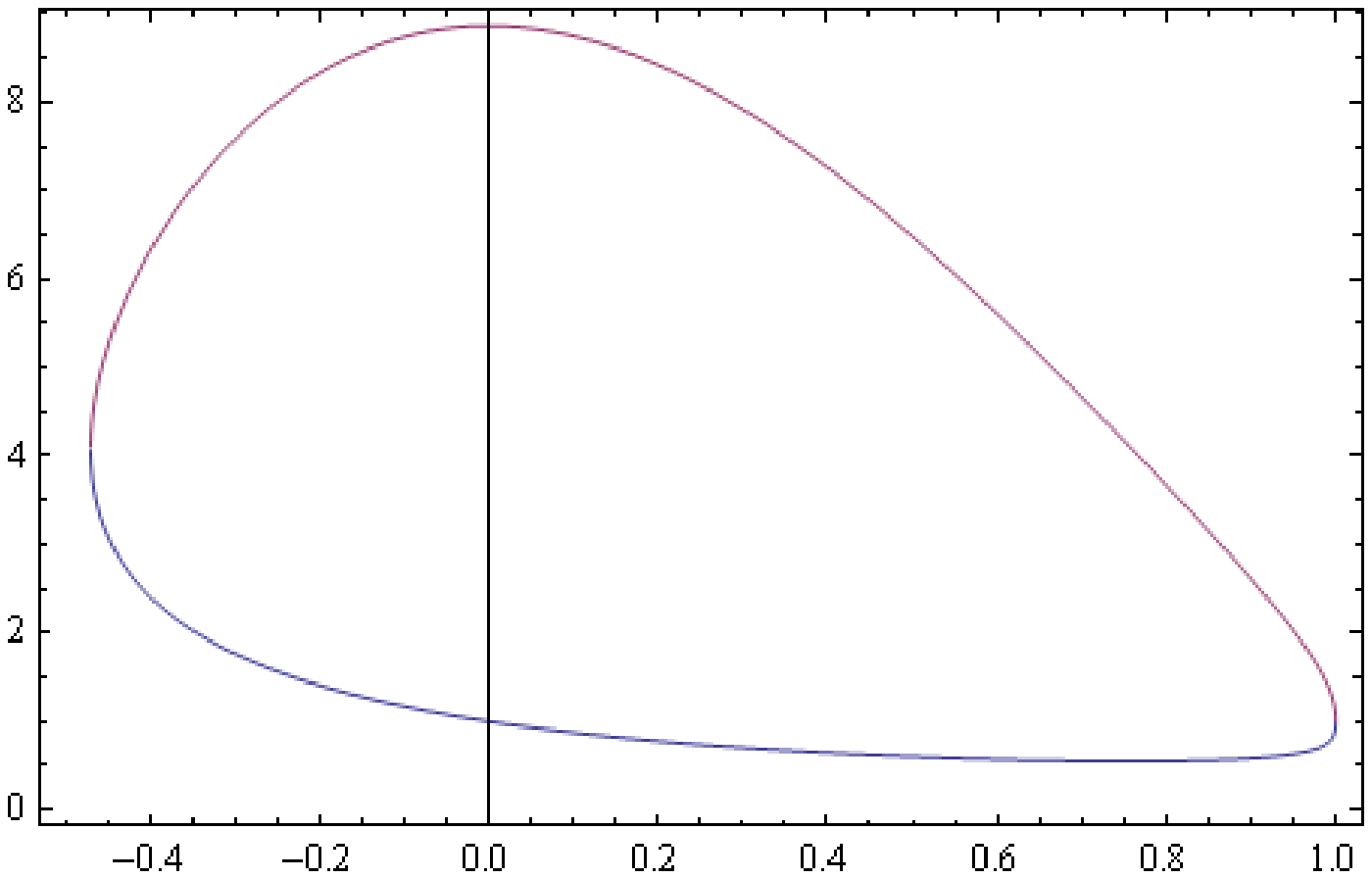}
\end{center}
\caption{Vacuum phase diagram: inside the plotted phase boundary $\mu
_{c}\left(  c_{1}\right)  /\Lambda_{\text{YM}}$ the theory is in its
strongly-coupled disordered phase. (Our treatment is approximately valid in
the range $\mu\gtrsim4\Lambda_{\text{YM}}$ and $c_{1}\in\left\{
-0.5,0.5\right\}  $).}%
\label{cl}%
\end{figure}

The above considerations imply that the solutions of Eq. (\ref{geq}) depend on
just two variational parameters, i.e. the RG scale $\mu\geq0$ which enters
through the coupling (\ref{g}) and $c_{1}<1$ which controls the leading
momentum dependence of the infrared covariance $G_{<}^{-1}\left(  k\right)  $.
Our next task will be to determine the critical line $\mu_{\text{c}}\left(
c_{1}\right)  $ in this parameter space, i.e. the subspace which joins the
points where the phase transition takes place and where thus the (dis-)order
parameter vanishes,
\begin{equation}
\bar{\xi}\left(  \mu_{\text{c}}\left(  c_{1}\right)  ,c_{1}\right)  =0.
\label{xi0}%
\end{equation}
This line can be found analytically\ since for $\xi=0$ the integrals
$\tilde{\imath}_{1}$, $\tilde{\imath}_{2}$ and $\tilde{j}_{2}$ in the gap
equation (\ref{geq}) simplify to (cf. App. \ref{lim})%
\begin{equation}
\tilde{\imath}\left(  c_{1}\right)  :=\tilde{\imath}_{1}\left(  0,c_{1}%
\right)  =\frac{\operatorname{arctanh}\sqrt{c_{1}}}{\sqrt{c_{1}}},\text{
\ \ \ \ }\tilde{\imath}_{2}\left(  0,c_{1}\right)  =\frac{\tilde{\imath
}\left(  c_{1}\right)  -1}{c_{1}}%
\end{equation}
and%
\begin{equation}
\tilde{j}_{2}\left(  0,c_{1}\right)  =\frac{1}{2}\left[  \frac{1}{1-c_{1}%
}+\tilde{\imath}\left(  c_{1}\right)  \right]  .
\end{equation}
In order to exhibit the $\mu$ dependence, we first resolve the $\xi=0$ gap
equation for $\gamma\left(  \mu\right)  $, which yields%
\begin{equation}
\frac{1}{\zeta_{\text{ct}}^{2}}\left(  \frac{\tilde{\imath}-1}{1-c_{1}}%
+\tilde{\imath}\left(  \tilde{\imath}-1\right)  \right)  \gamma^{2}%
+\frac{\tilde{\imath}}{\zeta_{\text{ct}}}\gamma-1=0.
\end{equation}
The two solutions of this equation can then be resolved for $\mu$ and combine
into the critical line $\mu_{\text{c}}\left(  c_{1}\right)  $. With
$\gamma\left(  \mu\right)  $ from Eq. (\ref{g}) it takes the explicit form
\begin{equation}
\frac{\mu_{\text{c,1,2}}\left(  c_{1}\right)  }{\Lambda_{\text{YM}}}%
=\exp\left[  \frac{48}{11}\frac{\left(  1-c_{1}\right)  \left[  1-\tilde
{\imath}\left(  c_{1}\right)  \right]  \left[  1+\left(  1-c_{1}\right)
\tilde{\imath}\left(  c_{1}\right)  \right]  }{\left(  1-c_{1}\right)
\tilde{\imath}\left(  c_{1}\right)  \pm\sqrt{5\tilde{\imath}^{2}\left(
c_{1}\right)  \left(  1-c_{1}\right)  ^{2}-4\left(  1-c_{1}\right)  \left[
1-c_{1}\tilde{\imath}\left(  c_{1}\right)  \right]  }}\right]  . \label{mu}%
\end{equation}
In order to facilitate the discussion of the vacuum phase structure encoded in
Eq. (\ref{mu}), we plot the phase boundary in Fig. \ref{cl}. It reveals that
Eq. (\ref{xi0}) is satisfied by two values of $\mu_{\text{c}}$ for each
$c_{1}$ for which a solution exists, except at the maximal and minimal values
where $\mu_{\text{c}}$ becomes unique. Vice versa, there are two $c_{1}$ for
each $\mu_{\text{c}}$ for which Eq. (\ref{xi0}) holds. The critical line
(\ref{mu}) covers the limited parameter ranges%
\begin{equation}
0.5\lesssim\frac{\mu_{\text{c}}}{\Lambda_{\text{YM}}}\lesssim8.86
\label{musolrg}%
\end{equation}
and%
\begin{equation}
-0.48\lesssim c_{1}<1.
\end{equation}
This prevents the minimal-energy solution $\bar{\xi}^{\ast}$ (to be determined
in Sec. \ref{res}) from attaining unacceptably large values of $\mu$ and
$\left\vert c_{1}\right\vert $. The normalizability condition $c_{1}<1$ is
automatically satisfied in the existence region of gap-equation solutions.
Equation (\ref{mu}) further shows that nontrivial (i.e. nonzero) solutions of
the gap equation exist only when the gauge coupling exceeds a critical value,
i.e. for
\begin{equation}
g^{2}\left(  \mu\right)  >g_{\text{c}}^{2}\left(  c_{1}\right)  ,
\end{equation}
as expected on physical grounds. The maximal critical coupling corresponds to
$c_{1}=0$:%
\begin{equation}
g_{\text{c}}^{2}\left(  c_{1}\right)  \leq g_{\text{c}}^{2}\left(
c_{1}=0\right)  =\frac{\pi^{2}}{N_{c}}.
\end{equation}
Only part of the critical line (\ref{mu}) is physically trustworthy, however.
Indeed, its validity range is limited to $\left\vert c_{1}\right\vert \ll1$
for which the $O\left(  c_{1}\right)  $ evaluation of the $4U$ correction
included in Eq. (\ref{mu}) is accurate, and to $\mu/\Lambda_{\text{YM}}%
\gtrsim5-6$ which justifies the $O\left(  g\right)  $ approximation. These two
restrictions eliminate most of the $\mu_{\text{c,1}}$ branch and the
lower-$\mu$ part of the $\mu_{\text{c,2}}$ branch. (Although the $c_{1}=0$
solution with $\mu_{\text{c}}/\Lambda_{\text{YM}}\rightarrow1$ and thus
$g^{2}\left(  \mu_{c}\right)  \rightarrow\infty$ invalidates the $O\left(
g\right)  $ treatment, it nevertheless demonstrates that the gap equation does
not reduce to $\gamma\tilde{\imath}_{1}\left(  \xi\right)  =1$ for
$c_{1}\rightarrow0$ if $\gamma\left(  \mu_{\text{c}}\right)  $ diverges simultaneously.)

\begin{figure}[ptb]
\begin{center}
\includegraphics[height = 9cm]{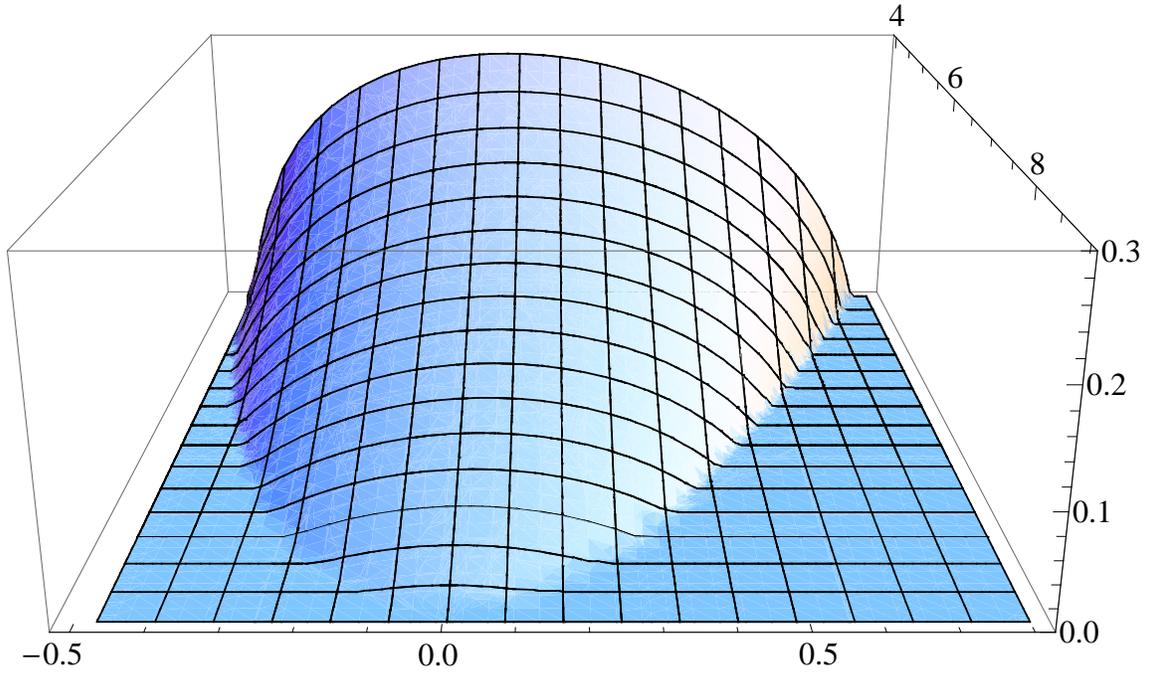}
\end{center}
\caption{The order-parameter solution $\bar{\xi}\left(  \mu,c_{1}\right)  $ of
the gap equation in the variable ranges $\mu\in\left\{  4,9\right\}
\Lambda_{\text{YM}}$ and $c_{1}\in\left\{  -0.5,0.8\right\}  .$}%
\label{gsoln}%
\end{figure}

In Fig. \ref{gsoln}, finally, we plot the numerically generated solution
$\bar{\xi}\left(  \mu,c_{1}\right)  $ of the gap equation (\ref{geq}) in the
physically reliable parameter range. As expected, the $\bar{\xi}\geq0$ region
is surrounded by the critical line (\ref{mu}) and exists for $0.5\lesssim
\mu/\Lambda_{\text{YM}}\lesssim8.86$. In addition, Fig. \ref{gsoln} reveals
that the (dis-)order parameter goes to zero continuously, i.e. that the
transition from the disordered to the ordered phase is of second order
\footnote{The first-order transition expected from the $\varepsilon$ expansion
and lattice simulations may be prvented by limitations of the mean-field
approximation \cite{kog95}.}.

\section{Vacuum energy density}

\label{vend}

We have now assembled all the infrared-mode information necessary to evaluate
the expectation value (\ref{hexp}) of the Yang-Mills Hamiltonian density
\begin{equation}
\mathcal{H}_{\text{YM}}=\frac{1}{2}\left(  E_{i}^{a}E_{i}^{a}+B_{i}^{a}%
B_{i}^{a}\right)  \label{hamYM}%
\end{equation}
in the generalized trial vacuum state (\ref{ginvvwf}). (We have adopted the
$A_{0}=0$ gauge, and defined the chromoelectric and -magnetic fields as
$E_{i}^{a}=F_{0i}^{a}$ and $B_{i}^{a}=\frac{1}{2}\varepsilon_{ijk}F_{jk}^{a}%
$.) Of course, the Hamiltonian (\ref{hamYM}) is formal and needs
renormalization. Since we are interested in its matrix elements between
Poincar\'{e}-invariant vacuum trial states with perturbative one-loop
corrections only, a regularization of the vacuum matrix elements by a momentum
cutoff $\Lambda_{\text{UV}}$ will be sufficient \cite{kog95}. In fact, the
$\Lambda_{\text{UV}}$ dependence of the vacuum energy density (\ref{hexp}) can
be completely removed by normal-ordering the Hamiltonian (\ref{hamYM}) with
respect to the noninteracting vacuum. This amounts to subtracting the
ultraviolet-divergent energy density%

\begin{equation}
\left\langle \mathcal{H}_{\text{YM}}\right\rangle _{0}=2\left(  N_{c}%
^{2}-1\right)  \frac{G_{0}^{-1}\left(  \vec{x},\vec{x}\right)  }{2}=2\left(
N_{c}^{2}-1\right)  \int\frac{d^{3}k}{\left(  2\pi\right)  ^{3}}\theta\left(
\Lambda_{\text{UV}}^{2}-\vec{k}^{2}\right)  \frac{\hbar\omega_{k}}{2}%
=\frac{N_{c}^{2}-1}{8\pi^{2}}\Lambda_{\text{UV}}^{4} \label{h0}%
\end{equation}
(where we wrote $\omega_{k}=k$ for the free gluon energy and reinstated
$\hbar$) of the noninteracting vacuum, i.e. the sum over the zero-point
energies of two transverse, massless vector modes, from $\left\langle
\mathcal{H}_{\text{YM}}\right\rangle $. (Recall that the momentum cutoffs are
imposed on integrals over the gauge-invariant $U$ modes and therefore do not
compromise (residual)\ gauge invariance.)

\subsection{Hard-mode contribution}

\label{huen}

According to our strategy for evaluating the trial energy density
$\left\langle \mathcal{H}_{\text{YM}}\right\rangle $ outlined in Sec.
\ref{strat}, the first step consists of integrating over the static gauge
fields and the hard modes $U_{>}$. In the following section, this will be done
for the chromoelectric $\left\langle E^{2}\right\rangle $ and -magnetic
$\left\langle B^{2}\right\rangle $ contributions separately. We start from the
intermediate matrix element%

\begin{align}
\left\langle \left\langle \left\langle E_{i}^{a}\left(  \vec{x}\right)
E_{j}^{b}\left(  \vec{y}\right)  \right\rangle \right\rangle \right\rangle  &
=\frac{\int D\vec{A}\psi_{0}\left[  \vec{A}^{U}\right]  \frac{i\delta}{\delta
A_{i}^{a}\left(  \vec{x}\right)  }\frac{i\delta}{\delta A_{j}^{b}\left(
\vec{y}\right)  }\psi_{0}\left[  \vec{A}\right]  }{\int D\vec{A}\psi\left[
\vec{A}^{U}\right]  \psi\left[  \vec{A}\right]  }=\delta_{ij}\delta^{ab}%
G^{-1}\left(  \vec{x},\vec{y}\right) \nonumber\\
&  -\int d^{3}z_{1}\int d^{3}z_{2}G^{-1}\left(  \vec{x}-\vec{z}_{1}\right)
G^{-1}\left(  \vec{y}-\vec{z}_{2}\right)  \left\langle \left\langle
\left\langle A_{i}^{a}\left(  \vec{z}_{1}\right)  A_{j}^{b}\left(  \vec{z}%
_{2}\right)  \right\rangle \right\rangle \right\rangle
\end{align}
($E_{i}^{a}=i\delta/\delta A_{i}^{a}$, cf. App. \ref{stacor}) for the
chromoelectic 2-point function. With the help of the expression (\ref{2p}) for
the gauge-field matrix element in terms of the functionals $a\left[  U\right]
$ and $\mathcal{M}\left[  U\right]  $ as defined in Eq. (\ref{ma}), this
specializes to
\begin{align}
\left\langle \left\langle \left\langle E^{2}\right\rangle \right\rangle
\right\rangle  &  :=\left\langle \left\langle \left\langle E_{i}^{a}\left(
\vec{x}\right)  E_{i}^{a}\left(  \vec{x}\right)  \right\rangle \right\rangle
\right\rangle =3\left(  N_{c}^{2}-1\right)  G^{-1}\left(  \vec{x},\vec
{x}\right) \nonumber\\
&  -\int d^{3}z_{1}d^{3}z_{2}G^{-1}\left(  \vec{x}-\vec{z}_{1}\right)
G^{-1}\left(  \vec{x}-\vec{z}_{2}\right)  \left[  \mathcal{M}_{ii}%
^{-1aa}\left(  \vec{z}_{1},\vec{z}_{2}\right)  +a_{i}^{a}\left(  \vec{z}%
_{1}\right)  a_{i}^{a}\left(  \vec{z}_{2}\right)  \right]  .
\end{align}
The analogous expression for the chromomagnetic contribution involves 2-, 3-
and 4-point functions of the gauge field. After rewriting them with $B_{i}%
^{a}=\varepsilon_{ijk}\left(  \partial_{j}A_{k}^{a}+\frac{g}{2}f^{abc}%
A_{j}^{b}A_{k}^{c}\right)  $ and Eqs. (\ref{2p}) -- (\ref{4p}), one finds%
\begin{align}
\left\langle \left\langle \left\langle B^{2}\right\rangle \right\rangle
\right\rangle  &  :=\left\langle \left\langle \left\langle B_{i}^{a}\left(
\vec{x}\right)  B_{i}^{a}\left(  \vec{x}\right)  \right\rangle \right\rangle
\right\rangle =\varepsilon_{ijk}\varepsilon_{ilm}\left.  \partial_{z_{1j}%
}\partial_{z_{2l}}\mathcal{M}_{km}^{-1aa}\left(  \vec{z}_{1},\vec{z}%
_{2}\right)  \right\vert _{\vec{z}_{i}=\vec{x}}+\varepsilon_{ijk}%
\varepsilon_{ilm}\partial_{j}a_{k}^{a}\partial_{l}a_{m}^{a}\nonumber\\
&  +g\varepsilon_{ijk}\varepsilon_{ilm}f^{abc}\left[  \mathcal{M}^{-1}\left(
\vec{x},\vec{x}\right)  _{lm}^{bc}\partial_{j}a_{k}^{a}\right.  +\partial
_{z_{1},j}\mathcal{M}^{-1}\left(  \vec{z}_{1},\vec{x}\right)  _{km}^{ac}%
a_{l}^{b}\nonumber\\
&  +\left.  \partial_{z_{1},j}\mathcal{M}^{-1}\left(  \vec{z}_{1},\vec
{x}\right)  _{kl}^{ab}a_{m}^{c}+\partial_{j}a_{k}^{a}a_{l}^{b}a_{m}%
^{c}\right]  _{z_{i}\rightarrow x}\nonumber\\
&  +\frac{g^{2}}{4}f^{abc}f^{ade}\left[  12\mathcal{M}^{-1bd}\mathcal{M}%
^{-1ce}+8\mathcal{M}^{-1bd}a_{i}^{c}a_{i}^{e}+2a_{i}^{b}a_{j}^{c}a_{i}%
^{d}a_{j}^{e}\right]
\end{align}
(where all spacial arguments are set to $\vec{x}$ after taking the derivatives).

Now we compute the hard-mode contributions to the chromoelectric and -magnetic
vacuum energies, as outlined in Sec. \ref{strat} and App. \ref{stacor}. This
amounts to evaluating the intermediate matrix elements (\ref{o2}) for $E^{2}$
and $B^{2}$, i.e.
\begin{equation}
\left\langle \left\langle E^{2},B^{2}\right\rangle \right\rangle :=\frac{\int
D\phi\left[  \left\langle \left\langle \left\langle E^{2},B^{2}\right\rangle
\right\rangle \right\rangle +O\left(  g\right)  \right]  \exp\left\{
-\Gamma_{\text{b},>}\left[  \phi\right]  \right\}  }{\int D\phi\exp\left\{
-\Gamma_{\text{b},>}\left[  \phi\right]  \right\}  }, \label{ebhm}%
\end{equation}
to leading order in $g$. The hard-mode action $\Gamma_{\text{b},>}\left[
\phi\right]  $ is defined in Eq. (\ref{gg}). Since it is bilinear in $\phi$,
the integrals in Eq. (\ref{ebhm}) are Gaussian and can be calculated
analytically, cf. Appendix \ref{stacor}. This yields%
\begin{align}
\left\langle \left\langle E^{2}\right\rangle \right\rangle  &  =\left(
N_{c}^{2}-1\right)  \int\frac{d^{3}k}{\left(  2\pi\right)  ^{3}}\theta\left(
\Lambda_{\text{UV}}^{2}-k^{2}\right)  G^{-1}\left(  k\right) \nonumber\\
&  +\frac{1}{2}\left(  N_{c}^{2}-1\right)  \int\frac{d^{3}k}{\left(
2\pi\right)  ^{3}}\theta\left(  \mu^{2}-k^{2}\right)  G^{-1}\left(  k\right)
+\left\langle \left\langle E^{2}\right\rangle \right\rangle _{U_{<}}+O\left(
g\right)  , \label{e2}%
\end{align}
which includes the contribution
\begin{equation}
\left\langle \left\langle E^{2}\right\rangle \right\rangle _{U_{<}}:=-\frac
{1}{4g^{2}}\int d^{3}z_{1}\int d^{3}z_{2}G_{<}^{-1}\left(  \vec{x}-\vec{z}%
_{1}\right)  G_{<}^{-1}\left(  \vec{x}-\vec{z}_{2}\right)  \left\langle
\left\langle L_{<,i}^{a}\left(  \vec{z}_{1}\right)  L_{<,i}^{a}\left(  \vec
{z}_{2}\right)  \right\rangle \right\rangle \label{e2u}%
\end{equation}
from the soft modes $U_{<}$, and%
\begin{equation}
\left\langle \left\langle B^{2}\right\rangle \right\rangle =\left(  N_{c}%
^{2}-1\right)  \int\frac{d^{3}k}{\left(  2\pi\right)  ^{3}}\theta\left(
\Lambda_{\text{UV}}^{2}-k^{2}\right)  k^{2}G\left(  k\right)  +O\left(
g\right)  . \label{b2h2}%
\end{equation}
With $\theta\left(  k^{2}-\mu^{2}\right)  G\left(  k\right)  =\theta\left(
k^{2}-\mu^{2}\right)  k^{-1}$ (cf. Eq. (\ref{gm1uv})) one finds the
$\Lambda_{\text{UV}}$ dependence of $\left\langle \left\langle E^{2}%
\right\rangle \right\rangle $ and $\left\langle \left\langle B^{2}%
\right\rangle \right\rangle $ to be identical. Hence the lowest-dimensional
gluon condensate $\left\langle F^{2}\right\rangle =2\left[  \left\langle
B^{2}\right\rangle -\left\langle E^{2}\right\rangle \right]  $ (to be
discussed in Sec. \ref{gcond})\ is $\Lambda_{\text{UV}}$-independent by itself
while the $\Lambda_{\text{UV}}$ dependence of $\left\langle \left\langle
\mathcal{H}_{\text{YM}}\right\rangle \right\rangle $ has to be removed by
renormalization. As noted above, at our level of approximation this amounts to
normal-ordering the Hamiltonian, i.e. to subtracting the non-interacting
vacuum expectation values $\left\langle E^{2}\right\rangle _{0}=\left\langle
B^{2}\right\rangle _{0}=\left(  N_{c}^{2}-1\right)  G_{m=0}^{-1}\left(
\vec{x},\vec{x}\right)  $. The results are%
\begin{equation}
\left\langle \left\langle \text{:}E^{2}\text{:}\right\rangle \right\rangle
=\left(  N_{c}^{2}-1\right)  \int\frac{d^{3}k}{\left(  2\pi\right)  ^{3}%
}\theta\left(  \mu^{2}-k^{2}\right)  \left[  \frac{3}{2}G^{-1}\left(
k\right)  -k\right]  +\left\langle \left\langle E^{2}\right\rangle
\right\rangle _{U_{<}}+O\left(  g\right)
\end{equation}
and%
\begin{equation}
\left\langle \left\langle \text{:}B^{2}\text{:}\right\rangle \right\rangle
=\left(  N_{c}^{2}-1\right)  \int\frac{d^{3}k}{\left(  2\pi\right)  ^{3}%
}\theta\left(  \mu^{2}-k^{2}\right)  \left[  k^{2}G\left(  k\right)
-k\right]  +O\left(  g\right)  .
\end{equation}
These expressions\ are indeed $\Lambda_{\text{UV}}$-independent and
exclusively receive contributions from momenta $k<\mu$. (Both properties have
to become manifest at this stage because they are unaffected by the remaining
integration over $U_{<}$.) Moreover, the above results provide a nontrivial
check of the fact that by restoring residual gauge invariance the integration
over $\phi$ has removed the longitudinal hard-mode contributions.

\subsection{Soft-mode contribution in the disordered phase}

It remains to calculate the chromoelectric soft-mode contribution (\ref{e2u})
to the vacuum energy density. Here the impact of the generalized IR covariance
(\ref{gm1}) becomes\ fully nonperturbative and the dependence on the
variational parameters enters partly through the solutions of the gap
equation. In order to evaluate Eq. (\ref{e2u}), we first perform the remaining
part of the $U_{<}$ integration (cf. Eq. (\ref{hsmvev})), i.e. we integrate
over $\Sigma,$ which yields
\begin{equation}
\left\langle E^{2}\right\rangle _{U_{<}}=-\frac{1}{4g^{2}}\int d^{3}z_{1}\int
d^{3}z_{2}G_{<}^{-1}\left(  \vec{x}-\vec{z}_{1}\right)  G_{<}^{-1}\left(
\vec{x}-\vec{z}_{2}\right)  \left\langle L_{<,i}^{a}\left(  \vec{z}%
_{1}\right)  L_{<,i}^{a}\left(  \vec{z}_{2}\right)  \right\rangle . \label{t}%
\end{equation}
(Recall that in the mean-field approximation of Sec. \ref{mfa} the $\Sigma$
integration amounts to evaluating $\left\langle \left\langle L_{<,i}%
^{a}\left(  \vec{z}_{1}\right)  L_{<,i}^{a}\left(  \vec{z}_{2}\right)
\right\rangle \right\rangle $ at the saddle-point solution $\bar{\Sigma
}\left(  \mu,c_{1}\right)  =\left[  \mu\bar{\xi}\left(  \mu,c_{1}\right)
\right]  ^{2}$ of the gap equation (\ref{geq}).) We further note that the
integrals over $G^{-1}\left(  \vec{x}-\vec{z}_{1}\right)  G^{-1}\left(
\vec{x}-\vec{z}_{2}\right)  $ in Eq. (\ref{t}) have most of their support at
distances $\left\vert \vec{z}_{1}-\vec{z}_{2}\right\vert <\mu^{-1}$ where
$U_{<}\left(  \vec{z}_{2}\right)  U_{<}^{\dagger}\left(  \vec{z}_{1}\right)
\simeq1$. This allows us to approximate
\begin{align}
\left\langle L_{<,i}^{a}\left(  \vec{z}_{1}\right)  L_{<,i}^{a}\left(  \vec
{z}_{2}\right)  \right\rangle  &  =2\left\langle tr\left\{  \partial_{i}%
U_{<}^{\dagger}\left(  \vec{z}_{2}\right)  U_{<}\left(  \vec{z}_{2}\right)
U_{<}^{\dagger}\left(  \vec{z}_{1}\right)  \partial_{i}U_{<}\left(  \vec
{z}_{1}\right)  \right\}  \right\rangle \label{ll}\\
&  \overset{\left\vert \vec{z}_{1}-\vec{z}_{2}\right\vert <\mu^{-1}}{\simeq
}2\left\langle tr\left\{  \partial_{i}U_{<}^{\dagger}\left(  \vec{z}%
_{2}\right)  \partial_{i}U_{<}\left(  \vec{z}_{1}\right)  \right\}
\right\rangle
\end{align}
(as in the $2U$ approximation (\ref{l2}) to the Lagrangian (\ref{lc})). From
Eq. (\ref{uu}) one then has%
\begin{equation}
\left\langle L_{<,i}^{a}\left(  \vec{z}_{1}\right)  L_{<,i}^{a}\left(  \vec
{z}_{2}\right)  \right\rangle \simeq2N_{c}^{2}\partial_{z_{2},i}\left[
\Delta^{-1}\left(  \vec{z}_{2}-\vec{z}_{1}\right)  +\frac{c_{1}\gamma}%
{\zeta\mu^{3}}\bar{\Delta}^{-1}\left(  \vec{z}_{2}-\vec{z}_{1}\right)
\right]  \overleftarrow{\partial}_{z_{1},i}%
\end{equation}
and with Eqs. (\ref{d}), (\ref{dm1(k)}) (specialized to $c_{n\geq2}=0$) and
(\ref{m}) further%
\begin{align}
\left\langle L_{<,i}^{a}\left(  \vec{z}_{1}\right)  L_{<,i}^{a}\left(  \vec
{z}_{2}\right)  \right\rangle  &  \simeq4N_{c}\pi^{2}\frac{\gamma}{\zeta}%
\mu^{2}\int\frac{d^{3}\kappa}{\left(  2\pi\right)  ^{3}}\kappa^{2}%
\theta\left(  1-\kappa^{2}\right)  e^{i\vec{\kappa}\mu\left(  \vec{z}_{2}%
-\vec{z}_{1}\right)  }\nonumber\\
&  \times\left[  \frac{1}{\kappa^{2}\left(  1-c_{1}\kappa^{2}\right)
+\bar{\xi}^{2}}+\frac{2c_{1}\gamma\zeta^{-1}\tilde{\imath}_{2}\left(  \bar
{\xi},c_{1}\right)  \kappa^{2}}{\left[  \kappa^{2}\left(  1-c_{1}\kappa
^{2}\right)  +\bar{\xi}^{2}\right]  ^{2}}\right]
\end{align}
which has the coincidence limit%
\begin{equation}
\left\langle L_{<,i}^{a}\left(  \vec{z}\right)  L_{<,i}^{a}\left(  \vec
{z}\right)  \right\rangle \simeq2N_{c}\frac{\gamma}{\zeta}\mu^{2}\left[
\tilde{\imath}_{2}\left(  \bar{\xi},c_{1}\right)  +2c_{1}\frac{\gamma}{\zeta
}\tilde{\imath}_{2}\left(  \bar{\xi},c_{1}\right)  \tilde{j}_{3}\left(
\bar{\xi},c_{1}\right)  \right]  . \label{2l}%
\end{equation}
With the explicit expression (\ref{gm1}) for the generalized IR\ covariance
(with $c_{i\geq2}=0$) Eq. (\ref{t}) evaluates further to
\begin{align}
\left\langle E^{2}\right\rangle _{U_{<}}  &  =-\frac{m_{g}^{2}}{4g^{2}}%
\int\frac{d^{3}k}{\left(  2\pi\right)  ^{3}}\theta\left(  \mu^{2}-\vec{k}%
^{2}\right)  \left(  1-c_{1}\frac{k^{2}}{\mu^{2}}\right)  \int\frac{d^{3}%
p}{\left(  2\pi\right)  ^{3}}\theta\left(  \mu^{2}-\vec{p}^{2}\right)  \left(
1-c_{1}\frac{p^{2}}{\mu^{2}}\right) \nonumber\\
&  \times\int d^{3}z_{1}\int d^{3}z_{2}e^{i\vec{k}\left(  \vec{x}-\vec{z}%
_{1}\right)  }e^{i\vec{p}\left(  \vec{x}-\vec{z}_{2}\right)  }\left\langle
L_{<,i}^{a}\left(  \vec{z}_{1}\right)  L_{<,i}^{a}\left(  \vec{z}_{2}\right)
\right\rangle \label{a}%
\end{align}
and, after inserting Eq. (\ref{2l}), assumes its final form%
\begin{equation}
\left\langle E^{2}\right\rangle _{U_{<}}=-\frac{N_{c}^{2}\zeta}{2\pi^{2}}%
\mu^{4}\left[  \tilde{\imath}_{2}-2c_{1}\tilde{\imath}_{3}+c_{1}^{2}%
\tilde{\imath}_{4}+2c_{1}\frac{\gamma}{\zeta}\tilde{\imath}_{2}\left(
\tilde{j}_{3}-2c_{1}\tilde{j}_{4}+c_{1}^{2}\tilde{j}_{5}\right)  \right]
\label{t2u}%
\end{equation}
where the integrals $\tilde{\imath}_{n},$ $\tilde{j}_{n}$ are understood to be
evaluated as $\tilde{\imath}_{n}\left(  \bar{\xi}\left(  \mu,c_{1}\right)
,c_{1}\right)  ,$ $\tilde{j}_{n}\left(  \bar{\xi}\left(  \mu,c_{1}\right)
,c_{1}\right)  $. (Note that these integrals receive most of their
contributions from vacuum modes with momenta $k\sim\mu$ (cf. App. \ref{stacor}).)

The rather complex parameter dependence of Eq. (\ref{t2u}) simplifies
considerably for small $\left\vert c_{1}\right\vert \ll1$. Indeed, after
specializing the IR mass parameter $\zeta$ to $\zeta_{\text{ct}}\left(
c_{1}\right)  =\left(  1-c_{1}\right)  ^{-1}=1+c_{1}+O\left(  c_{1}%
^{2}\right)  $ (which ensures a continuous covariance, cf. Eq. (\ref{zct}))
and using the small-$c_{1}$ expansions (\ref{inexp}), (\ref{jnexp}) of the
integrals $\tilde{\imath}_{n}$ and $\tilde{j}_{n}$, the soft-mode contribution
$\varepsilon_{U_{<}}=\left\langle E^{2}\right\rangle _{U_{<}}/2$ to the energy
density becomes
\begin{align}
\varepsilon_{U_{<}}\left(  \mu,c_{1},\zeta_{\text{ct}};\bar{\xi}\right)   &
=-\frac{N_{c}^{2}}{4\pi^{2}}\mu^{4}\left\{  \tilde{\imath}_{2}\left(  \bar
{\xi}\right)  +c_{1}\left[  \tilde{\imath}_{2}\left(  \bar{\xi}\right)
-2\tilde{\imath}_{3}\left(  \bar{\xi}\right)  +\tilde{j}_{4}\left(  \bar{\xi
}\right)  +2\gamma\tilde{\imath}_{2}\left(  \bar{\xi}\right)  \tilde{j}%
_{3}\left(  \bar{\xi}\right)  \right]  \right\}  +O\left(  c_{1}^{2}\right) \\
&  \simeq-\frac{N_{c}^{2}}{4\pi^{2}}\mu^{4}\left(  1+c_{1}\right)
\tilde{\imath}_{2}\left(  \bar{\xi}\right)
\end{align}
for $\left\vert c_{1}\right\vert \ll1$. (The last expression provides a very
reasonable approximation for $\left\vert c_{1}\right\vert \lesssim0.1$.) With
$c_{1}=0$ and $m_{\text{g}}=\mu$ (and approximating $N_{c}^{2}-1\sim N_{c}%
^{2}$), finally, the above expression reduces to the result
\begin{equation}
\varepsilon_{U_{<}}\left(  \mu,c_{1}=0,\zeta=1;\bar{\xi}\right)  =-\frac
{N_{c}^{2}}{4\pi^{2}}\mu^{4}\tilde{\imath}_{2}\left(  \bar{\xi}\right)
=-\frac{N_{c}^{2}}{4\pi^{2}}\mu^{4}\left[  \frac{1}{3}-\bar{\xi}^{2}\left(
1-\bar{\xi}\arctan\frac{1}{\bar{\xi}}\right)  \right]
\end{equation}
of Ref. \cite{kog95}, as it should.

\subsection{Energy density in the disordered phase (i.e. for $\mu\ll
\Lambda_{\text{UV}}$)}

The (normalized) integral over the auxiliary field $\Sigma$ turns the matrix
element $\left\langle \left\langle \mathcal{H}_{\text{YM}}\right\rangle
\right\rangle $ into the trial vacuum energy density $\left\langle
\mathcal{H}_{\text{YM}}\right\rangle $ (cf. Eq. (\ref{hsmvev})). This integral
is nontrivial only for the $U_{<}$ and hence $\xi$ dependent part (\ref{e2u})
of the integrand (\ref{e2}). Its evaluation in the previous section yielded
Eq. (\ref{t2u}). Combining these results and separating the complete vacuum
energy density $\varepsilon=E/V=\left\langle \mathcal{H}_{\text{YM}%
}\right\rangle $ into hard and soft contributions,%
\begin{equation}
\varepsilon\left(  \mu,c_{1},\zeta;\bar{\xi}\right)  =\left\langle
\mathcal{H}_{\text{YM}}\right\rangle =\varepsilon_{>}\left(  \mu\right)
+\varepsilon_{<}\left(  \mu,c_{1},\zeta;\bar{\xi}\right)  \label{etot}%
\end{equation}
(with $\zeta$ still unspecified), one finds%
\begin{equation}
\varepsilon_{>}\left(  \mu\right)  =\frac{1}{2}\left(  N_{c}^{2}-1\right)
\int\frac{d^{3}k}{\left(  2\pi\right)  ^{3}}\left[  \theta\left(
\Lambda_{\text{UV}}^{2}-k^{2}\right)  -\theta\left(  \mu^{2}-k^{2}\right)
\right]  \left[  G_{>}^{-1}\left(  k\right)  +k^{2}G_{>}\left(  k\right)
\right]
\end{equation}
and%
\begin{equation}
\varepsilon_{<}\left(  \mu,c_{1},\zeta;\bar{\xi}\right)  =\frac{1}{2}\left(
N_{c}^{2}-1\right)  \int\frac{d^{3}k}{\left(  2\pi\right)  ^{3}}\theta\left(
\mu^{2}-k^{2}\right)  \left[  \frac{3}{2}G_{<}^{-1}\left(  k\right)
+k^{2}G_{<}\left(  k\right)  \right]  +\frac{1}{2}\left\langle E^{2}%
\right\rangle _{U_{<}}\left(  \mu,c_{1},\zeta;\bar{\xi}\right)  . \label{es}%
\end{equation}
The energy density $\varepsilon_{>}$ of the hard modes involves only
$G_{>}^{-1}\left(  k\right)  =k=$ $k^{2}G_{>}\left(  k\right)  $ and evaluates
further to%
\begin{equation}
\varepsilon_{>}\left(  \mu\right)  =2\left(  N_{c}^{2}-1\right)  \int
\frac{d^{3}k}{\left(  2\pi\right)  ^{3}}\left[  \theta\left(  \Lambda
_{\text{UV}}^{2}-k^{2}\right)  -\theta\left(  \mu^{2}-k^{2}\right)  \right]
\frac{k}{2}=\frac{N_{c}^{2}-1}{8\pi^{2}}\left(  \Lambda_{\text{UV}}^{4}%
-\mu^{4}\right)  , \label{eg}%
\end{equation}
which is the (regularized) zero-point energy density of the two transverse,
\emph{massless} vector modes with energy $\omega\left(  k\right)  =k$ (recall
that the integration over$\ \phi$ has removed the longitudinal-mode
contribution) and reflects the built-in asymptotic freedom of the $k>\mu$
modes. As anticipated, the subtraction of the free vacuum energy density
(\ref{h0}) in the course of normal-ordering cancels its $\Lambda_{\text{UV}}$ dependence.

The soft-mode contribution is mainly of nonperturbative origin and therefore
structurally more involved. Inserting the covariance (\ref{gm}) and its
inverse (\ref{gk}) (for $c_{n\geq2}=0,$ as before, and $\zeta=m_{\text{g}}%
/\mu$) into Eq. (\ref{es}) yields%
\begin{equation}
\varepsilon_{<}=\frac{N_{c}^{2}-1}{4\pi^{2}}\left[  \frac{\zeta}{2}\left(
1-\frac{3c_{1}}{5}\right)  -\frac{1}{\zeta c_{1}^{2}}\left(  1+\frac{c_{1}}%
{3}-\frac{\operatorname{arctanh}\sqrt{c_{1}}}{\sqrt{c_{1}}}\right)  \right]
\mu^{4}+\frac{\left\langle E^{2}\right\rangle _{U_{<}}}{2}, \label{epss}%
\end{equation}
so that the total energy density (\ref{etot}), after dropping the constant
zero-point contribution $\left(  N_{c}^{2}-1\right)  \Lambda_{\text{UV}}%
^{4}/\left(  8\pi^{2}\right)  $, becomes
\begin{equation}
\varepsilon=\frac{N_{c}^{2}-1}{4\pi^{2}}\frac{\mu^{4}}{\zeta c_{1}^{2}}\left[
-\frac{3\left(  3\zeta c_{1}-5\zeta+5\right)  \zeta c_{1}^{2}+10c_{1}+30}%
{30}+\frac{\operatorname{arctanh}\sqrt{c_{1}}}{\sqrt{c_{1}}}\right]
+\frac{\left\langle E^{2}\right\rangle _{U_{<}}}{2}.
\end{equation}
After further approximating $N_{c}^{2}-1\simeq N_{c}^{2}$, ensuring the
continuity of $G_{<}\left(  k\right)  $ by imposing Eq. (\ref{zct}) and using
Eq. (\ref{t2u}), we obtain our final expression for the energy density in the
disordered phase,%
\begin{align}
\bar{\varepsilon}\left(  \mu,c_{1}\right)  :=  &  \text{ }\varepsilon\left(
\mu,c_{1},\zeta_{\text{ct}}\left(  c_{1}\right)  ;\bar{\xi}\left(  \mu
,c_{1}\right)  \right) \nonumber\\
=  &  -\frac{N_{c}^{2}}{4\pi^{2}}\mu^{4}\left[  \frac{4c_{1}^{3}+10c_{1}%
^{2}-50c_{1}+30}{30c_{1}^{2}\left(  1-c_{1}\right)  }-\frac{1-c_{1}}{c_{1}%
^{2}}\frac{\operatorname{arctanh}\sqrt{c_{1}}}{\sqrt{c_{1}}}\right.
\nonumber\\
&  +\left.  \frac{\tilde{\imath}_{2}-2c_{1}\tilde{\imath}_{3}+c_{1}^{2}%
\tilde{\imath}_{4}+2\gamma c_{1}\left(  1-c_{1}\right)  \tilde{\imath}%
_{2}\left(  \tilde{j}_{3}-2c_{1}\tilde{j}_{4}+c_{1}^{2}\tilde{j}_{5}\right)
}{1-c_{1}}\right]  , \label{e}%
\end{align}
where the integrals $\tilde{\imath}_{n},$ $\tilde{j}_{n}$ are evaluated at
$\bar{\xi}\left(  \mu,c_{1}\right)  $. For $c_{1}\rightarrow1$ several terms
of the energy density (\ref{e}) diverge and prevent the vacuum from
encountering the limiting instability.\ For $\left\vert c_{1}\right\vert \ll
1$, on the other hand, the rather complex $c_{1}$ dependence of the full
vacuum energy density (\ref{e}) simplifies to
\begin{equation}
\bar{\varepsilon}\left(  \mu,c_{1}\right)  =\frac{N_{c}^{2}}{4\pi^{2}}\mu
^{4}\left\{  \frac{1}{5}-\tilde{\imath}_{2}\left(  \bar{\xi}\right)  -\left[
\frac{1}{7}+\tilde{\imath}_{4}\left(  \bar{\xi}\right)  -2\tilde{\imath}%
_{3}\left(  \bar{\xi}\right)  +2\gamma\tilde{\imath}_{2}\left(  \bar{\xi
}\right)  \tilde{j}_{3}\left(  \bar{\xi}\right)  \right]  c_{1}\right\}
+O\left(  c_{1}^{2}\right)  \label{ea}%
\end{equation}
(where the solution $\bar{\xi}$ of the gap equation is evaluated at $c_{1}%
=0$). For $c_{1}=0$ and $\zeta=1$, finally, it reduces as expected to the
energy density of Ref. \cite{kog95},%
\begin{equation}
\varepsilon\left(  \mu,c_{1}=0,\zeta=1;\bar{\xi}\left(  \mu,c_{1}=0\right)
\right)  =\frac{N_{c}^{2}}{4\pi^{2}}\mu^{4}\left[  -\frac{2}{15}+\bar{\xi}%
^{2}\left(  1-\bar{\xi}\arctan\frac{1}{\bar{\xi}}\right)  \right]  .
\end{equation}
The full energy density (\ref{e}) is plotted in Fig. \ref{ed} and will be
discussed further in Sec. \ref{res}.

\subsection{Energy density in the ordered phase (i.e. for $\mu\ggg
\Lambda_{\text{YM}}$)}

\label{edo}

As argued above, the mean-field approximation is reliable in the disordered
phase, i.e. for those $\mu<\mu_{c}\leq8.86\Lambda_{\text{YM}}$ (cf. Eq.
\ref{musolrg}) for which $\bar{\xi}$ is not too small. In order to get a more
complete picture of the vacuum energy density and its $\mu$ dependence,
however, one has to evaluate the soft-mode contributions in the ordered phase
as well, i.e. for $\mu\ggg\Lambda_{\text{YM}}$ where the mean-field
approximation breaks down. However, in this phase the Yang-Mills coupling
becomes small \footnote{Numerical studies \cite{kog82} show that the phase
transition in the nonlinear $\sigma$-model (\ref{L0}) (i.e. for $c_{n}=0$) is
strongly first order and occurs when $\left\langle U\right\rangle \gtrsim0.5$,
indicating that perturbation theory should remain qualitatively reliable down
to the transition point.}, i.e.%
\begin{equation}
g^{2}\left(  \mu\ggg\Lambda_{\text{YM}}\right)  \ll1,
\end{equation}
so that the soft-mode dynamics can instead be treated perturbatively. After
restricting ourselves in accordance with our previous approximation scheme to
the $2U$ contributions in Eqs. (\ref{L0}) and (\ref{l2}) (again with
$c_{i\geq2}=0$), this dynamics becomes
\begin{equation}
\Gamma_{<}^{\left(  2U\right)  }\left[  U_{<}\right]  =-\frac{m_{\text{g}}%
}{2g^{2}}\int d^{3}ztr\left\{  U_{<}^{\dagger}\left(  \partial^{2}+\frac
{c_{1}}{\mu^{2}}\partial^{4}\right)  U_{<}\right\}  .
\end{equation}
Since in the ordered phase fluctuations $\varphi_{<}^{a}$ around $U_{<}\sim1$
are small, one can furthermore truncate the weak-coupling expansion of $U_{<}%
$, i.e.
\begin{equation}
U_{<}=\exp\left(  ig\varphi_{<}^{a}\lambda^{a}\right)  =1+ig\varphi_{<}%
^{a}\lambda^{a}+O\left(  g^{2}\right)  ,
\end{equation}
which to $O\left(  g^{2}\right)  $ yields (with $tr\left\{  \lambda^{a}%
\lambda^{b}\right\}  =2\delta^{ab}$ and denoting the Fourier transform of
$\varphi$ as $\tilde{\varphi}$) the bilinear action
\begin{equation}
\Gamma_{<}^{\left(  2U\right)  }\left[  \varphi_{<}\right]  =\frac{1}{2}%
\int\frac{d^{3}k}{\left(  2\pi\right)  ^{3}}\left[  \tilde{\varphi}_{<}%
^{a}\left(  -\vec{k}\right)  2m_{\text{g}}\left(  k^{2}-\frac{c_{1}}{\mu^{2}%
}k^{4}\right)  \tilde{\varphi}_{<}^{a}\left(  \vec{k}\right)  +...\right]  .
\end{equation}
From this \textquotedblleft kinetic\textquotedblright\ term one reads off the
static $k<\mu$ propagator of the $\varphi$ modes (which contains, in contrast
to the soft-mode propagator (\ref{d}), (\ref{dm1(k)}) in the disordered phase
and as a consequence of Goldstone's theorem, no mass term) as
\begin{equation}
\left\langle \varphi_{<}^{a}\left(  \vec{z}_{1}\right)  \varphi_{<}^{b}\left(
\vec{z}_{2}\right)  \right\rangle =\frac{\delta^{ab}}{2m_{\text{g}}}\int
\frac{d^{3}k}{\left(  2\pi\right)  ^{3}}\theta\left(  \mu^{2}-k^{2}\right)
\frac{e^{-i\vec{k}\left(  \vec{z}_{1}-\vec{z}_{2}\right)  }}{k^{2}-\frac
{c_{1}}{\mu^{2}}k^{4}}.
\end{equation}
The correlator (\ref{ll}), which determines the chromoelectric vacuum energy
density due to the soft modes, then becomes (with $\delta^{aa}=N_{c}^{2}-1$)
\begin{align}
\left\langle L_{<,i}^{a}\left(  \vec{z}_{1}\right)  L_{<,i}^{a}\left(  \vec
{z}_{2}\right)  \right\rangle  &  \simeq2\left\langle tr\left\{  \partial
_{i}U_{<}^{\dagger}\left(  \vec{z}_{1}\right)  \partial_{i}U_{<}\left(
\vec{z}_{2}\right)  \right\}  \right\rangle \simeq4g^{2}\left\langle
\partial_{i}\varphi_{<}^{a}\left(  \vec{z}_{1}\right)  \partial_{i}\varphi
_{<}^{a}\left(  \vec{z}_{2}\right)  \right\rangle \\
&  =2\left(  N_{c}^{2}-1\right)  \frac{g^{2}\mu^{2}}{\zeta}\int\frac
{d^{3}\kappa}{\left(  2\pi\right)  ^{3}}\frac{\theta\left(  1-\kappa
^{2}\right)  }{1-c_{1}\kappa^{2}}e^{-i\mu\vec{\kappa}\left(  \vec{z}_{1}%
-\vec{z}_{2}\right)  }.
\end{align}
After inserting this result into Eq. (\ref{a}) one obtains%
\begin{align}
\left\langle E^{2}\right\rangle _{U_{<}}  &  =-\left(  N_{c}^{2}-1\right)
\frac{\zeta\mu^{4}}{2}\int\frac{d^{3}\kappa}{\left(  2\pi\right)  ^{3}}%
\theta\left(  1-\kappa^{2}\right)  \left(  1-c_{1}\kappa^{2}\right) \\
&  =-\frac{N_{c}^{2}-1}{4\pi^{2}}\zeta\mu^{4}\left(  \frac{1}{3}-\frac{1}%
{5}c_{1}\right)  .
\end{align}
Hence Eq. (\ref{epss}) yields the infrared contributions%
\begin{equation}
\varepsilon_{<}=\frac{N_{c}^{2}-1}{4\pi^{2}}\frac{\mu^{4}}{\zeta c_{1}^{2}%
}\left(  \frac{-3\zeta^{2}c_{1}^{3}+5\zeta^{2}c_{1}^{2}-5c_{1}-15}{15}%
+\frac{\operatorname{arctanh}\sqrt{c_{1}}}{\sqrt{c_{1}}}\right)
\end{equation}
to the energy density which specialize with $\zeta_{\text{ct}}\left(
c_{1}\right)  =\left(  1-c_{1}\right)  ^{-1}$ to%
\begin{equation}
\varepsilon_{<}=\frac{N_{c}^{2}-1}{4\pi^{2}}\mu^{4}\frac{1-c_{1}}{c_{1}^{2}%
}\left[  \frac{-8c_{1}^{3}+25c_{1}-15}{15\left(  1-c_{1}\right)  ^{2}}%
+\frac{\operatorname{arctanh}\sqrt{c_{1}}}{\sqrt{c_{1}}}\right]  . \label{el}%
\end{equation}
Together with the hard-mode contribution (\ref{eg}) (which is identical in
both vacuum phases) and after discarding the zero-point contribution, finally,
the total vacuum energy density in the ordered phase becomes%
\begin{align}
\varepsilon\left(  \mu,c_{1}\right)   &  =\frac{N_{c}^{2}-1}{4\pi^{2}}\mu
^{4}\frac{1-c_{1}}{c_{1}^{2}}\left[  -\frac{c_{1}^{3}+15c_{1}^{2}-50c_{1}%
+30}{30\left(  1-c_{1}\right)  ^{2}}+\frac{\operatorname{arctanh}\sqrt{c_{1}}%
}{\sqrt{c_{1}}}\right] \label{elmu}\\
&  =\frac{N_{c}^{2}-1}{4\pi^{2}}\mu^{4}\left[  \frac{1}{30}+\frac{8}{105}%
c_{1}+\frac{32}{315}c_{1}^{2}+O\left(  c_{1}^{3}\right)  \right]  .
\end{align}

Several properties of Eq. (\ref{elmu}) are noteworthy. First of all, in the
$c_{1}\rightarrow0$ limit it reduces, as expected, to the result of Ref.
\cite{kog95} (cf. their Eq. (4.24)),
\begin{equation}
\varepsilon\left(  \mu,c_{1}=0\right)  =\frac{N_{c}^{2}-1}{120\pi^{2}}\mu^{4}.
\end{equation}
The total energy density diverges for $c_{1}\rightarrow1$ where the vacuum
wave functionals would become unnormalizable, furthermore, which prevents
$c_{1}$ from growing beyond unity during energy minimization. Most
importantly, however, for $\mu\gg\Lambda_{\text{YM}}$ the negative hard-mode
contribution to the energy density (i.e. Eq. (\ref{eg}) without the
$\Lambda_{\text{UV}}^{4}$ term) is overcome by the positive contribution
(\ref{el}) from the soft modes. Indeed, the (large $\mu$) vacuum energy
density (\ref{elmu}) is positive for all $c_{1}<1$ and hence a monotonically
\emph{increasing} function of $\mu$. In addition, Eq. (\ref{elmu}) increases
monotonically with $c_{1}$ in the range $-2<c_{1}<1$ which includes those
$\left\vert c_{1}\right\vert \ll1$ values for which our perturbative $O\left(
c_{1}\right)  $ treatment is reliable (and thus our variational result
$c_{1}^{\ast}\simeq0.15$ to be determined in Sec. \ref{res}). Under these
circumstances it is reasonable to expect that the perturbative result
(\ref{elmu}) remains at least qualitatively reliable for $\mu$ values down to
the phase transition \cite{kog95} at $\mu_{\text{c}}\lesssim8.86\Lambda
_{\text{YM}}$.

\section{Variational analysis}

\label{vmin}

We are now prepared to minimize the vacuum energy density in our
trial-functional family according to the Rayleigh-Ritz variational principle.
Above we found that (inside the validity range of our approximations) the
energy density (\ref{e}) in the strongly-coupled disordered phase decreases
monotonically with increasing $\mu$ up to the phase transition, whereas its
counterpart (\ref{elmu}) in the ordered weak-coupling phase monotonically
increases both with $\mu$ and $c_{1}$. The combination of these results
indicates that the vacuum energy density attains its minimum at the phase
boundary in the disordered phase, where the number of massless particles
becomes maximal. Indeed, the number $N_{c}^{2}-1$ of massless Goldstone modes
in the ordered phase (at very large $\mu$ all other modes have $m\sim
\Lambda_{\text{UV}}$ and are too massive to contribute significantly) doubles
at the transition where they are joined by degenerate parity partners. Hence
the massless particles generate roughly twice the internal energy at the
second-order transition, which becomes maximal at the critical $\mu_{\text{c}%
}^{\ast}$ \cite{kog95}. (This argument is not affected by the higher-gradient
interactions (\ref{lc}) since those share the symmetries of the leading term
(\ref{L0}).)

Our program for the following subsections will be as follows: after
determining the energy minimum at the boundary of the disordered phase
quantitatively (Sec. \ref{res}), we evaluate the associated, four-dimensional
gluon condensate in Sec. \ref{gcond}. The resulting vacuum field distribution
and its physical interpretation are discussed in Sec. \ref{phimpl}, and its
impact on the phase structure of the soft-mode dynamics is subject of Sec.
\ref{phstr}. In Sec. \ref{cng2} we comment on the qualitative role of the
higher-gradient interactions in the soft-mode Lagrangian (\ref{lc}). Finally,
in Sec. \ref{out}, we mention several options for improvements and future
applications of the gauge-invariant variational framework.

\subsection{Vacuum energy minimization}

\label{res}

\begin{figure}[ptb]
\begin{center}
\includegraphics[height = 9cm]{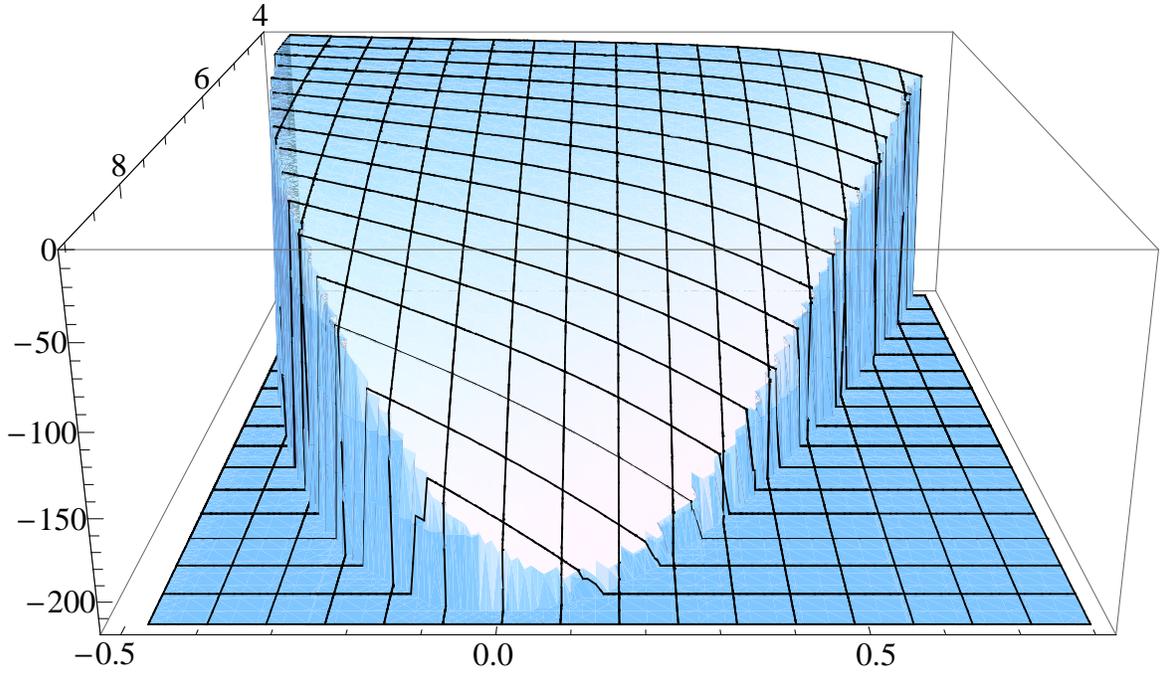}
\end{center}
\caption{The energy density $\bar{\varepsilon}\left(  \mu,c_{1}\right)  $ of
the vacuum field solution $\bar{\xi}\left(  \mu,c_{1}\right)  $ in the
parameter ranges $\mu\in\left\{  4,9\right\}  \Lambda_{\text{YM}}$ and
$c_{1}\in\left\{  -0.5,0.8\right\}  .$ Note the minimum of the energy surface
at $c_{1}\simeq0.15$.}%
\label{ed}%
\end{figure}

In Fig. \ref{ed} the energy density (\ref{e}) of the disordered vacuum phase
(calculated on the basis of the numerical solution of the gap equation, cf.
Fig. \ref{gsoln}) is plotted as a function of $\mu$ and $c_{1}$. The plot
range of $\mu$ includes those regions in which (i)\ the one-loop evaluation of
the hard-mode contributions remains reasonably accurate (corresponding roughly
to $\mu\geq4\Lambda_{\text{YM}}$), (ii) the system stays in the disordered
phase where the nontrivial solution $\bar{\xi}$ of the gap equation exists,
and (iii) the mean-field approximation remains approximately valid. Together
with Eq. (\ref{musolrg}) these conditions require
\begin{equation}
4\Lambda_{\text{YM}}\lesssim\mu\leq8.86\Lambda_{\text{YM}}.
\end{equation}
For illustrative purposes the plot range of $c_{1}$ is chosen to cover almost
the full existence range $-0.48\lesssim c_{1}\lesssim0.95$ of gap-equation
solutions, on the other hand, although our perturbative treatment of the $4U$
interactions becomes questionable close to the upper limit.

\begin{figure}[ptb]
\begin{center}
\includegraphics[height = 9cm]{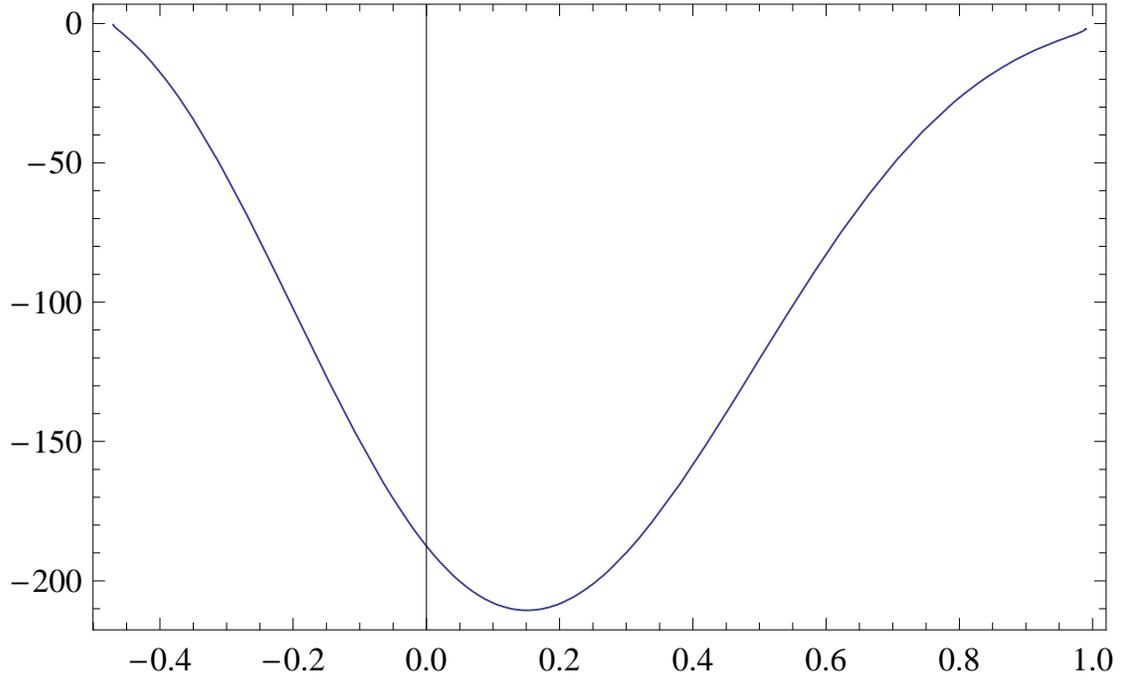}
\end{center}
\caption{The energy density $\bar{\varepsilon}\left(  c_{1}\right)  \equiv
\bar{\varepsilon}\left(  \mu_{c}\left(  c_{1}\right)  ,c_{1}\right)  $ at the
phase transition point $\mu_{c}$ in units of $\Lambda_{\text{YM}}$.}%
\label{enc}%
\end{figure}

The arguably most prominent qualitative feature of the vacuum energy density
$\bar{\varepsilon}$ $\left(  \mu,c_{1}\right)  $ in the disordered phase is
its monotonic decrease with both $\mu$ and $c_{1}$ which continues until the
nontrivial solution of the gap equation ceases to exist at the second-order
phase transition (cf. Fig. \ref{ed}). This essential feature manifests itself
already in the linearization (\ref{ea}) of the $c_{1}$ dependence around
$c_{1}=0$. As a consequence, the energy of the disordered vacuum is minimized
at $\bar{\xi}=0_{+},$ i.e. at the disorder-order phase transition, for each
admissible value of $c_{1}$. In order to find the precise minimum of
$\bar{\varepsilon}$ and the corresponding parameter values $\mu^{\ast}$ and
$c_{1}^{\ast}$, we plot the energy density along the critical line, i.e.%
\begin{equation}
\bar{\varepsilon}\left(  c_{1}\right)  :=\bar{\varepsilon}\left(
\mu_{\text{c}}\left(  c_{1}\right)  ,c_{1}\right)  ,
\end{equation}
in the range $c_{1}\in\left[  -0.48,1\right]  $ where $\bar{\xi}\left(
\mu_{\text{c}}\left(  c_{1}\right)  ,c_{1}\right)  =0$. As can be read off
from Fig. \ref{enc}, the minimum of $\bar{\varepsilon}\left(  c_{1}\right)  $
is attained at
\begin{equation}
c_{1}^{\ast}\simeq0.15\text{ \ \ \ \ \ with \ \ \ \ \ }\mu^{\ast}%
=\mu_{\text{c}}\left(  c_{1}^{\ast}\right)  =8.61\Lambda_{\text{YM}}.
\label{res1}%
\end{equation}
Evaluating $\alpha_{\text{YM}}\left(  \mu^{\ast}\right)  =g_{\text{YM}}%
^{2}\left(  \mu^{\ast}\right)  /\left(  4\pi\right)  \simeq0.27$ (to one loop)
at this scale confirms that the running coupling remains small enough to
justify the perturbative treatment of the hard modes. (It exceeds
$\alpha_{\text{YM}}\left(  \mu_{\text{c}}\left(  c_{1}=0\right)  \right)
\simeq0.26$ by less than 2\%.) Similarly, the resulting $c_{1}^{\ast}\ll1$
justifies our perturbative treatment of the $4U$ contributions. In addition,
the rather small curvature of the critical line (\ref{mu}) around its maximum
\begin{equation}
\mu_{\text{c}}^{(\text{max)}}=\mu_{\text{c}}\left(  c_{1}=0\right)
=\exp\left(  \frac{24}{11}\right)  \simeq8.86\Lambda_{\text{YM}} \label{mcmax}%
\end{equation}
at $c_{1}=0$ causes our dynamical IR mass scale $\mu^{\ast}\simeq
8.6\Lambda_{\text{YM}}$ to be only about 3\% smaller than $\mu_{\text{c}%
}^{(\text{max)}}$. This has important consequences for the value of the gluon
condensate to be discussed in Sec. \ref{gcond} and for other vacuum scales.
The corresponding IR gluon mass, e.g., turns out to be%
\begin{equation}
m_{\text{g}}^{\ast}=\frac{\mu_{\text{c}}\left(  c_{1}^{\ast}\right)  }%
{1-c_{1}^{\ast}}\simeq10.14\Lambda_{\text{YM}}.
\end{equation}
(For $\Lambda_{\text{YM}}\simeq0.15$ GeV \cite{kog95} this value is somewhat
larger than the potentially related mass $m_{\text{LG}}\sim1.1$ GeV extracted
from the intermediate-momentum behavior of the Landau-gauge gluon propagator
on the lattice (second reference of \cite{sil04}).)

The quantitative improvement of the vacuum description due to our generalized
trial functional basis can be measured in terms of the achieved variational
bound on the Yang-Mills vacuum energy density. Comparing the minimum of our
trial energy density (\ref{e}),
\begin{equation}
\bar{\varepsilon}\left(  c_{1}^{\ast}\right)  \simeq-210.59\Lambda_{\text{YM}%
}^{4},
\end{equation}
to the value $\bar{\varepsilon}\left(  c_{1}=0\right)  \simeq-187.52\Lambda
_{\text{YM}}^{4}$ generated by the lowest-order covariance $G_{<}^{-1}\left(
k\right)  =\mu$ shows that the $c_{1}$ corrections reduce the vacuum energy
density by about 11\%. Hence variational optimization in our extended trial
space has produced a rather substantial improvement of the vacuum functional.
(The massive gluon propagator (\ref{g0}), on the other hand, is energetically
disfavored: its IR behavior is approximately reproduced by $c_{1}=-1/2$ and
yields $\bar{\varepsilon}\left(  c_{1}=-1/2\right)  \simeq0$. We expect
sizeable corrections to the $O\left(  c_{1}\right)  $ treatment of the $4U$
interactions for $c_{1}=-1/2$, though, and further note that Eq. (\ref{g0})
does not match to the UV covariance (\ref{gm1uv}) at $k=\mu$, in contrast to
our IR covariance.)

\subsection{Gluon condensate}

\label{gcond}

Gluon condensates, i.e. the vacuum expectation values of gauge-invariant local
operators composed of gluon fields, are among the key amplitudes which
characterize the Yang-Mills ground state. The most important gluon condensate,
which dominates the power corrections in the operator product expansion e.g.
of the glueball correlators \cite{for05}, is the expectation value of the
lowest-, i.e. four-dimensional gluonic operator $F^{2}\equiv F_{\mu\nu}%
^{a}F^{a,\mu\nu}=-2\left(  E_{i}^{a}E_{i}^{a}-B_{i}^{a}B_{i}^{a}\right)  $. In
our trial state, and to lowest order in the gauge coupling, this condensate
becomes
\begin{align}
\left\langle F^{2}\right\rangle  &  =2\left[  \left\langle B^{2}\right\rangle
-\left\langle E^{2}\right\rangle \right] \label{gcnd}\\
&  =2\left(  N_{c}^{2}-1\right)  \int\frac{d^{3}k}{\left(  2\pi\right)  ^{3}%
}\theta\left(  k^{2}-\Lambda_{\text{UV}}^{2}\right)  \left[  k^{2}G\left(
k\right)  -\frac{3}{2}G^{-1}\left(  k\right)  \right] \nonumber\\
&  +\left(  N_{c}^{2}-1\right)  \int\frac{d^{3}k}{\left(  2\pi\right)  ^{3}%
}\left[  \theta\left(  k^{2}-\Lambda_{\text{UV}}^{2}\right)  -\theta\left(
k^{2}-\mu^{2}\right)  \right]  G^{-1}\left(  k\right)  -2\left\langle
E^{2}\right\rangle _{U_{<}} \label{gc2}%
\end{align}
where we used Eqs. (\ref{e2}) -- (\ref{b2h2}) and again performed the $\Sigma$
integrals in the saddle-point approximation. As anticipated above, the
hard-mode (i.e. $k>\mu$) contributions to the chromo-electric and -magnetic
parts cancel, which indicates that the gluon condensate (\ref{gc2}) is
renormalized at $\mu$. After inserting the IR covariance (\ref{gm}) and the
propagator (\ref{gk}) (with $c_{n\geq2}=0$, as before) the condensate becomes
\begin{align}
\left\langle F^{2}\right\rangle  &  =2\left(  N_{c}^{2}-1\right)  \int
\frac{d^{3}k}{\left(  2\pi\right)  ^{3}}\theta\left(  k^{2}-\mu^{2}\right)
\left[  k^{2}G_{<}\left(  k\right)  -\frac{3}{2}G_{<}^{-1}\left(  k\right)
\right]  -2\left\langle E^{2}\right\rangle _{U_{<}}\label{gcir}\\
&  =-\frac{N_{c}^{2}-1}{\pi^{2}}\frac{\mu^{4}}{\zeta c_{1}^{2}}\left[
1+\frac{c_{1}}{3}+\frac{1}{2}\zeta^{2}c_{1}^{2}\left(  1-\frac{3}{5}%
c_{1}\right)  -\frac{\operatorname{arctanh}\sqrt{c_{1}}}{\sqrt{c_{1}}}\right]
-2\left\langle E^{2}\right\rangle _{U_{<}}.
\end{align}
At the border of the disordered phase, where $\bar{\xi}\left(  \mu_{c}\left(
c_{1}\right)  ,c_{1}\right)  =0$ and where the energy becomes minimal (cf.
Sec. \ref{res}), we furthermore have from the $\xi\rightarrow0$ limit of Eq.
(\ref{t2u}) and the expressions for the integrals $\tilde{\imath}_{n}\left(
0,c_{1}\right)  $ and $\tilde{j}_{n}\left(  0,c_{1}\right)  $ given in App.
\ref{lim} that%
\begin{equation}
\left\langle E^{2}\right\rangle _{U_{<}}\left(  \mu_{c}\left(  c_{1}\right)
,\bar{\xi}\left(  \mu_{c}\left(  c_{1}\right)  ,c_{1}\right)  =0\right)
=-\frac{\left(  N_{c}^{2}-1\right)  \zeta}{2\pi^{2}}\mu^{4}\left[  \frac{1}%
{3}-\frac{c_{1}}{5}-\frac{2\gamma}{3\zeta}\left(  1-\frac
{\operatorname{arctanh}\sqrt{c_{1}}}{\sqrt{c_{1}}}\right)  \right]  .
\end{equation}
Combining the above results and specializing as before to $\zeta_{\text{ct}%
}=\left(  1-c_{1}\right)  ^{-1}$ then results in our final expression%
\begin{equation}
\left\langle F^{2}\right\rangle =-\frac{N_{c}^{2}-1}{\pi^{2}}\mu^{4}\left[
\frac{7c_{1}^{3}-20\gamma^{\ast}c_{1}^{3}+15c_{1}^{2}+20\gamma^{\ast}c_{1}%
^{2}-50c_{1}+30}{30c_{1}^{2}\left(  1-c_{1}\right)  }-\left(  \frac{1-c_{1}%
}{c_{1}^{2}}+\frac{2\gamma^{\ast}}{3}\right)  \frac{\operatorname{arctanh}%
\sqrt{c_{1}}}{\sqrt{c_{1}}}\right]  \label{f}%
\end{equation}
($\gamma^{\ast}=g^{2}\left(  \mu^{\ast}\right)  N_{c}/\pi^{2}\simeq1.012$) for
the gluon condensate. For small $c_{1}$ Eq. (\ref{f}) expands into powers of
$c_{1}$ as%
\begin{equation}
\left\langle F^{2}\right\rangle =\frac{N_{c}^{2}-1}{\pi^{2}}\mu^{4}\left[
\frac{1}{30}-\left(  \frac{13}{105}-\frac{2}{9}\gamma^{\ast}\right)
c_{1}+O\left(  c_{1}^{2}\right)  \right]  ,
\end{equation}
which shows that $\left\langle F^{2}\right\rangle $ grows with $c_{1}$ when
$c_{1}\ll1$. In the uncorrected case, with $c_{1}=0$ and the corresponding
value $\alpha_{\text{YM}}\left(  \mu_{c}\left(  c_{1}=0\right)  \right)
/\pi=1/\left(  4N_{c}\right)  $ or $\gamma^{\ast}\left(  \mu_{c}\left(
c_{c}=0\right)  \right)  =1$ (cf. Eq. (\ref{dimlp})) for the gauge coupling as
well as Eq. (\ref{mcmax}) for the IR scale, and furthermore setting $N_{c}%
^{2}-1\sim N_{c}^{2}$, $N_{c}=3$, one finds \cite{kog95}
\begin{equation}
\left\langle \frac{\alpha}{\pi}F^{2}\right\rangle ^{\left(  \text{KK}\right)
}=\frac{N_{c}}{120\pi^{2}}\mu_{c}^{4}\left(  c_{c}=0\right)  \simeq
15.\allowbreak62\Lambda_{\text{YM}}^{4}. \label{gckk}%
\end{equation}
Our improved value of the gluon condensate, on the other hand, is obtained by
inserting the energy minimizing values $c_{1}^{\ast}\simeq0.151$ and
$\mu^{\ast}=8.606\Lambda_{\text{YM}}$ from the last section and the
corresponding coupling $\gamma^{\ast}\simeq1.012$. The result
\begin{equation}
\left\langle \frac{\alpha}{\pi}F^{2}\right\rangle =20.87\Lambda_{\text{YM}%
}^{4}\simeq0.011\text{ GeV}^{4} \label{gc}%
\end{equation}
is about 25\% larger than the uncorrected value (\ref{gckk}). In the second
equation above we have specialized to $\Lambda_{\text{YM}}\simeq0.15$ GeV,
which allows for comparison with the value range $\left\langle \left(
\alpha/\pi\right)  F^{2}\right\rangle =0.0080-0.024$ GeV$^{4}$ obtained from
QCD sum rules (for references see e.g. Ref. \cite{for05}). (We refrain from
using the currently preferred value for $\Lambda_{\text{QCD}}$ which contains
quark contributions.) Our condensate value prediction (\ref{gc}) lies
comfortably within this standard range.

\subsection{Vacuum field distribution and quasigluon velocity}

\label{phimpl}

\begin{figure}[ptb]
\begin{center}
\includegraphics[height = 9cm]{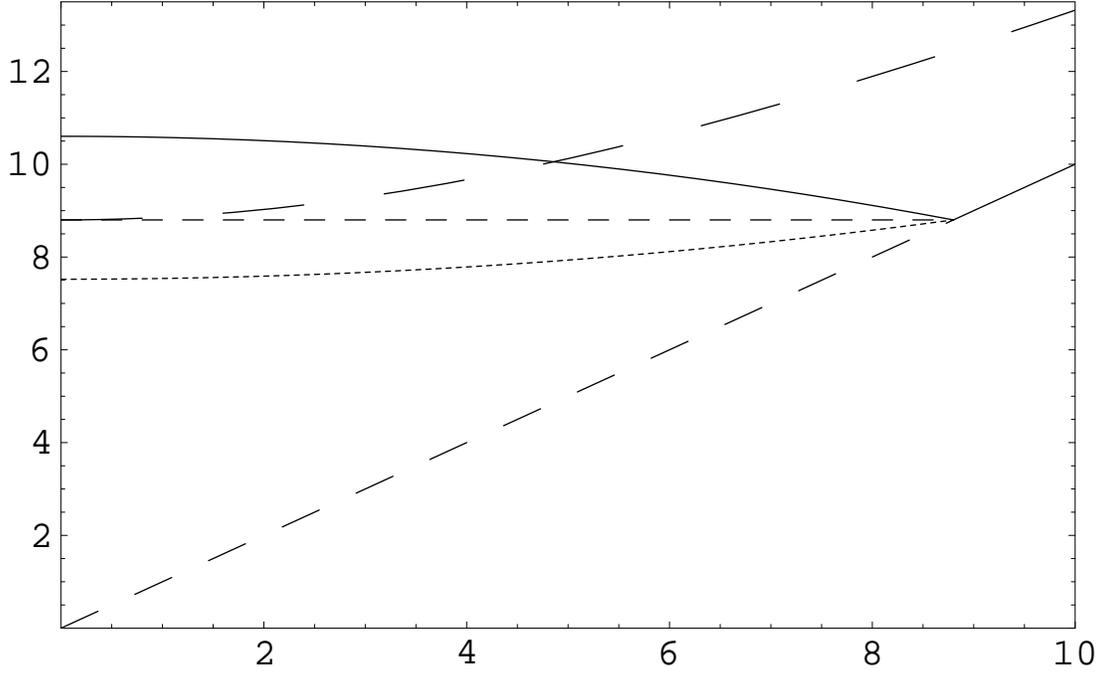}
\end{center}
\caption{The generalized IR covariance $G_{<}^{-1}\left(  k\right)
=\mu\left(  1-c_{1}k^{2}/\mu^{2}\right)  /\left(  1-c_{1}\right)  $ for
$c_{1}=0.17$ (full line) and $c_{1}=-0.17$ (dotted), together with the UV
covariance $G_{>}^{-1}\left(  k\right)  =k$, and in comparison with
$G_{0}^{-1}\left(  k\right)  =k$ (medium dashed), $G_{\text{0,}m_{g}=\mu}%
^{-1}\left(  k\right)  =\sqrt{k^{2}+\mu^{2}}$ (long dashed) and $G_{\text{KK}%
}^{-1}\left(  k\right)  =\mu$ (short dashed) (all curves in units of
$\Lambda_{\text{YM}}$ and for $\mu=8.8\Lambda_{\text{YM}}$).}%
\label{epdel}%
\end{figure}

Our variationally optimized wave functional determines the distribution of the
gauge-field modes $A_{i}\left(  k\right)  $ and thus contains new information
about the field composition in the vacuum state. In our approximation, the IR
covariance (\ref{gm}) with $c_{n\geq2}=0$ and Eq. (\ref{mgc}) takes the form
\begin{equation}
G_{<}^{-1}\left(  k\right)  =\frac{\mu}{1-c_{1}}\left(  1-c_{1}\frac{k^{2}%
}{\mu^{2}}\right)  \theta\left(  \mu^{2}-k^{2}\right)  . \label{disprel}%
\end{equation}
By construction, the covariance (\ref{disprel}) is non-negative in the
physical parameter range $\mu>0,$ $c_{1}<1$ and thus yields a normalizable
wave functional. In addition, $G_{<}^{-1}\left(  k\right)  $ becomes larger
(smaller) for positive (negative) values of $c_{1}.$ This holds for each
$k<\mu$ independently, but gets more pronounced for smaller $k$ (cf. Fig.
\ref{epdel}). Hence the Gaussian weight factors
\begin{equation}
\exp\left[  -\left(  2\pi\right)  ^{-3}d^{3}kG_{<}^{-1}\left(  k\right)
A^{2}\left(  k\right)  /2\right]  ,
\end{equation}
which (when multiplied for all $k<\mu$) make up the IR part (\ref{gir}) of the
unprojected core functional, become narrower (wider) in Fourier-mode space
\footnote{The contributions of the Fourier modes to the vacuum wave functional
are of course also affected by the ($c_{1}$ dependent) normalization
$Z^{-1/2}=\left(  \int D\vec{A}\Psi\left[  \vec{A}\right]  \Psi\left[  \vec
{A}\right]  \right)  ^{-1/2}$. This overall factor does not change the
relative weight of the $A_{i}^{a}\left(  k\right)  $ for different $k$,
however.}. As a consequence, large $A_{i}^{a}\left(  k\right)  $ contributions
to the IR part (\ref{gir}) of the unnormalized Gaussian vacuum functional, and
thus to the functional integrands\ of any amplitude, are increasingly
suppressed (enhanced) towards smaller $k\in\left[  0,\mu\right]  $.

Hence our positive result $c_{1}^{\ast}=0.15$ reshapes the vacuum field
population in a specific way: relative to the uncorrected case $c_{1}\equiv0$
the attractive IR interactions generate a depletion of the
ultralong-wavelength $k\rightarrow0$ modes and an enhancement of the $k\sim
\mu$ modes. This effect helps to prevent the Savvidy instability which
constant chromomagnetic fields experience in the Yang-Mills vacuum
\cite{sav77}. It also contributes to generating the expected average
wavelength
\begin{equation}
\lambda\sim\Lambda_{\text{YM}}^{-1}%
\end{equation}
of the vacuum fields. In contrast, the massive vector covariance (\ref{g0})
with $c_{1}=-1/2$ would populate the medium-soft modes with $k\lesssim\mu$
less strongly (cf. Fig. \ref{epdel}) although this is energetically
disfavored. (When comparing the overall weight of the $k<\mu$ modes associated
with different covariances, the $G^{-1}$ dependent normalization factor
$\mathcal{N}_{G}$ has to be taken into account as well.) Since the
higher-derivative soft-mode interactions in Eq. (\ref{lc}) do not affect the
$k=0$ modes, the above findings underline the importance of finite-momentum IR
modes in shaping the vacuum structure.

Regarding Eq. (\ref{disprel}) as the dispersion relation $\omega\left(
k\right)  $ of \textquotedblleft quasigluon\textquotedblright\ modes\ in the
vacuum, one may further relate the parameter $c_{1}$ to the dimensionless
quasigluon group velocity%
\begin{equation}
\vec{v}\left(  \vec{k}\right)  =\frac{\partial G_{<}^{-1}\left(  \vec
{k}\right)  }{\partial\vec{k}}=-2c_{1}\frac{m_{\text{g}}}{\mu}\frac{\vec{k}%
}{\mu}\overset{m_{\text{g}}=\mu\left(  1-c_{1}\right)  ^{-1}}{\longrightarrow
}-\frac{2c_{1}}{1-c_{1}}\frac{\vec{k}}{\mu}.
\end{equation}
Since our approximate dispersion relation (\ref{disprel}) is quadratic in the
momentum, one may further define a (momentum-independent) effective kinetic
gluon mass%
\begin{equation}
\overline{m}_{\text{g}}=-\frac{\mu^{2}}{2c_{1}m_{\text{g}}}=-\frac{1-c_{1}%
}{2c_{1}}\mu\label{mkin}%
\end{equation}
which relates velocity and momentum as $\vec{k}=\overline{m}_{\text{g}}\vec
{v}$. Note that $\overline{m}_{\text{g}}$ is negative for $0>c_{1}>1$. With
$v\left(  k\right)  :=\left\vert \vec{v}\left(  \vec{k}\right)  \right\vert $
one further has
\begin{equation}
\left\vert c_{1}\right\vert =\frac{\mu}{2m_{\text{g}}}v\left(  \mu\right)
\end{equation}
which shows that $c_{1}$ is related to the quasigluon group velocity at the IR
renormalization point $k=\mu$. Specializing to a continuous covariance by
identifying $\mu/m_{\text{g}}=1-c_{1}$ then yields%
\begin{equation}
\left\vert c_{1}\right\vert =\frac{v\left(  \mu\right)  }{v\left(  \mu\right)
+2}.
\end{equation}
This relation implies among other things that the velocity diverges toward the
bound $c_{1}\rightarrow1$, and that the admissible velocity range is
restricted to $\infty>v\left(  \mu\right)  =2\left\vert c_{1}\right\vert
/\left(  1-c_{1}\right)  \geq-2$. Our result $c_{1}^{\ast}=0.15$ corresponds
to $v^{\ast}:=v\left(  \mu^{\ast}\right)  =\allowbreak0.35$.

The positive sign of $c_{1}^{\ast}$ implies that the effective \emph{kinetic}
mass (\ref{mkin}) is negative, i.e. that the quasigluon velocity in the vacuum
is opposite to its momentum. Hence these quasigluons decelerate when an
external force is applied, in stark contrast to the behavior of free gluons.
(An interesting and related issue is the vacuum response to external electric
and magnetic background fields, cf.\ e.g. Ref. \cite{hei00}.) In this respect
our IR dispersion relation is reminiscent of optical phonon branches in
periodic structures over a Brillouin zone, or of the dispersion of electrons
moving in such structures. In the latter example, the effect is generated by
the lattice exerting a large retarding force on the electron that counteracts
and overcomes the applied force. Of course, in our aperiodic momentum space no
Bloch-type oscillations are generated by external color-electric fields.
Nevertheless, quasigluons (with their small scattering amplitudes) or quarks
could show a negative differential resistance, in remarkable contrast to the
impact of a simple infrared mass term of the type (\ref{g0}). The origin of
these effects in the Yang-Mills dynamics should be investigated further. On a
cautionary note, however, one has to keep in mind that not all properties of
the gauge-\emph{variant} core functional (\ref{ga}) play a physical role (cf.
Sec. \ref{gcf} and Ref. \cite{zar98}).

\subsection{Phase structure and transition order}

\label{phstr}

It is instructive to analyze some additional features of the $c_{1}\neq0$
interactions which have direct impact both on the phase structure of the
soft-mode dynamics (\ref{L0}), (\ref{lc}) and on the vacuum energy minimum. To
begin with, we recall that the vacuum instability for $c_{1}\geq1$ is encoded
in\textbf{ }$c_{1}\rightarrow1$ singularities of the energy density and in
zeros of the amplitude entering the gap equation. Hence the constraint
$c_{1}<1$ does not have to be imposed by hand but is automatically enforced by
the diverging vacuum energy density and by the nontrivial saddle-point
solution ceasing to exist. Since the energy density (\ref{e}) decreases
monotonically with $\mu$, furthermore, its variational minimization could
drive $\mu$ far below the validity range of the perturbative integration over
the hard modes (cf. Sec. \ref{huen}), were it not stopped at a sufficiently
large $\mu^{\ast}$ by the vanishing of the gap-equation solution $\bar{\xi
}\left(  \mu,c_{1}\right)  $, i.e. by the order-disorder phase transition.
This behavior generalizes to the $c_{1}$ dependence. Indeed, although the
energy density (\ref{e}) monotonically decreases with $c_{1}$ as well (cf.
Fig. \ref{ed}), the eventual vanishing of $\bar{\xi}$ prevents the variational
solution to attain $c_{1}$ values \ too close to unity where our $O\left(
c_{1}\right)  $ treatment of the $4U$ interactions would break down. Hence the
bounded existence range of $\bar{\xi}\left(  \mu,c_{1}\right)  >0$ solutions,
inside the critical line of Fig. \ref{cl}, is indispensable both for our
approximations to remain valid and for the quantitatively reasonable size
prediction of the gluon condensate (\ref{gc}) and other vacuum scales to emerge.

Another issue meriting discussion is the order of the transition between the
strongly- and weakly-coupled vacuum phases. Above we found the (dis)order
parameter $\bar{\xi}$ to vanish continuously at the phase boundary, which
indicates a second-order transition. The same order was encountered in the
Ref. \cite{kog95}, although lattice simulations predict a first-order
transition \cite{kog82}. In Ref. \cite{kog95} this mismatch was argued to be a
shortcoming of the mean-field approximation (which should break down close to
the transition). Nevertheless, one may wonder whether the mean-field
approximation could generate a first-order transition when the
higher-derivative interactions corresponding to $c_{n}\neq0$ are taken into
account. One could imagine, for example, that the lowest-energy gap-equation
solution $\bar{\xi}>0$ may cease to exist for increasing $\mu$ before reaching
zero and/or that a competing solution $\bar{\xi}\simeq0$ may become
energetically favorable before any of them vanishes. This raises the question
whether more than one solution of the gap equation (\ref{geq}) could exist. In
principle, this is indeed possible. An example can be constructed by treating
the dominant $2U$ part of the $c_{1}\neq0$ interactions perturbatively to
$O\left(  c_{1}\right)  $, instead of resumming it as we did in Eq.
(\ref{dm1(k)}). In this case two solution branches of the gap equation would
indeed emerge. One of them is not continuous in the $c_{1}=0$ limit, though,
and the energy competition with the other one turns out to be unable to
generate a first-order transition. In our case such scenarios are excluded
from the outset, furthermore, since the nontrivial solution of our
gap-equation is unique.

\subsection{Impact of the higher-gradient corrections}

\label{cng2}

From Sec. \ref{excor} onward, we have restricted our practical calculations to
contributions from the dominant higher-gradient interactions, associated with
the low-momentum coupling $c_{1}$. Hence a few comments on the expected impact
of the subleading higher-gradient corrections (governed by the $c_{n\geq2}$
couplings in the soft-mode Lagrangian (\ref{lc})) may be useful. All $c_{n}$
dependence originates from the expansion (\ref{gm1}) of the infrared
covariance. Through the generating functional (\ref{gf}) it then enters the
soft-mode matrix elements (cf. e.g. Eq. (\ref{uu})) and thereby any amplitude,
both in terms of the propagators $\Delta^{-1}$ and $\bar{\Delta}^{-1}$ (cf.
Eqs. (\ref{d}) and (\ref{m})) and by virtue of the perturbative expansion
(\ref{vapr}) of the $4U$ contributions. In order to take the contributions
from the $2\left(  n+1\right)  $-gradient interactions with $n\geq2$ into
account, one therefore has (i) to include the explicit $c_{n}$ dependence from
Eq. (\ref{gm}) in the expressions for the chromo-electric (\ref{e2}) and
-magnetic (\ref{b2h2}) expectation values, (ii) to continue the expansion of
the $4U$ contributions (\ref{vapr}) to higher $c_{n}$, and (iii)\ to
generalize the integrals (\ref{in}) and (\ref{jn}), which summarize the impact
of the $c_{n}$ contributions on the $2U$ part of the soft-mode amplitudes, to
\begin{align}
\tilde{\imath}_{n}\left(  \xi,c_{1},c_{2},c_{3,...}\right)   &  :=\int_{0}%
^{1}\frac{\kappa^{2n}}{\kappa^{2}\left(  1-c_{1}\kappa^{2}+c_{2}\kappa
^{4}-c_{3}\kappa^{6}+...\right)  +\xi^{2}}d\kappa\text{
\ \ \ \ \ \ \ \ \ \ \ \ \ \ \ }n\geq1,\label{i}\\
\tilde{j}_{n}\left(  \xi,c_{1},c_{2},c_{3,...}\right)   &  :=\int_{0}^{1}%
\frac{\kappa^{2n}}{\left[  \kappa^{2}\left(  1-c_{1}\kappa^{2}+c_{2}\kappa
^{4}-c_{3}\kappa^{6}+...\right)  +\xi^{2}\right]  ^{2}}d\kappa\text{
\ \ \ \ \ \ \ \ \ \ \ \ }n\geq1. \label{j}%
\end{align}
(The monotonicity properties of the integrals (\ref{i}), (\ref{j}) are
straightforward generalizations of those discussed in App. \ref{int}.)

In Eqs. (\ref{i}) and (\ref{j}) the low-momentum constants $c_{n}$ appear
multiplied by $2n$ powers of the dimensionless momentum $\kappa\in\left[
0,1\right]  .$ This provides the basis for a consistent power counting.
Indeed, the contributions from larger $n$ are systematically suppressed and at
least parametrically dominated by the four-derivative interactions associated
with $c_{1}$. Since the $c_{n}$ are furthermore bounded by the normalizability
constraints (\ref{cbnds}) and since our optimal value $c_{1}^{\ast}\simeq0.15$
keeps already the dominant contribution rather small, one may expect that a
variational treatment of the $c_{n\geq2}$ contributions would result in even
substantially smaller corrections. (Vacuum energy, gluon condensate and other
matrix elements could therefore be estimated by expanding the integrals
(\ref{i}) and (\ref{j}) around $c_{n\geq2}=0$, incidentally, with the
expansion coefficients determined by the integrals given in App. \ref{int}.)
With the above considerations in mind, we have not pursued the quantitative
evaluation of $c_{n\geq2}$ contributions in this paper, although our framework
is fully equipped to take them into account. Another reason for neglecting
these corrections was that their inclusion would reduce the transparency of
our explorative study. (The visualization of the vacuum energy density and its
minimum in Fig. \ref{ed}, for instance, is possible only if $c_{n\geq2}=0$.)

\subsection{Further physical implications and applications}

\label{out}

We conclude our analysis by briefly reviewing several additional features of
the resulting vacuum description. To begin with the UV physics, we note that
the one-loop Yang-Mills $\beta$ function can be recovered almost completely
from the dynamics implemented in the UV covariance (\ref{gm1uv}) (up to a
missing $\allowbreak6\%$ correction due to color screening \cite{bro99}). As
already mentioned, the screening contribution may be accounted for as well,
e.g. by generalizing the tensor structure of the UV covariance \cite{dia98}.
(We note in passing that Rayleigh-Schr\"{o}dinger perturbation theory, even to
higher orders of the gauge coupling, could be set up directly in our
wave-functional basis (\ref{ginvvwf}) since it renders Gauss' law manifest.)

Turning to the IR physics content, we start by emphasizing that dimensional
transmutation emerges naturally from gauge-invariant wave functionals
(\ref{ginvvwf}) with Gaussian cores (\ref{ga}), and that it generates the
infrared mass scale and the crucial mass gap dynamically. Since a Gaussian
becomes the exact ground state of both massless and \emph{massive} vector
fields in the non-interacting limit, furthermore, one may expect it to yield
an adequate description whenever the main net effect of the interactions is to
create a quasi-particle-like selfenergy. Similarly, it was argued in Ref.
\cite{kog95} that the Gaussian approximation should provide a reasonable
description of Yang-Mills vacuum physics as long as the latter is dominated by
a single gluon condensate (which QCD sum-rule analyses suggest especially in
the glueball sector \cite{nar98,for05}).

An important advantage of the gauge-projected wave functionals (\ref{ginvvwf})
is that they incorporate the nontrivial topology of the gauge group and
thereby of the gauge fields. In particular, they contain instanton
effects\ \cite{bro299,for06} which play an especially important role in the
spin-0 glueball sector \cite{sch98,sch95,for05} and emerge even in basic
holographic duals for Yang-Mills theory \cite{for08}. Furthermore, the
functional (\ref{ginvvwf}) comprises large classes of additional,
gauge-invariant IR degrees of freedom which are often of topological origin as
well \cite{for06}. Those are typically collective excitations gathered by
gauge orbits of vacuum fields which were found to dominate the saddle-point
expansion of the generating soft-mode functional (\ref{gf}) and to provide
gauge-invariant physical interpretations e.g. for merons and Faddeev-Niemi
knots \cite{for06,for07}.

Another crucial consequence of the infrared fields populating the Yang-Mills
vacuum, confinement, is characterized by the emergence of a linear potential
between sufficiently far separated, static quark sources. Despite the rather
direct access to the interquark potential provided in the Schr\"{o}dinger
representation \cite{zar98,dia98} and heuristic arguments given in Refs.
\cite{kog95,zar98,dia98}, however, it remains to be shown that gauge-averaged
wave functionals (\ref{ginvvwf}) with a Gaussian core of the type (\ref{ga})
are capable of generating the confinement-induced area law for large,
spacelike Wilson loops. Several cautionary lessons from 3-dimensional, compact
electrodynamics can be found in Ref. \cite{kov99}. (The behavior of the
adjoint Wilson loop would be of interest as well.) Explorations of confinement
in Coulomb gauge encouragingly found a Gaussian wave functional, divided by
the square root of the Faddeev-Popov\ determinant, to generate a linear
heavy-quark potential \cite{rei07}.

In this context, one should further emphasize that confinement (as well as
other crucial vacuum features) is expected to emerge in our framework only
after gauge projection. In fact, Feynman had argued almost three decades ago
\cite{fey81} that for a mass gap to be generated in a non-Abelian gauge
theory, all field configurations with non-vanishing support at
\textquotedblleft large\textquotedblright\ $A$ should either be
gauge-equivalent to others with support only at \textquotedblleft
small\textquotedblright\ $A$ or damped by their magnetic field energy. This
argument is suggestive because then, in analogy with finite-dimensional
quantum mechanics, the wave functionals of the ground and first-excited states
cannot arbitrarily reduce their energy difference by favoring long-wavelength
potentials\ $A$ extending over all of space. As a consequence, the energy or
(in the static case) mass gap cannot be closed. It would be important to check
whether a gauge-projected Gaussian wave functional can realize this mechanism.

In order to simplify the analysis and to focus on the still rather poorly
understood infrared gluon sector of QCD, we have restricted our investigation
to pure Yang-Mills theory without quarks. As the quenched approximation to
lattice QCD, this provides stable and unmixed glueball states which set a
benchmark for sorting out the QCD glueball spectrum. (For a first attempt to
study glueball states on the basis of a gauge-invariant vacuum wave functional
see Ref. \cite{gri02}.) However, static quark sources could be
straightforwardly implemented into the gauge-invariant vacuum functional
\cite{zar98,kov99} and even the inclusion of light, dynamical quarks adds no
conceptual problems, although a nonperturbative treatment of their
interactions with the gluon sector appears challenging. (For a variational
study of fermions in the 1+1 dimensional Abelian Schwinger model see Ref.
\cite{bro00}.)

We close this section by briefly discussing the improvement potential of our
approach and by mentioning a few additional applications. The $O\left(
g\right)  $ treatment of the energy density and the $O\left(  c_{1}\right)  $
treatment of the $4U$ interactions could be systematically refined. In
addition, one may evaluate the corrections due to higher-gradient
contributions with couplings $c_{n\geq2}$ in the effective soft-mode action
(\ref{lc}). For the gauge group SU$\left(  2\right)  $, the mean-field
analysis of the integral over the auxiliary field $\Sigma$ could alternatively
be performed in quaternionic variables, as suggested in Ref. \cite{dia98}. If
this formulation could be generalized to higher $N_{c}$, it would keep the
fluctuations around the mean field under control, and the corresponding
saddle-point approximation may in fact become exact in the large-$N_{c}$ limit
\cite{pol87}. One could further generalize the trial functional basis by
including both longitudinal and transverse mode contributions (potentially
non-analytic) to the covariance, or even allow for a non-diagonal color
structure. Finally, one may consider to solve the nonlinear soft-mode dynamics
(\ref{smd}) and its phase structure exactly on a lattice.

A particularly interesting class of Yang-Mills amplitudes are (equal-time)
glueball correlation functions. They contain information on the whole spectrum
and the decay constants of glueballs which was first explored variationally in
Ref. \cite{gri02} and is accessible in our framework as well. It could e.g. be
analyzed by comparison with results from lattice simulations \cite{def92}, the
operator product expansion \cite{nar98,for05} and holographic strong-coupling
duals \cite{for08}. The pseudoscalar glueball correlator contains the
topological susceptibility (in the zero-momentum limit), furthermore, whose
evaluation would complement the analysis of the topological vacuum structure
mentioned above. Another interesting type of amplitude which may become at
least approximately accessible are the Bethe-Salpeter amplitudes of glueballs
for which e.g. lattice \cite{def92} and instanton liquid model \cite{sch95}
results are available. A further natural and interesting extension of our
analysis would be its generalization to finite temperature, as pioneered in
Ref. \cite{kog02} for the minimal trial-functional family of Ref.
\cite{kog95}. This would open up several important applications, including the
study of the deconfinement phase transition \cite{gri03} as well as of
properties of the currently intensely debated and potentially strongly coupled
gluon plasma at temperatures of up to 2-3 times the critical temperature
\cite{shu04}. In the longer run, the potential of our Minkowski space
formulation to describe quantum nonequilibrium physics should be exploited,
too, e.g. by applying it to currently much discussed nonequilibrium processes
in the early universe as well as in the aftermath of ultrahigh-energy nuclear
collisions (for a brief introduction see Sec. V of Ref. \cite{far09}).
Hopefully, some of the above applications may also provide guidance for how to
improve the Gaussian approximation (\ref{ga}) to the core wave functional.

\section{Summary and conclusions}

\label{suc}

We have implemented and studied a gauge-invariant variational approximation
scheme for the Yang-Mills vacuum wave functional. Our approach is based on
minimizing the Yang-Mills Hamiltonian in a gauge-projected Gaussian
trial-functional space that allows for an essentially general momentum
distribution of the soft vacuum fields. This amounts to a substantial
improvement upon previously used trial states which populated all infrared
modes with identical, i.e. momentum-independent strength. Our extension of the
trial basis rests on a controlled expansion of the infrared-mode dispersion
relation (or \textquotedblleft covariance\textquotedblright), which
characterizes the vacuum wave functional, into powers of the soft momenta
divided by the dynamically generated mass scale. The momentum dependence can
then be systematically approximated by truncating the series, which allows us
to restrict the quantitative part of our study to the leading-order
corrections. After complementing the generalized soft vacuum-mode population
by the leading perturbative ultraviolet modes, finally, one ends up with the
largest gauge-invariant trial space for the Yang-Mills vacuum wave functional
(in 3+1 dimensions) which is currently available and which keeps the
variational analysis analytically manageable.

The multifaceted vacuum physics emerging from energy minimization in this
extended trial space turns out to be partly shaped by a subtle interplay
between different parts of the dynamics. To begin with, the leading-order
momentum distribution of the infrared modes is characterized by just one
variational parameter. After translating the dynamical content of the wave
functional into an effective Lagrangian for collective\ sets of soft
gauge-field orbits, this parameter reappears as the coupling constant of the
dominant higher-gradient interactions which modify the low-momentum dispersion
of the vacuum modes. Moreover, it acquires an intuitive physical
interpretation in terms of the group velocity of gauge-field quasiparticles in
the vacuum. While this velocity vanishes in the unimproved ground state, one
of our main findings is that it becomes finite in the improved one. The novel
dispersion relation further reveals that the infrared quasigluons in the
vacuum decelerate when a force is applied to them. In other words, their
\textquotedblleft effective kinetic mass\textquotedblright\ is negative
(analogous situations are encountered in several condensed matter systems) and
their color conductivity can increase with increasing strength of an external
color-electric field, implying a negative differential resistance of the vacuum.

The variationally optimized value of the higher-gradient coupling turns out to
be positive. This sign may at first appear surprising since it is opposite to
the one induced by the massive vector propagator which was previously adopted
as a model for the infrared covariance. Its physical impact is quite
intuitive, however, since it implies that soft gauge-field modes with larger
momenta populate the vacuum more strongly than their longest-wavelength
counterparts. Vacuum fields with constant color-magnetic field strength are
thus energetically disfavored, in particular, which prevents the onset of the
Savvidy instability. An energy barrier against larger quasigluon velocities
preserves the normalizability of the vacuum wave functional and keeps the
variationally determined value of the quasigluon velocity small. In addition,
it implies the existence of a saddle-point solution for the disorder-parameter
field at a dynamically generated mass scale which is consistent with lattice
results for the Yang-Mills scale. (As a practical side benefit, the relatively
small optimized coupling-parameter value also justifies a leading-order
perturbative treatment of the four-soft-mode interactions.)

Moreover, the above findings indicate that the dimensional transmutation
mechanism is an almost quantitatively robust feature of gauge-projected
Gaussian core functionals (instead of emerging accidentally from an
oversimplified trial space). As a consequence of this rather nontrivial result
the lowest-dimensional gluon condensate, which scales with the fourth power of
the dynamically generated infrared mass scale, acquires a magnitude well
inside the range predicted by other sources. This observation further supports
the conjecture that Gaussian trial functionals provide a reasonable vacuum
description in dynamical situations where just one vacuum condensate
dominates. While this holds e.g. for the Cooper-pair condensate in the BCS
theory of superconductivity, it is less well established in Yang-Mills theory
where the lowest-dimensional gluon condensate provides the natural candidate.
The main reasons for the robustness of the vacuum scales turn out to be that
the nontrivial solution of the gap equation is unique and that the disordered
vacuum phase exists only in a limited domain of all variational parameters.
Even the soft-mode dynamics' order-disorder phase transition, where the vacuum
energy attains its minimum, remains squarely of second order. This is in
contrast to the first order predicted by lattice simulations and probably a
shortcoming of the employed mean-field approximation near the transition. In
any case, the improved dispersion generates a considerably larger attraction
among the variationally optimized vacuum fields and lowers the vacuum energy
density by more than $10\%$. This indicates that the nonvanishing quasigluon
velocity provides a rather substantial overall improvement of the vacuum description.

Nevertheless, the gauge-invariant variational framework remains analytically
tractable, the computational costs are comparable to those of fully
gauge-fixed approaches, and the rather high degree of transparency owed to the
explicit representation of the vacuum state is maintained. In combination,
these qualities suggest considerable further potential of our approach as an
intuitive theoretical laboratory for gaining insights into the nonperturbative
real-time physics of Yang-Mills theories. The established framework provides,
in particular, a privileged testing ground for the impact of topological
vacuum excitations and their relation to the gauge symmetry. Indeed, both
aspects are manifest in our effective soft-mode dynamics and should be
explored further. In addition, our approach seems well suited to chart the
maximal physics content which gauge-projected Gaussian core functionals can
accommodate. In the longer run, this would allow to clarify the principal
limitations of the Gaussian approximation and may help to trigger new ideas
for going beyond it in a both systematic and analytically manageable way.

\begin{acknowledgments}
The author would like to thank Dietmar Ebert for a careful reading of the
manuscript. Financial support from the Funda\c{c}\~{a}o de Amparo \`{a}
Pesquisa do Estado de S\~{a}o Paulo (FAPESP) and from the Deutsche
Forschungsgemeinschaft (DFG), as well as the hospitality of the Abdus Salam
International Centre for Theoretical Physics (ICTP) in Trieste, Italy, are
gratefully acknowledged.
\end{acknowledgments}

\appendix

\section{Evaluation of vacuum expectation values}

\label{stacor}

In the following appendix we review the general approach to calculating vacuum
matrix elements in gauge-projected trial functionals of the type
(\ref{ginvvwf}), following Ref. \cite{kog95}. We recast this framework into
the language of generating functionals and also derive several expressions
which will be needed in the main text.

To begin with, consider local gauge-invariant operators $\mathcal{O}\left(
A,E\right)  $ composed of static gauge fields $A_{i}^{a}\left(  \vec
{x}\right)  $ and their canonically conjugate momenta $-E_{i}^{a}\left(
\vec{x}\right)  $ (where $E$ is the chromoelectric field) in the Hamiltonian
formulation of SU$\left(  N_{c}\right)  $ Yang-Mills theory, adopting temporal
(or Weyl) gauge. In the Schr\"{o}dinger coordinate representation, the
canonical commutation relation $\left[  E_{i}^{a}\left(  \vec{x}\right)
,A_{j}^{b}\left(  \vec{y}\right)  \right]  =i\delta^{ab}\delta_{ij}\delta
^{3}\left(  \vec{x}-\vec{y}\right)  $ then implies that $E_{i}^{a}\left(
\vec{x}\right)  =i\delta/\delta A_{i}^{a}\left(  \vec{x}\right)  $. Our aim is
to calculate the (trial)\ vacuum expectation value of $\mathcal{O}$, i.e. the
matrix element
\begin{equation}
\left\langle \mathcal{O}\left(  A,E\right)  \right\rangle =\frac{\int D\vec
{A}\Psi_{0}^{\ast}\left[  \vec{A}\right]  \mathcal{O}\left(  \vec{A}^{a}%
,\frac{i\delta}{\delta\vec{A}^{a}}\right)  \Psi_{0}\left[  \vec{A}\right]
}{\int D\vec{A}\Psi_{0}^{\ast}\left[  \vec{A}\right]  \Psi_{0}\left[  \vec
{A}\right]  }, \label{me}%
\end{equation}
where the unrestricted linear measure $D\vec{A}$ comprises all gauge orbits
and topological charges. (Recall that Gauss'\ law is enforced as a constraint
on the physical states and in particular on the vacuum wave functional
$\Psi_{0}$, cf. Sec. \ref{ginv}.)

\subsection{Exact integration over the gauge field}

After inserting the (residual) gauge-invariant, unnormalized wave functional
(\ref{ginvvwf}) into Eq. (\ref{me}) (and interchanging the order of the
$A,\bar{U}$ and $\tilde{U}$ integrations), the matrix element becomes%
\begin{equation}
\left\langle \mathcal{O}\left(  A,E\right)  \right\rangle =\frac{\int D\bar
{U}\int D\tilde{U}\int D\vec{A}\psi_{0}\left[  \vec{A}^{\bar{U}}\right]
\mathcal{O}\left(  \vec{A}^{a},\frac{i\delta}{\delta\vec{A}^{a}}\right)
\psi_{0}\left[  \vec{A}^{\tilde{U}}\right]  }{\int D\bar{U}\int D\tilde{U}\int
D\vec{A}\psi_{0}\left[  \vec{A}^{\bar{U}}\right]  \psi_{0}\left[  \vec
{A}^{\tilde{U}}\right]  }. \label{o1}%
\end{equation}
Since $\mathcal{O}$ is gauge-invariant, the integrand in the numerator can
only depend on the relative transformation $U:=\tilde{U}^{-1}\bar{U}$. Indeed,
after changing the integration variable $A\rightarrow A^{\prime}=A^{\tilde{U}%
}$ (for fixed $\tilde{U}$), one has
\begin{equation}
\int D\vec{A}\psi_{0}\left[  \vec{A}^{\bar{U}}\right]  \mathcal{O}\left(
\vec{A},\frac{i\delta}{\delta\vec{A}}\right)  \psi_{0}\left[  \vec{A}%
^{\tilde{U}}\right]  =\int D\vec{A}^{\prime}\psi_{0}\left[  A^{\prime\tilde
{U}^{-1}\bar{U}}\right]  \mathcal{O}\left(  \vec{A}^{\prime},\frac{i\delta
}{\delta\vec{A}^{\prime}}\right)  \psi_{0}\left[  A^{\prime}\right]
\end{equation}
and the analogous expression for the denominator. Making use of $\mathcal{O}%
\left(  A\right)  =\mathcal{O}\left(  A^{\prime}\right)  $ and $DA=DA^{\prime
}$, renaming $A^{\prime}$ to $A$ and employing the translational invariance
$D\bar{U}=DU$ of the Haar measure then yields%
\begin{equation}
\int D\bar{U}\int D\vec{A}\psi_{0}\left[  \vec{A}^{\bar{U}}\right]
\mathcal{O}\left(  \vec{A},\frac{i\delta}{\delta\vec{A}}\right)  \psi
_{0}\left[  \vec{A}^{\tilde{U}}\right]  =\int DU\int D\vec{A}\psi_{0}\left[
\vec{A}^{U}\right]  \mathcal{O}\left(  \vec{A},\frac{i\delta}{\delta\vec{A}%
}\right)  \psi_{0}\left[  \vec{A}\right]
\end{equation}
(and the analogous expression for the denominator) which does not depend on
$\tilde{U}$ since the latter was absorbed into the integration variable $A$.
Hence the factored (infinite) gauge group volumes $\int D\tilde{U}$ in
numerator and denominator cancel, leaving us with
\begin{equation}
\left\langle \mathcal{O}\left(  A,E\right)  \right\rangle =\frac{\int DU\int
D\vec{A}\psi_{0}\left[  \vec{A}^{U}\right]  \mathcal{O}\left(  \vec{A}%
^{a},\frac{i\delta}{\delta\vec{A}^{a}}\right)  \psi_{0}\left[  \vec{A}\right]
}{\int DU\int D\vec{A}\psi_{0}\left[  \vec{A}^{U}\right]  \psi_{0}\left[
\vec{A}\right]  }.
\end{equation}
Now it is useful to define an effective bare action $\Gamma_{b}\left[
U\right]  $ and the associated nonlocal gauge-field correlators as
\begin{align}
\exp\left\{  -\Gamma_{\text{b}}\left[  U\right]  \right\}   &  :=\int D\vec
{A}\psi_{0}\left[  \vec{A}^{U}\right]  \psi_{0}\left[  \vec{A}\right]  =:\int
D\vec{A}e^{-\gamma\left[  A,U\right]  },\label{gamb}\\
\left\langle \left\langle \left\langle \vec{A}...\vec{A}...\vec{E}...\vec
{E}\right\rangle \right\rangle \right\rangle \exp\left\{  -\Gamma_{\text{b}%
}\left[  U\right]  \right\}   &  :=\int D\vec{A}\psi_{0}\left[  \vec{A}%
^{U}\right]  \vec{A}...\vec{A}...\frac{i\delta}{\delta\vec{A}}...\frac
{i\delta}{\delta\vec{A}}\psi_{0}\left[  \vec{A}\right]  , \label{ame}%
\end{align}
where the triple-bracket $\left\langle \left\langle \left\langle
...\right\rangle \right\rangle \right\rangle $ denotes (nondiagonal) matrix
elements between $U$-rotated and unrotated core-functional states. (Note the
symmetry $\Gamma_{\text{b}}\left[  U\right]  =\Gamma_{\text{b}}\left[
g_{L}^{\dagger}Ug_{R}\right]  $ with $g_{L,R}\in$ SU$\left(  N_{c}\right)  $
which originates from the translational invariance of the two group
integrations in Eq. (\ref{o1}).) With the relation
\begin{equation}
\mathcal{O}\left(  \vec{A}^{a},\frac{i\delta}{\delta\vec{A}^{a}\left(  \vec
{x}\right)  }\right)  \psi_{0}\left[  \vec{A}\right]  \equiv\mathcal{\bar{O}%
}\left[  A\right]  \psi_{0}\left[  \vec{A}\right]  ,
\end{equation}
which replaces the (quasi-) local, differential operator $\mathcal{O}$ by a
nonlocal functional $\mathcal{\bar{O}}$\ of the $A$ fields only, one then
arrives at%
\begin{equation}
\left\langle \mathcal{O}\left(  A,E\right)  \right\rangle =\frac{\int DU\int
D\vec{A}\psi_{0}\left[  \vec{A}^{U}\right]  \mathcal{\bar{O}}\left[  A\right]
\psi_{0}\left[  \vec{A}\right]  }{\int DU\int D\vec{A}\psi_{0}\left[  \vec
{A}^{U}\right]  \psi_{0}\left[  \vec{A}\right]  }=\frac{\int DU\left\langle
\left\langle \left\langle \mathcal{O}\left(  A,E\right)  \right\rangle
\right\rangle \right\rangle \exp\left\{  -\Gamma_{\text{b}}\left[  U\right]
\right\}  }{\int DU\exp\left\{  -\Gamma_{\text{b}}\left[  U\right]  \right\}
}. \label{o3}%
\end{equation}
Hence the calculation of the matrix element is reduced to evaluating the
vacuum expectation value of the expressions (\ref{ame}), specialized to
contain $A$ fields only, in the $U$ field dynamics determined by
$\Gamma_{\text{b}}$.

Owing to the Gaussian nature of the wave functional (\ref{ga}), the effective
bare action defined in Eq. (\ref{gamb}) and hence the matrix elements
(\ref{ame}) can be calculated exactly. In order to render the $U$ dependence
explicit, we write
\begin{equation}
\gamma\left[  A,U\right]  :=-\ln\psi_{0}^{\ast}\left[  \vec{A}^{U}\right]
\psi_{0}\left[  \vec{A}\right]  =\frac{1}{2}\left[  \text{ }\left(
A^{U}\right)  ^{T}G^{-1}A^{U}+\text{ }A^{T}G^{-1}A\right]  \label{gam}%
\end{equation}
(in a triple matrix notation for the color, vector and spacial
\textquotedblleft indices\textquotedblright) where the rotated gauge field
\begin{equation}
\left(  \vec{A}^{U}\right)  ^{a}=S^{ab}\left(  \vec{x}\right)  \vec{A}%
^{b}+g^{-1}\vec{L}^{a}\left(  \vec{x}\right)  \label{au}%
\end{equation}
is expressed in terms of the homogeneous and inhomogeneous transformation
functions
\begin{equation}
S^{ab}\left(  \vec{x}\right)  =\frac{1}{2}tr\left\{  \lambda^{a}U^{\dagger
}\left(  \vec{x}\right)  \lambda^{b}U\left(  \vec{x}\right)  \right\}
,\ \ \ \ \ \vec{L}^{a}\left(  \vec{x}\right)  =itr\left\{  \lambda
^{a}U^{\dagger}\left(  \vec{x}\right)  \vec{\partial}U\left(  \vec{x}\right)
\right\}  .
\end{equation}
After inserting the representation (\ref{au}) into Eq. (\ref{gam}), one finds
that $\gamma$ contains an $A$ independent piece and a term which is bilinear
in $A$,
\begin{equation}
\gamma\left[  A,U\right]  =\frac{1}{2}\left(  A+a\right)  ^{T}\mathcal{M}%
\left(  A+a\right)  +\frac{1}{2g_{\text{b}}^{2}}L^{T}\mathcal{D}L.
\end{equation}
Above we have defined \cite{kog95}%
\begin{align}
\mathcal{M}  &  :=S^{T}G^{-1}S+G^{-1}=\mathcal{M}^{T},\text{ \ \ \ \ }\vec
{a}:=g^{-1}\mathcal{M}^{-1}S^{T}G^{-1}\vec{L},\label{ma}\\
\mathcal{D}  &  :=G^{-1}-G^{-1}S\mathcal{M}^{-1}S^{T}G^{-1}=\left(
G+SGS^{-1}\right)  ^{-1}. \label{dd}%
\end{align}
($G^{-1}$ and $\mathcal{M}$ are symmetric in $a,b$, $i,j$ and $\vec{x},\vec
{y}$, and transposition of $\vec{a}$ is understood to invert the direction of
the derivatives in $G^{-1}$ as well.) After performing the Gaussian
integration over $A$, we arrive at the explicit form
\begin{align}
\Gamma_{\text{b}}\left[  U\right]   &  =\int d^{3}x\int d^{3}y\mathcal{L}%
_{\text{b}}\left(  \vec{x},\vec{y}\right) \nonumber\\
&  =-\ln\int D\vec{A}\exp\left\{  -\gamma\left[  A,U\right]  \right\}
=\frac{1}{2}Tr\ln\left(  \mathcal{M}\right)  +\frac{1}{2g_{\text{b}}^{2}}%
L^{T}\mathcal{D}L \label{gb}%
\end{align}
of the bare action which is a nonlocal, nonlinear $\sigma$-model. (The
regularization by the ultraviolet cutoff $\Lambda_{\text{UV}}$ is implicit in
the definition of the wave functional.) This dynamics sums up the
self-interaction of the \textquotedblleft background\textquotedblright\ field
$U$ which the gauge fields induce through the vector $\vec{a}\left[  U\right]
$ and the propagator $\mathcal{M}^{-1}\left[  U\right]  $. In the following we
will omit the first term $\propto Tr\ln\left(  \mathcal{M}\right)  $ in Eq.
(\ref{gb}) since it is of $O\left(  g^{2}\right)  $ relative to the second one
and does not contain $U$ derivatives \cite{kog95,bro99}.

The connected (equal-time) $n$-point functions at fixed $U$,
\begin{equation}
\left\langle \left\langle \left\langle A_{i_{1}}^{a_{1}}\left(  \vec{z}%
_{1}\right)  A_{i_{2}}^{a_{2}}\left(  \vec{z}_{2}\right)  ...A_{i_{n}}^{a_{n}%
}\left(  \vec{z}_{n}\right)  \right\rangle \right\rangle \right\rangle
=\frac{\int D\vec{A}\psi_{0}\left[  \vec{A}^{U}\right]  A_{i_{1}}^{a_{1}%
}\left(  \vec{z}_{1}\right)  A_{i_{2}}^{a_{2}}\left(  \vec{z}_{2}\right)
...A_{i_{n}}^{a_{n}}\left(  \vec{z}_{n}\right)  \psi_{0}\left[  \vec
{A}\right]  }{\int D\vec{A}\psi_{0}\left[  \vec{A}^{U}\right]  \psi_{0}\left[
\vec{A}\right]  },
\end{equation}
specialize the definition (\ref{ame}) exclusively to $A$ fields and can now be
evaluated by means of the generating functional%
\begin{align}
\int D\vec{A}\psi_{0}\left[  \vec{A}^{U}\right]  \psi_{0}\left[  \vec
{A}\right]  \exp\left(  -\vec{J}\vec{A}\right)   &  =\int D\vec{A}\exp\left(
-\gamma\left[  \vec{A},U\right]  -\vec{J}\vec{A}\right) \\
&  =\exp\left(  \frac{1}{2}\vec{J}\mathcal{M}^{-1}\vec{J}+\vec{J}\vec
{a}\right)  \exp\left(  -\Gamma_{\text{b}}\left[  U\right]  \right)
\end{align}
as
\begin{equation}
\left\langle \left\langle \left\langle A_{i_{1}}^{a_{1}}\left(  \vec{z}%
_{1}\right)  A_{i_{2}}^{a_{2}}\left(  \vec{z}_{2}\right)  ...A_{i_{n}}^{a_{n}%
}\left(  \vec{z}_{n}\right)  \right\rangle \right\rangle \right\rangle
=\left.  \frac{\left(  -1\right)  ^{n}\delta^{n}\exp\left(  \frac{1}{2}\vec
{J}\mathcal{M}^{-1}\vec{J}+\vec{J}\vec{a}\right)  }{\delta J_{i_{1}}^{a_{1}%
}\left(  \vec{z}_{1}\right)  ...\delta J_{i_{n}}^{a_{n}}\left(  \vec{z}%
_{n}\right)  }\right\vert _{\vec{J}=0}.
\end{equation}
This expression reproduces the Wick expansion for correlators of a vector
field $\vec{A}$ with nontrivial propagator $\mathcal{M}^{-1}$ in the
background of the $U$ and $\vec{a}$ fields. (In contrast to free or mean-field
correlators, it does not factorize $2n$-point functions into products of $n$
2-point functions, however, and also generates finite $\left(  2n+1\right)
$-point functions.) Explicit expressions for the 2, 3 and 4-point functions
needed in the main text are%
\begin{equation}
\left\langle \left\langle \left\langle A_{i_{1}}^{a_{1}}\left(  \vec{z}%
_{1}\right)  A_{i_{2}}^{a_{2}}\left(  \vec{z}_{2}\right)  \right\rangle
\right\rangle \right\rangle =\mathcal{M}_{i_{1}i_{2}}^{-1a_{1}a_{2}}\left(
\vec{z}_{1},\vec{z}_{2}\right)  +a_{i_{1}}^{a_{1}}\left(  \vec{z}_{1}\right)
a_{i_{2}}^{a_{2}}\left(  \vec{z}_{2}\right)  \label{2p}%
\end{equation}
and (in shorthand notation for arguments and indices)%
\begin{equation}
\left\langle \left\langle \left\langle A\left(  1\right)  A\left(  2\right)
A\left(  3\right)  \right\rangle \right\rangle \right\rangle =-\left[
\mathcal{M}^{-1}\left(  2,3\right)  a\left(  1\right)  +\mathcal{M}%
^{-1}\left(  1,3\right)  a\left(  2\right)  +\mathcal{M}^{-1}\left(
1,2\right)  a\left(  3\right)  +a\left(  1\right)  a\left(  2\right)  a\left(
3\right)  \right]  \label{3p}%
\end{equation}
as well as%
\begin{align}
\left\langle \left\langle \left\langle A\left(  1\right)  A\left(  2\right)
A\left(  3\right)  A\left(  4\right)  \right\rangle \right\rangle
\right\rangle =  &  \mathcal{M}^{-1}\left(  1,2\right)  \mathcal{M}%
^{-1}\left(  3,4\right)  +\mathcal{M}^{-1}\left(  1,3\right)  \mathcal{M}%
^{-1}\left(  2,4\right)  +\mathcal{M}^{-1}\left(  1,4\right)  \mathcal{M}%
^{-1}\left(  2,3\right) \nonumber\\
&  +\mathcal{M}^{-1}\left(  1,2\right)  a\left(  3\right)  a\left(  4\right)
+\mathcal{M}^{-1}\left(  2,3\right)  a\left(  1\right)  a\left(  4\right)
+\mathcal{M}^{-1}\left(  3,4\right)  a\left(  1\right)  a\left(  2\right)
\nonumber\\
&  +\mathcal{M}^{-1}\left(  1,4\right)  a\left(  2\right)  a\left(  3\right)
+\mathcal{M}^{-1}\left(  1,3\right)  a\left(  2\right)  a\left(  4\right)
+\mathcal{M}^{-1}\left(  2,4\right)  a\left(  1\right)  a\left(  3\right)
\nonumber\\
&  +a\left(  1\right)  a\left(  2\right)  a\left(  3\right)  a\left(
4\right)  . \label{4p}%
\end{align}
Hence the above results reduce the calculation of the matrix element
(\ref{o1}) to integrating over the nonlocally interacting $U\left(  \vec
{x}\right)  $\ fields according to Eq. (\ref{o3}). This integral will be
performed in the next two sections.

\subsection{Perturbative integration over the hard $U$ modes}

\label{puint}

Since we are interested in amplitudes with soft external momenta $\left\vert
\vec{p}_{i}\right\vert \ll\mu$ (where $\mu$ is a typical hadronic scale, e.g.
the lowest glueball mass), the integrations over $U$ in Eq. (\ref{o3}) can be
approximately done in two steps \cite{kog95}: after integrating out the hard
modes $U_{>}$ (containing momenta $k\geq\mu$) perturbatively, one performs the
integral over the remaining soft modes $U_{<}$ (which contain the momenta
$k<\mu$) in the saddle-point approximation. In preparation for this procedure,
we factor the integration variable $U$ in a group-structure preserving manner
as%
\begin{equation}
U\left(  \vec{x}\right)  =U_{<}\left(  \vec{x}\right)  U_{>}\left(  \vec
{x}\right)  ,\text{ \ \ \ \ }U_{>}\left(  \vec{x}\right)  =\exp\left(
-ig\phi^{a}\left(  \vec{x}\right)  \lambda^{a}/2\right)  . \label{split}%
\end{equation}
Note that the soft modes vary only weakly over distances\ $\left\vert \vec
{x}-\vec{y}\right\vert <\mu^{-1}$, i.e. $U_{<}\left(  \vec{x}\right)  \simeq
U_{<}\left(  \vec{y}\right)  $. Since asymptotic freedom is manifest in the
high-momentum covariance (\ref{gm1uv}), the bare coupling $g_{b}$ is small at
the large cutoff scale $\Lambda_{\text{UV}}\gg\Lambda_{\text{YM}}$ where the
theory is defined. Hence the hard modes can be integrated perturbatively e.g.
by Wilson's momentum-shell technique \cite{wil74}, down to the infrared scale
$\mu$ chosen such that the renormalized coupling $g\left(  \mu\right)  $
remains sufficiently small. The functional measure factorizes according to Eq.
(\ref{split}) as
\begin{equation}
DU=DU_{<}DU_{>}\propto DU_{<}D\phi^{a}.
\end{equation}
As a consequence, the bare action becomes a functional of the $\phi$ and
$U_{<}$ fields,%
\begin{equation}
\Gamma_{\text{b}}\left[  \phi,U_{<}\right]  =\Gamma_{\text{b},>}\left[
\phi\right]  +\Gamma_{\text{b},<>}\left[  \phi,U_{<}\right]  +\Gamma
_{\text{b},<}\left[  U_{<}\right]  ,
\end{equation}
and Eq. (\ref{o3}) for the matrix element turns into%
\begin{equation}
\left\langle \mathcal{O}\left(  A,E\right)  \right\rangle =\frac{\int
DU_{<}\int D\phi\left\langle \left\langle \left\langle \mathcal{O}\left(
A,E\right)  \right\rangle \right\rangle \right\rangle \exp\left\{
-\Gamma_{\text{b}}\left[  \phi,U_{<}\right]  \right\}  }{\int DU_{<}\int
D\phi\exp\left\{  -\Gamma_{\text{b}}\left[  \phi,U_{<}\right]  \right\}  }.
\label{o5}%
\end{equation}
As long as $g^{2}\left(  \mu\right)  \ll1$, the $\phi$ field can be integrated
perturbatively to one loop in the bare coupling (where the integrals over loop
momenta are regularized by the cutoff $\Lambda_{\text{UV}}$). The result has
the form%
\begin{equation}
\left\langle \left\langle \mathcal{O}\left(  A,E\right)  \right\rangle
\right\rangle \exp\left\{  -\Gamma_{<}\left[  U_{<}\right]  \right\}  :=\int
D\phi\left\langle \left\langle \left\langle \mathcal{O}\left(  A,E\right)
\right\rangle \right\rangle \right\rangle \exp\left\{  -\Gamma_{\text{b}%
}\left[  \phi,U_{<}\right]  \right\}
\end{equation}
where we defined the soft-mode action $\Gamma_{<}\left[  U_{<}\right]  $ as
\begin{equation}
\exp\left\{  -\Gamma_{<}\left[  U_{<}\right]  \right\}  :=\int D\phi
\exp\left\{  -\Gamma_{\text{b}}\left[  \phi,U_{<}\right]  \right\}  ,
\label{smadef}%
\end{equation}
made use of\ $\Gamma_{\text{b},<>}\left[  U_{<},\phi\right]  =0$ to leading
order in the coupling, and assumed that the (one-loop) anomalous dimension of
$\mathcal{O}$ vanishes. (This is appropriate for the conserved Yang-Mills
Hamiltonian density (\ref{hamYM}) which we consider in the bulk of the paper.)
The expression (\ref{o5}) for the matrix element is then further reduced to
the soft-mode integral%
\begin{equation}
\left\langle \mathcal{O}\left(  A,E\right)  \right\rangle =\frac{\int
DU_{<}\left\langle \left\langle \mathcal{O}\left(  A,E\right)  \right\rangle
\right\rangle \exp\left\{  -\Gamma_{<}\left[  U_{<}\right]  \right\}  }{\int
DU_{<}\exp\left\{  -\Gamma_{<}\left[  U_{<}\right]  \right\}  }.
\end{equation}

In order to exhibit the (approximate) $U_{<}$ dependence of $\Gamma_{<}\left[
U_{<}\right]  $ and $\left\langle \left\langle \mathcal{O}\left(  A,E\right)
\right\rangle \right\rangle $, we now exploit the smallness of the bare
coupling to truncate the expansion of $U_{>}$ in powers of $g.$ Up to
corrections of $O\left(  g^{3}\right)  $ one finds%
\begin{align}
U_{>}  &  =\exp\left(  -ig\phi^{a}\frac{\lambda^{a}}{2}\right)  =1-ig\phi
^{a}\frac{\lambda^{a}}{2}-\frac{g^{2}}{8}\left(  \phi^{a}\lambda^{a}\right)
^{2}+O\left(  g^{3}\right)  ,\\
\lambda_{>,i}^{a}  &  =\partial_{i}\phi^{a}+O\left(  g^{3}\right)  ,\text{
\ \ \ \ }S_{>}^{ab}=\delta^{ab}+gf^{abc}\phi^{c}-\frac{g^{2}}{2}f^{ace}%
f^{bde}\phi^{c}\phi^{d}+O\left(  g^{3}\right)  ,\\
\lambda_{i}^{a}  &  =S_{>}^{ab}\lambda_{<,i}^{b}+\lambda_{>,i}^{a}%
=\lambda_{<,i}^{a}+\partial_{i}\phi^{a}+gf^{abc}\lambda_{<,i}^{b}\phi
^{c}-\frac{g^{2}}{2}f^{ace}f^{bde}\lambda_{<,i}^{b}\phi^{c}\phi^{d}+O\left(
g^{3}\right)  .
\end{align}
(Above we have rescaled the Cartan-Maurer form (\ref{L}) as $\lambda_{i}%
^{a}:=L_{i}^{a}/g$ in order to facilitate explicit power counting in $g$. The
spacial index should avoid confusion with the SU$\left(  N_{c}\right)  $
generators.) To the same order, the operator (\ref{dd}) expands as
\begin{align}
\mathcal{D}^{ab}\left(  \vec{x},\vec{y}\right)   &  =\frac{G^{-1}\left(
\vec{x}-\vec{y}\right)  }{2}\delta^{ab}\left\{  1+\frac{g^{2}}{N^{2}-1}\left[
\phi^{a}\left(  \vec{x}\right)  \phi^{a}\left(  \vec{x}\right)  -2\phi
^{a}\left(  \vec{x}\right)  \phi^{a}\left(  \vec{y}\right)  +\phi^{a}\left(
\vec{y}\right)  \phi^{a}\left(  \vec{y}\right)  \right]  \right\} \nonumber\\
&  +O\left(  g^{3}\right)  .
\end{align}
(Note that $\mathcal{D}$ is independent of $U_{<}$, as a consequence of
$S_{<}GS_{<}^{-1}\simeq G$ which holds because $S_{<}$ varies slowly over
distances $\left\vert \vec{x}-\vec{y}\right\vert \lesssim\mu^{-1}$ over which
the decaying $G$ remains non-negligible.) Inserting the above expansions into
the bare action (\ref{gb}), one finds the associated Lagrangian
\begin{equation}
\mathcal{L}_{\text{b}}\left(  \vec{x},\vec{y}\right)  =\mathcal{L}%
_{\text{b},>}\left(  \vec{x},\vec{y}\right)  +\mathcal{L}_{\text{b},<}\left(
\vec{x},\vec{y}\right)  +\mathcal{L}_{\text{b},<>}\left(  \vec{x},\vec
{y}\right)  +O\left(  g^{3}\right)  \label{ldecomp}%
\end{equation}
with the bilinear parts
\begin{align}
\mathcal{L}_{\text{b},>}\left(  \vec{x},\vec{y}\right)   &  =\frac{1}%
{4}\partial_{i}\phi^{a}\left(  \vec{x}\right)  G^{-1}\left(  \vec{x}-\vec
{y}\right)  \partial_{i}\phi^{a}\left(  \vec{y}\right)  ,\label{lhm}\\
\mathcal{L}_{\text{b},<}\left(  \vec{x},\vec{y}\right)   &  =\frac{1}%
{4}\lambda_{<,i}^{a}\left(  \vec{x}\right)  G^{-1}\left(  \vec{x}-\vec
{y}\right)  \lambda_{<,i}^{a}\left(  \vec{y}\right)
\end{align}
and the hard-soft-mode interactions (after using $G^{-1}\left(  \vec{x}%
-\vec{y}\right)  =G^{-1}\left(  \vec{y}-\vec{x}\right)  $ and anticipating the
exchange symmetry $\vec{x}\leftrightarrow\vec{y}$ after integration over
$\vec{x}$ and $\vec{y}$)
\begin{align}
\mathcal{L}_{\text{b},<>}\left(  \vec{x},\vec{y}\right)  =  &  \frac{g}%
{2}f^{abc}\lambda_{<,i}^{b}\left(  \vec{x}\right)  G^{-1}\left(  \vec{x}%
-\vec{y}\right)  \phi^{c}\left(  \vec{x}\right)  \partial_{i}\phi^{a}\left(
\vec{y}\right) \nonumber\\
&  +\frac{g^{2}}{8}\left(  \frac{N_{c}\delta^{ab}\delta^{cd}}{N^{2}%
-1}-2f^{ace}f^{bde}\right) \nonumber\\
&  \times\lambda_{<,i}^{b}\left(  \vec{x}\right)  G^{-1}\left(  \vec{x}%
-\vec{y}\right)  \lambda_{<,i}^{a}\left(  \vec{y}\right)  \left[  \phi
^{c}\left(  \vec{x}\right)  \phi^{d}\left(  \vec{x}\right)  -\phi^{c}\left(
\vec{x}\right)  \phi^{d}\left(  \vec{y}\right)  \right]  .
\end{align}
(Note that $\mathcal{L}_{\text{b}}$ is invariant under $\phi\rightarrow-\phi
$.) To leading order in $g$ the hard and soft modes decouple, and the
higher-order interactions $\mathcal{L}_{\text{b},<>}$ can be treated
perturbatively. (The $O\left(  g\right)  $ term in $\mathcal{L}_{\text{b},<>}$
contributes neither to the $\beta$-function nor to the soft-mode action.)
Writing further
\begin{equation}
\Gamma_{\text{b}}\left[  U\right]  =\Gamma_{\text{b},<}\left[  U_{<}\right]
+\Gamma_{\text{b},>}\left[  \phi\right]  +g\Gamma_{1}\left[  \phi
,U_{<}\right]  +g^{2}\Gamma_{2}\left[  \phi,U_{<}\right]  +O\left(
g^{3}\right)
\end{equation}
renders the bare-coupling dependence of the action explicit. In terms of the
kinetic operator
\begin{equation}
K^{ab}\left(  \vec{x}-\vec{y}\right)  \equiv\frac{1}{2}\delta^{ab}\left[
\partial_{yi}\partial_{xi}G^{-1}\left(  \vec{x}-\vec{y}\right)  \right]
_{p^{2}>\mu^{2}}%
\end{equation}
of the Lagrangian (\ref{lhm}), furthermore, the hard-mode action becomes%
\begin{align}
\Gamma_{\text{b},>}\left[  \phi\right]   &  =\int d^{3}x\int d^{3}%
y\mathcal{L}_{>,\text{b}}\left(  \vec{x},\vec{y}\right)  =\frac{1}{2}\int
d^{3}x\int d^{3}y\phi^{a}\left(  \vec{x}\right)  K^{ab}\left(  \vec{x}-\vec
{y}\right)  \phi^{b}\left(  \vec{y}\right)  +O\left(  g\right) \\
&  =\frac{1}{4}\frac{d^{3}k}{\left(  2\pi\right)  ^{3}}\left[  \theta\left(
\Lambda_{\text{UV}}^{2}-k^{2}\right)  -\theta\left(  \mu^{2}-k^{2}\right)
\right]  \tilde{\phi}^{a}\left(  \vec{k}\right)  G^{-1}\left(  k\right)
k^{2}\tilde{\phi}^{a}\left(  -\vec{k}\right)  \label{gg}%
\end{align}
from which one reads off the static $\phi^{a}$ propagator%
\begin{align}
\left\langle \phi^{a}\left(  \vec{x}\right)  \phi^{b}\left(  \vec{y}\right)
\right\rangle  &  \equiv\frac{\int D\phi_{\mu<k<\Lambda_{\text{UV}}}\phi
^{a}\left(  \vec{x}\right)  \phi^{b}\left(  \vec{y}\right)  \exp\left\{
-\Gamma_{\text{b},>}\left[  \phi\right]  \right\}  }{\int D\phi\exp\left\{
-\Gamma_{\text{b},>}\left[  \phi\right]  \right\}  }=K^{-1ab}\left(  \vec
{x}-\vec{y}\right) \\
&  =\int\frac{d^{3}k}{\left(  2\pi\right)  ^{3}}\left[  \theta\left(
\Lambda_{\text{UV}}^{2}-k^{2}\right)  -\theta\left(  \mu^{2}-k^{2}\right)
\right]  e^{i\vec{k}\left(  \vec{x}-\vec{y}\right)  }\frac{2G_{>}\left(
k\right)  }{k^{2}}\delta^{ab}.
\end{align}
The nonstandard kinetic term of the high-momentum $\sigma$ model (\ref{gg})
thus generates a propagator with large-$k$ asymptotics $K^{-1}\left(
k\right)  \propto G_{>}\left(  k\right)  /k^{2}\sim1/k^{3}$ (instead of the
standard $1/k^{2}$ behavior which impedes perturbative renormalizability in
three dimensions). Hence the high-momentum dynamics (\ref{gg}) is
renormalizable\ (the tadpole diagram diverges only logarithmically) and
asymptotically free, with a perturbative $\beta$ function very similar to that
of Yang-Mills theory \cite{bro99}.

From the definition (\ref{smadef}) and the decomposition (\ref{ldecomp}) of
the bare action one then has
\begin{align}
\exp\left\{  -\Gamma_{<}\left[  U_{<}\right]  \right\}   &  =\exp\left\{
-\Gamma_{\text{b},<}\left[  U_{<}\right]  \right\}  \int D\phi\exp\left\{
-\Gamma_{\text{b},>}\left[  \phi\right]  -\Gamma_{\text{b},<>}\left[
U_{<},\phi\right]  \right\} \\
&  =\exp\left\{  -\Gamma_{\text{b},<}\left[  U_{<}\right]  -\left\langle
\Gamma_{\text{b},<>}\left[  U_{<},\phi\right]  \right\rangle _{\Gamma
_{\text{b,}>}}+O\left(  g^{2}\right)  \right\}
\end{align}
(where the average $\left\langle ...\right\rangle _{\Gamma_{\text{b,}>}}$ over
$\phi$ is weighted by $\exp\left(  -\Gamma_{\text{b},>}\left[  \phi\right]
\right)  $) which yields the renormalized soft-mode action%
\begin{align}
\Gamma_{<}\left[  U_{<}\right]   &  =-\Gamma_{\text{b},<}\left[  U_{<}\right]
-\left\langle \Gamma_{\text{b},<>}\left[  \phi,U_{<}\right]  \right\rangle
_{\Gamma_{\text{b},>}}+O\left(  g^{2}\right) \\
&  =\frac{1}{4g^{2}\left(  \mu\right)  }\int d^{3}x\int d^{3}yL_{<,i}%
^{a}\left(  \vec{x}\right)  G_{<}^{-1}\left(  \vec{x}-\vec{y}\right)
L_{<,i}^{a}\left(  \vec{y}\right)  +O\left(  g^{2}\right)  . \label{gsm}%
\end{align}
To the considered order (and for $G_{>}\left(  k\right)  =k^{-1}$),
$\Gamma_{<}$ has therefore the same form as the bare soft-mode action
$\Gamma_{\text{b},<}$, but with the bare coupling replaced by the renormalized
one,
\begin{equation}
g\left(  \mu\right)  =g_{\text{b}}+\frac{g_{\text{b}}^{3}N_{c}}{\left(
2\pi\right)  ^{2}}\ln\frac{\Lambda_{\text{UV}}}{\mu}+O\left(  g_{\text{b}}%
^{5}\right)  ,
\end{equation}
which was calculated in Ref. \cite{bro99}. This renormalization of the
soft-mode interactions compensates for the removal of the hard $U_{>}$ modes
from amplitudes with external momenta $p^{2}\ll\mu^{2}$. Analogously, the
$\phi$ mode contributions%
\begin{equation}
\left\langle \left\langle \mathcal{O}\left(  A,E\right)  \right\rangle
\right\rangle =\frac{\int D\phi\left\langle \left\langle \left\langle
\mathcal{O}\left(  A,E\right)  \right\rangle \right\rangle \right\rangle
\exp\left\{  -\Gamma_{\text{b}}\left[  U_{<},\phi\right]  \right\}  }{\int
D\phi\exp\left\{  -\Gamma_{\text{b}}\left[  U_{<},\phi\right]  \right\}  }%
\end{equation}
can be integrated out perturbatively by rewriting
\begin{align}
\left\langle \left\langle \mathcal{O}\left(  A,E\right)  \right\rangle
\right\rangle  &  =\frac{\exp\left\{  -\Gamma_{\text{b},<}\left[
U_{<}\right]  \right\}  }{\exp\left\{  -\Gamma_{<}\left[  U_{<}\right]
\right\}  }\int D\phi\left\langle \left\langle \left\langle \mathcal{O}\left(
A,E\right)  \right\rangle \right\rangle \right\rangle \exp\left\{
-\Gamma_{\text{b},<>}\left[  U_{<},\phi\right]  \right\}  \exp\left\{
-\Gamma_{\text{b},>}\left[  \phi\right]  \right\} \\
&  \simeq\frac{\int D\phi\left\langle \left\langle \left\langle \mathcal{O}%
\left(  A,E\right)  \right\rangle \right\rangle \right\rangle \exp\left\{
-\Gamma_{\text{b},>}\left[  \phi\right]  \right\}  }{\int D\phi\exp\left\{
-\Gamma_{\text{b},>}\left[  \phi\right]  \right\}  }.
\end{align}
(To the considered order the anomalous dimension of $\mathcal{O}\left(
A,E\right)  $ arising from the perturbative interactions $\Gamma_{\text{b}%
,<>}\left[  U_{<},\phi\right]  $ can be neglected. This approximation becomes
exact for the operator considered in the main text, i.e. the conserved
Yang-Mills Hamiltonian density (\ref{hamYM}) in temporal gauge.) After
introducing another generating functional%
\begin{align}
z_{>}\left[  j\right]   &  :=\int D\phi\exp\left[  -\Gamma_{\text{b},>}\left[
\phi\right]  -\int d^{3}xj^{a}\left(  \vec{x}\right)  \phi^{a}\left(  \vec
{x}\right)  \right] \\
&  =z_{>}\left[  0\right]  \exp\left[  \frac{1}{2}\int d^{3}x\int d^{3}%
yj^{a}\left(  \vec{x}\right)  K^{-1ab}\left(  \vec{x}-\vec{y}\right)
j^{b}\left(  \vec{y}\right)  \right]  ,
\end{align}
the $\phi^{a}$ fields in $\left\langle \left\langle \left\langle
\mathcal{O}\left(  A,E\right)  \right\rangle \right\rangle \right\rangle $ can
be replaced by derivatives $-\delta/\delta j^{a}$ with respect to the source
$j^{a}$, which yields%
\begin{equation}
\left\langle \left\langle \mathcal{O}\left(  A,E\right)  \right\rangle
\right\rangle \simeq\left.  \frac{1}{z_{>}\left[  0\right]  }\left\langle
\left\langle \left\langle \mathcal{O}\left(  U_{<},\frac{-\delta}{\delta
j}\right)  \right\rangle \right\rangle \right\rangle z_{>}\left[  j\right]
\right\vert _{_{j=0}} \label{o4}%
\end{equation}
and shows that this intermediate matrix element is a functional of $G^{-1}$
and $U_{<}$ only.

\subsection{Saddle-point integration over the soft $U$ modes}

\label{spint}

With the result (\ref{o4}) for the intermediary matrix element $\left\langle
\left\langle \mathcal{O}\left(  A,E\right)  \right\rangle \right\rangle $ at
hand, it remains to perform the functional integral
\begin{equation}
\left\langle \mathcal{O}\left(  A,E\right)  \right\rangle =\frac{\int
DU_{<}\left\langle \left\langle \mathcal{O}\left(  A,E\right)  \right\rangle
\right\rangle \exp\left\{  -\Gamma_{<}\left[  U_{<}\right]  \right\}  }{\int
DU_{<}\exp\left\{  -\Gamma_{<}\left[  U_{<}\right]  \right\}  }%
\end{equation}
over the soft modes $U_{<}$. In order to prepare for this last step, we
integrate over a complex matrix field $V$ and represent the unitarity
constraint $U_{<}U_{<}^{\dagger}=1$ by inserting a delta functional. The
latter is rendered explicit by the additional integration over an auxiliary
Hermitian field $\Sigma\left(  \vec{x}\right)  $ which acts as a Lagrange
multiplier, as explained in Sec. \ref{gfun}. The result is%
\begin{equation}
\left\langle \mathcal{O}\left(  A,E\right)  \right\rangle =\frac{\int
D\Sigma\int DV\left\langle \left\langle \mathcal{O}\left(  A,E\right)
\right\rangle \right\rangle \exp\left\{  -\Gamma_{<}\left[  V\right]
-\Gamma_{\Sigma}\left[  V,\Sigma\right]  \right\}  }{\int D\Sigma\int
DV\exp\left\{  -\Gamma_{<}\left[  V\right]  -\Gamma_{\Sigma}\left[
V,\Sigma\right]  \right\}  } \label{mefin}%
\end{equation}
where the integration contour for $\Sigma\left(  \vec{x}\right)  $ runs
parallel to the imaginary axis and where%
\begin{equation}
\Gamma_{\Sigma}\left[  V,\Sigma\right]  \equiv\frac{m_{g}}{2g^{2}}\int
d^{3}xtr\left\{  \Sigma\left(  V^{\dagger}V-1\right)  \right\}  .
\end{equation}
(Note that $\Sigma$ is Hermitian since $VV^{\dagger}$ is. The unimodularity
constraint $\det U_{<}=1$ could additionally be implemented by integrating
over a minimally coupled U$\left(  1\right)  $ gauge field \cite{kog95}. We
refrain from doing so because the impact of the difference between SU$\left(
N_{c}\right)  $ and U$\left(  N_{c}\right)  $ is of order $1/N_{c}^{2}$.) The
matrix element (\ref{mefin}) can then be obtained from the soft-mode
generating functional (\ref{zsigv}), i.e.
\begin{equation}
Z\left[  j,j^{\dagger}\right]  =\int D\Sigma\int DV\exp\left[  -\Gamma
_{<}\left[  V\right]  -\Gamma_{\Sigma}\left[  V,\Sigma\right]  -\int
d^{3}ztr\left\{  jV^{\dagger}+j^{\dagger}V\right\}  \right]  , \label{zjjd}%
\end{equation}
after replacing the $U_{<}$ and $U_{<}^{\dagger}$ fields in $\left\langle
\left\langle \mathcal{O}\left(  A,E\right)  \right\rangle \right\rangle $ by
derivatives with respect to the matrix sources $j^{\dagger}$ and $j$:
\begin{equation}
\left\langle \mathcal{O}\left(  A,E\right)  \right\rangle \simeq\left.
\frac{1}{Z\left[  0,0\right]  }\left\langle \left\langle \mathcal{O}\left(
\frac{-\delta}{\delta j},\frac{-\delta}{\delta j^{\dagger}}\right)
\right\rangle \right\rangle Z\left[  j,j^{\dagger}\right]  \right\vert
_{_{j,j^{\dagger}=0}}. \label{o6}%
\end{equation}
The unconstrained integral over $V$ in Eq. (\ref{zjjd}) is Gaussian and can be
done exactly (cf. Sec. \ref{gfun}), and the remaining integral over $\Sigma$
can be performed in the saddle-point (or mean-field) approximation, as
detailed in Sec. \ref{mfa}. To leading order, this just amounts to
substituting the solution $\bar{\Sigma}$ of the gap equation (\ref{gapeq})
into the integrand of Eq. (\ref{zjjd}).

Given an explicit expression for the covariance, as in our case Eq.
(\ref{gm1}), the soft-mode action $\Gamma_{<}\left[  U_{<}\right]  $ (cf. Eq.
(\ref{smd})) and the amplitudes (\ref{o6}) are thus uniquely determined
functionals of $G_{<}^{-1}$. In the (trial) vacuum expectation value
$\left\langle \mathcal{H}_{\text{YM}}\right\rangle $ of the Yang-Mills
Hamiltonian, in particular, $G_{<}^{-1}$ plays the role of a variational trial
function. By means of the parametrization (\ref{gm}), $\left\langle
\mathcal{H}_{\text{YM}}\right\rangle $ becomes a function of the variational
parameters $\mu$, $m_{g}$, and $\left\{  c_{i}\right\}  $ with respect to
which it has to be minimized according to the Rayleigh-Ritz procedure. (The
bare cutoff $\Lambda_{\text{UV}}\gg\Lambda_{\text{YM}}$, on the other hand,
has been traded for $\Lambda_{\text{YM}}$ which will be fixed from lattice
data, cf. Sec. \ref{vmin}.)

\section{Coincidence limit integrals}

\label{int}

The coincidence limit of the soft-mode correlators calculated in Secs.
\ref{ircf} and \ref{excor} is (for $c_{n\geq2}=0$) determined by integrals of
the type%
\begin{align}
\tilde{\imath}_{n}\left(  \xi,c_{1}\right)   &  :=\int_{0}^{1}\frac
{\kappa^{2n}}{\kappa^{2}\left(  1-c_{1}\kappa^{2}\right)  +\xi^{2}}%
d\kappa\text{ \ \ \ \ \ \ \ \ \ \ \ \ \ \ \ }n\geq1,\label{in}\\
\tilde{j}_{n}\left(  \xi,c_{1}\right)   &  :=\int_{0}^{1}\frac{\kappa^{2n}%
}{\left[  \kappa^{2}\left(  1-c_{1}\kappa^{2}\right)  +\xi^{2}\right]  ^{2}%
}d\kappa\text{ \ \ \ \ \ \ \ \ \ \ \ \ }n\geq1 \label{jn}%
\end{align}
with $c_{1}<1$ (which ensures normalizability of the vacuum wave functional).
In the following we list pertinent properties of these integrals and derive
analytic expressions to be used e.g. for the numerical solution of the gap
equation and for the evaluation of the vacuum energy. To start with, we note
the obvious relations
\begin{align}
\tilde{\imath}_{n}\left(  \xi,c_{1}\right)   &  >\tilde{\imath}_{n+1}\left(
\xi,c_{1}\right)  ,\\
\tilde{j}_{n}\left(  \xi,c_{1}\right)   &  >\tilde{j}_{n+1}\left(  \xi
,c_{1}\right)
\end{align}
which hold point by point. Furthermore, for the physically required $c_{1}<1$
all integrals (\ref{in}), (\ref{jn}) are non-negative. They decrease
monotonically with increasing $\xi\geq0$ and increase monotonically with
increasing $c_{1}$ in the allowed range $-\infty<c_{1}<1$ (where
$1-c_{1}\kappa^{2}\geq0$ decreases with increasing $c_{1}$ for any $\kappa
\in\left[  0,1\right]  $). All integrals are regular at $c_{1}=0$ and at
$\xi=0$, with the exception of $\tilde{j}_{1}$ which contains a single pole at
$\xi=0$ (see below). Due to the factors of $\kappa^{2}$ in the numerator, all
integrands have most of their support close to $\kappa\rightarrow1$, i.e. at
momenta $k\sim\mu$.

\subsection{Analytical expressions}

An efficient strategy for evaluating the integrals $\tilde{\imath}_{n}\left(
\xi,c_{1}\right)  $ analytically starts from the identity
\begin{equation}
\frac{1}{\kappa^{2}\left(  1-c_{1}\kappa^{2}\right)  +\xi^{2}}=\frac{1}%
{\sqrt{1+4c_{1}\xi^{2}}}\left(  \frac{1}{\kappa^{2}-\kappa_{2}^{2}}-\frac
{1}{\kappa^{2}-\kappa_{1}^{2}}\right)  \label{id1}%
\end{equation}
which exhibits poles at the positions%
\begin{equation}
\kappa_{1,2}^{2}\left(  \xi,c_{1}\right)  =\frac{1}{2c_{1}}\left(  1\pm
\sqrt{1+4c_{1}\xi^{2}}\right)  .\label{pp}%
\end{equation}
Hence the evaluation of%
\begin{equation}
\tilde{\imath}_{n}\left(  \xi,c_{1}\right)  =\frac{1}{\sqrt{1+4c_{1}\xi^{2}}%
}\int_{0}^{1}\kappa^{2n}\left(  \frac{1}{\kappa^{2}-\kappa_{2}^{2}}-\frac
{1}{\kappa^{2}-\kappa_{1}^{2}}\right)  d\kappa
\end{equation}
is reduced to the calculation of one-pole integrals. The simple identity%
\begin{equation}
\frac{\kappa^{2n}}{\kappa^{2}-x^{2}}=x^{2}\frac{\kappa^{2\left(  n-1\right)
}}{\kappa^{2}-x^{2}}+\kappa^{2\left(  n-1\right)  }%
\end{equation}
implies the recursion relation
\begin{equation}
\int_{0}^{1}\frac{\kappa^{2n}}{\kappa^{2}-x^{2}}d\kappa=\frac{1}{2n-1}%
+x^{2}\int_{0}^{1}\frac{\kappa^{2\left(  n-1\right)  }}{\kappa^{2}-x^{2}%
}d\kappa
\end{equation}
and allows to reduce the $\tilde{\imath}_{n}\left(  \xi,c_{1}\right)  $ to
analytic expressions involving only%
\begin{equation}
\int_{0}^{1}\frac{1}{\kappa^{2}-x^{2}}d\kappa=-\frac{1}{2x}\ln\frac{x+1}%
{x-1}=-\frac{1}{x}\operatorname{arctanh}\left(  \frac{1}{x}\right)  \text{
\ \ \ \ (for }x^{2}\not \in \left]  0,1\right[  \text{)}.
\end{equation}
(Note that our $\kappa_{1.2}^{2}$ ensure $x^{2}\not \in \left]  0,1\right[  $
inside the $\kappa$ integration range (except at $\xi=0$).) From the recursion
relation one then finds (for $x^{2}\not \in \left]  0,1\right[  $)%
\begin{align}
\int_{0}^{1}\frac{\kappa^{2}}{\kappa^{2}-x^{2}}d\kappa &
=1-x\operatorname{arctanh}\left(  \frac{1}{x}\right)  ,\\
\int_{0}^{1}\frac{\kappa^{4}}{\kappa^{2}-x^{2}}d\kappa &  =\frac{1}{3}%
+x^{2}-x^{3}\operatorname{arctanh}\left(  \frac{1}{x}\right)  ,\\
\int_{0}^{1}\frac{\kappa^{6}}{\kappa^{2}-x^{2}}d\kappa &  =\frac{1}{5}%
+\frac{1}{3}x^{2}+x^{4}-x^{5}\operatorname{arctanh}\left(  \frac{1}{x}\right)
,...
\end{align}
and so on. Combining the above results one obtains for the first four
$\tilde{\imath}_{n}$, which are needed in the main text, the following
expressions:
\begin{align}
\tilde{\imath}_{1}\left(  \xi,c_{1}\right)   &  =\frac{\kappa_{1}%
\operatorname{arctanh}\left(  \frac{1}{\kappa_{1}}\right)  -\kappa
_{2}\operatorname{arctanh}\left(  \frac{1}{\kappa_{2}}\right)  }%
{\sqrt{1+4c_{1}\xi^{2}}},\\
\tilde{\imath}_{2}\left(  \xi,c_{1}\right)   &  =\frac{\kappa_{2}^{2}\left[
1-\kappa_{2}\operatorname{arctanh}\left(  \frac{1}{\kappa_{2}}\right)
\right]  -\kappa_{1}^{2}\left[  1-\kappa_{1}\operatorname{arctanh}\left(
\frac{1}{\kappa_{1}}\right)  \right]  }{\sqrt{1+4c_{1}\xi^{2}}},\\
\tilde{\imath}_{3}\left(  \xi,c_{1}\right)   &  =\frac{\kappa_{2}^{2}\left[
\frac{1}{3}+\kappa_{2}^{2}-\kappa_{2}^{3}\operatorname{arctanh}\left(
\frac{1}{\kappa_{2}}\right)  \right]  -\kappa_{1}^{2}\left[  \frac{1}%
{3}+\kappa_{1}^{2}-\kappa_{1}^{3}\operatorname{arctanh}\left(  \frac{1}%
{\kappa_{1}}\right)  \right]  }{\sqrt{1+4c_{1}\xi^{2}}},\\
\tilde{\imath}_{4}\left(  \xi,c_{1}\right)   &  =\frac{\kappa_{2}^{2}\left[
\frac{1}{5}+\frac{\kappa_{2}^{2}}{3}+\kappa_{2}^{4}-\kappa_{2}^{5}%
\operatorname{arctanh}\left(  \frac{1}{\kappa_{2}}\right)  \right]
-\kappa_{1}^{2}\left[  \frac{1}{5}+\frac{\kappa_{1}^{2}}{3}+\kappa_{1}%
^{4}-\kappa_{1}^{5}\operatorname{arctanh}\left(  \frac{1}{\kappa_{1}}\right)
\right]  }{\sqrt{1+4c_{1}\xi^{2}}}.
\end{align}
(Note that the main $\xi$ and $c_{1}$ dependence of these expressions
originates from the pole positions $\kappa_{1,2}\left(  \xi,c_{1}\right)  $,
cf. Eq. (\ref{pp}).)

Analytical solutions for the integrals $\tilde{j}_{n}\left(  \xi,c_{1}\right)
$ can either be derived from those for the $\tilde{\imath}_{n}\left(
\xi,c_{1}\right)  $ by using the identity
\begin{equation}
\tilde{j}_{n}\left(  \xi,c_{1}\right)  =-\frac{1}{2\xi}\frac{\partial
}{\partial\xi}\tilde{\imath}_{n}\left(  \xi,c_{1}\right)  ,
\end{equation}
or again by direct integration of the pole decomposition. Following the latter
path, we have from the square of relation (\ref{id1})%
\begin{equation}
\tilde{j}_{n}\left(  \xi,c_{1}\right)  =\frac{1}{1+4c_{1}\xi^{2}}\int_{0}%
^{1}\left(  \frac{\kappa^{2n}}{\left(  \kappa^{2}-\kappa_{1}^{2}\right)  ^{2}%
}-\frac{2\kappa^{2n}}{\left(  \kappa^{2}-\kappa_{1}^{2}\right)  \left(
\kappa^{2}-\kappa_{2}^{2}\right)  }+\frac{\kappa^{2n}}{\left(  \kappa
^{2}-\kappa_{2}^{2}\right)  ^{2}}\right)  d\kappa
\end{equation}
which we evaluate further with%
\begin{align}
\int_{0}^{1}\frac{\kappa^{2}}{\left(  \kappa^{2}-x^{2}\right)  ^{2}}d\kappa &
=\frac{1}{2}\frac{1}{x^{2}-1}-\frac{1}{2x}\operatorname{arctanh}\frac{1}{x},\\
\int_{0}^{1}\frac{\kappa^{2}}{\left(  \kappa^{2}-x_{1}^{2}\right)  \left(
\kappa^{2}-x_{2}^{2}\right)  }d\kappa &  =\frac{x_{2}\operatorname{arctanh}%
\frac{1}{x_{2}}-x_{1}\operatorname{arctanh}\frac{1}{x_{1}}}{x_{1}^{2}%
-x_{2}^{2}}%
\end{align}
and%
\begin{align}
\int_{0}^{1}\frac{\kappa^{4}}{\left(  \kappa^{2}-x^{2}\right)  ^{2}}d\kappa &
=\frac{1}{2}\frac{3x^{2}-2}{x^{2}-1}-\frac{3}{2}x\text{ }%
\operatorname{arctanh}\frac{1}{x},\\
\int_{0}^{1}\frac{\kappa^{4}}{\left(  \kappa^{2}-x_{1}^{2}\right)  \left(
\kappa^{2}-x_{2}^{2}\right)  }d\kappa &  =1-\frac{x_{1}^{3}%
\operatorname{arctanh}\frac{1}{x_{1}}-x_{2}^{3}\operatorname{arctanh}\frac
{1}{x_{2}}}{x_{1}^{2}-x_{2}^{2}}%
\end{align}
as well as
\begin{align}
\int_{0}^{1}\frac{\kappa^{6}}{\left(  \kappa^{2}-x^{2}\right)  ^{2}}d\kappa &
=\frac{2+10x^{2}-15x^{4}}{6\left(  1-x^{2}\right)  }-\frac{5}{2}%
x^{3}\operatorname{arctanh}\frac{1}{x},\\
\int_{0}^{1}\frac{\kappa^{6}}{\left(  \kappa^{2}-x_{1}^{2}\right)  \left(
\kappa^{2}-x_{2}^{2}\right)  }d\kappa &  =\frac{1}{3}+\left(  x_{1}^{2}%
+x_{2}^{2}\right)  -\frac{x_{1}^{5}\operatorname{arctanh}\frac{1}{x_{1}}%
-x_{2}^{5}\operatorname{arctanh}\frac{1}{x_{2}}}{x_{1}^{2}-x_{2}^{2}}%
\end{align}
(all for $x^{2}\not \in \left]  0,1\right[  $), and so on.

This yields
\begin{align}
\tilde{j}_{1}\left(  \xi,c_{1}\right)   &  =\frac{1}{1+4c_{1}\xi^{2}}\left[
\frac{1}{2}\frac{1}{\kappa_{1}^{2}-1}+\frac{1}{2}\frac{1}{\kappa_{2}^{2}%
-1}\right.  +\left(  \frac{2\kappa_{1}}{\kappa_{1}^{2}-\kappa_{2}^{2}}%
-\frac{1}{2\kappa_{1}}\right)  \operatorname{arctanh}\frac{1}{\kappa_{1}%
}\nonumber\\
&  -\left.  \left(  \frac{2\kappa_{2}}{\kappa_{1}^{2}-\kappa_{2}^{2}}+\frac
{1}{2\kappa_{2}}\right)  \operatorname{arctanh}\frac{1}{\kappa_{2}}\right]
\end{align}
and%
\begin{align}
\tilde{j}_{2}\left(  \xi,c_{1}\right)   &  =\frac{1}{1+4c_{1}\xi^{2}}\left[
\frac{1}{2}\frac{3\kappa_{1}^{2}-2}{\kappa_{1}^{2}-1}\right.  +\frac{1}%
{2}\frac{3\kappa_{2}^{2}-2}{\kappa_{2}^{2}-1}-2\nonumber\\
&  -\left(  \frac{3}{2}\kappa_{1}\text{ }-\frac{2\kappa_{1}^{3}}{\kappa
_{1}^{2}-\kappa_{2}^{2}}\right)  \operatorname{arctanh}\frac{1}{\kappa_{1}%
}-\left.  \left(  \frac{3}{2}\kappa_{2}\text{ }+\frac{2\kappa_{2}^{3}}%
{\kappa_{1}^{2}-\kappa_{2}^{2}}\right)  \operatorname{arctanh}\frac{1}%
{\kappa_{2}}\right]
\end{align}
and%
\begin{align}
\tilde{j}_{3}\left(  \xi,c_{1}\right)   &  =\frac{1}{1+4c_{1}\xi^{2}}\left[
\frac{1}{6}\frac{15\kappa_{1}^{4}-10\kappa_{1}^{2}-2}{\kappa_{1}^{2}%
-1}\right.  +\frac{1}{6}\frac{15\kappa_{2}^{4}-10\kappa_{2}^{2}-2}{\kappa
_{2}^{2}-1}-\frac{2}{3}\frac{\kappa_{1}^{2}+3\kappa_{1}^{4}-\kappa_{2}%
^{2}-3\kappa_{2}^{4}}{\kappa_{1}^{2}-\kappa_{2}^{2}}\nonumber\\
&  -\left(  \frac{5}{2}\kappa_{1}^{3}\text{ }-\frac{2\kappa_{1}^{5}}%
{\kappa_{1}^{2}-\kappa_{2}^{2}}\right)  \operatorname{arctanh}\frac{1}%
{\kappa_{1}}-\left.  \left(  \frac{5}{2}\kappa_{2}^{3}+\frac{2\kappa_{2}^{5}%
}{\kappa_{1}^{2}-\kappa_{2}^{2}}\right)  \operatorname{arctanh}\frac{1}%
{\kappa_{2}}\right]
\end{align}
and%
\begin{align}
\tilde{j}_{4}\left(  \xi,c_{1}\right)   &  =\frac{1}{1+4c_{1}\xi^{2}}\left[
\frac{1}{30}\frac{6+14\kappa_{1}^{2}+70\kappa_{1}^{4}-105\kappa_{1}^{6}%
}{1-\kappa_{1}^{2}}\right.  +\frac{1}{30}\frac{6+14\kappa_{2}^{2}+70\kappa
_{2}^{4}-105\kappa_{2}^{6}}{1-\kappa_{2}^{2}}\nonumber\\
&  -\frac{2}{15}\frac{3\kappa_{1}^{2}+5\kappa_{1}^{4}+15\kappa_{1}^{6}%
-3\kappa_{2}^{2}-5\kappa_{2}^{4}-15\kappa_{2}^{6}}{\kappa_{1}^{2}-\kappa
_{2}^{2}}\nonumber\\
&  -\left(  \frac{7}{2}\kappa_{1}^{5}\text{ }-\frac{2\kappa_{1}^{7}}%
{\kappa_{1}^{2}-\kappa_{2}^{2}}\right)  \operatorname{arctanh}\frac{1}%
{\kappa_{1}}-\left.  \left(  \frac{7}{2}\kappa_{2}^{5}+\frac{2\kappa_{2}^{7}%
}{\kappa_{1}^{2}-\kappa_{2}^{2}}\right)  \operatorname{arctanh}\frac{1}%
{\kappa_{2}}\right]
\end{align}
and finally%
\begin{align}
\tilde{j}_{5}\left(  \xi,c_{1}\right)   &  =\frac{1}{1+4c_{1}\xi^{2}}\left[
\frac{1}{70}\frac{10+18\kappa_{1}^{2}+42\kappa_{1}^{4}+210\kappa_{1}%
^{6}-315\kappa_{1}^{8}}{1-\kappa_{1}^{2}}\right.  +\frac{1}{70}\frac
{10+18\kappa_{2}^{2}+42\kappa_{2}^{4}+210\kappa_{2}^{6}-315\kappa_{2}^{8}%
}{1-\kappa_{2}^{2}}\nonumber\\
&  -\frac{2}{105}\frac{15\kappa_{1}^{2}+21\kappa_{1}^{4}+35\kappa_{1}%
^{6}+105\kappa_{1}^{8}-15\kappa_{2}^{2}-21\kappa_{2}^{4}-35\kappa_{2}%
^{6}-105\kappa_{2}^{8}}{\kappa_{1}^{2}-\kappa_{2}^{2}}\nonumber\\
&  -\left(  \frac{9}{2}\kappa_{1}^{7}\text{ }-\frac{2\kappa_{1}^{9}}%
{\kappa_{1}^{2}-\kappa_{2}^{2}}\right)  \operatorname{arctanh}\frac{1}%
{\kappa_{1}}-\left.  \left(  \frac{9}{2}\kappa_{2}^{7}+\frac{2\kappa_{2}^{9}%
}{\kappa_{1}^{2}-\kappa_{2}^{2}}\right)  \operatorname{arctanh}\frac{1}%
{\kappa_{2}}\right]
\end{align}
(the $\kappa_{1,2}\left(  \xi,c_{1}\right)  $ are defined in Eq.
(\ref{pp}))\ which we needed to evaluate the matrix elements derived in the
main text.

\subsection{Limits}

\label{lim}

When analyzing the soft-mode contributions to gap equation and energy density,
it is useful to consider the $\xi\rightarrow0$ and $c_{1}\rightarrow
0$\ limits. This boils down to finding the limits of the $\tilde{\imath}_{n}$
and $\tilde{j}_{n}$ integrals which we provide in the present section. The
$\xi\rightarrow0$ limits are (for $c_{1}\leq1$, as before)
\begin{align}
\tilde{\imath}_{n}\left(  \xi=0,c_{1}\right)   &  =\int_{0}^{1}\frac
{\kappa^{2n-2}}{1-c_{1}\kappa^{2}}d\kappa=\frac{1}{2n-1}\text{ }_{2}%
F_{1}\left(  1,n-\frac{1}{2},n+\frac{1}{2},c_{1}\right) \\
&  \overset{c_{1}\rightarrow0}{\longrightarrow}\frac{1}{2n-1}%
\end{align}
for $n>1/2$ and
\begin{align}
\tilde{j}_{n}\left(  \xi=0,c_{1}\right)   &  =\int_{0}^{1}\frac{\kappa^{2n-4}%
}{\left(  1-c_{1}\kappa^{2}\right)  ^{2}}d\kappa=\frac{1}{2n-3}\text{ }%
_{2}F_{1}\left(  2,n-\frac{3}{2},n-\frac{1}{2},c_{1}\right) \\
&  \overset{c_{1}\rightarrow0}{\longrightarrow}\frac{1}{2n-3}%
\end{align}
for $n>3/2$. (The hypergeometric functions $_{2}F_{1}\left(  a,b,c,z\right)  $
are defined e.g. in Ref. \cite{abr}.) Note that only $\tilde{j}_{1}$, which
does not appear in the coincidence limit of the soft-mode correlation
functions, contains a $\xi\rightarrow0$ divergence (cf. Eq. (\ref{j1})). The
$\xi\rightarrow0$ limit of the $\tilde{\imath}_{n}$ integrals can be
reexpressed via continued partial integration as%
\begin{equation}
\tilde{\imath}_{n}\left(  \xi=0,c_{1}\right)  =-\frac{1}{c_{1}^{n-1}}\left(
\sum_{k=0}^{n-2}\frac{c_{1}^{k}}{2k+1}-\frac{\operatorname{arctanh}\sqrt
{c_{1}}}{\sqrt{c_{1}}}\right)
\end{equation}
which yields
\begin{align}
\tilde{\imath}_{1}\left(  \xi=0,c_{1}\right)   &  =\frac
{\operatorname{arctanh}\sqrt{c_{1}}}{\sqrt{c_{1}}}\overset{c\rightarrow
0}{\longrightarrow}1,\\
\tilde{\imath}_{2}\left(  \xi=0,c_{1}\right)   &  =-\frac{1}{c_{1}}\left(
1-\frac{\operatorname{arctanh}\sqrt{c_{1}}}{\sqrt{c_{1}}}\right)  ,
\end{align}
etc.. Furthermore,%
\begin{equation}
\tilde{j}_{1}\left(  \xi,c_{1}\right)  \overset{\xi\rightarrow0}%
{\longrightarrow}\frac{\pi}{4}\frac{1}{\xi}-\frac{1}{2}\left(  \frac{2-3c_{1}%
}{1-c_{1}}-3\sqrt{c_{1}}\operatorname{arctanh}\sqrt{c_{1}}\right)
-\frac{15\pi}{8}c_{1}\xi+O\left(  \xi^{2}\right)  \label{j1}%
\end{equation}
as well as%
\begin{align}
\tilde{j}_{2}\left(  \xi=0,c_{1}\right)   &  =\frac{1}{2}\left(  \frac
{1}{1-c_{1}}+\frac{\operatorname{arctanh}\sqrt{c_{1}}}{\sqrt{c_{1}}}\right)
,\\
\tilde{j}_{3}\left(  \xi=0,c_{1}\right)   &  =\frac{1}{2c_{1}}\left(  \frac
{1}{1-c_{1}}-\frac{\operatorname{arctanh}\sqrt{c_{1}}}{\sqrt{c_{1}}}\right)
,\\
\tilde{j}_{4}\left(  \xi=0,c_{1}\right)   &  =\frac{1}{2c_{1}^{2}}\left(
\frac{3-2c_{1}}{1-c_{1}}-3\frac{\operatorname{arctanh}\sqrt{c_{1}}}%
{\sqrt{c_{1}}}\right)  ,\\
\tilde{j}_{5}\left(  \xi=0,c_{1}\right)   &  =\frac{1}{6c_{1}^{3}}\left(
\frac{15-10c_{1}-2c_{1}^{2}}{1-c_{1}}-15\frac{\operatorname{arctanh}%
\sqrt{c_{1}}}{\sqrt{c_{1}}}\right)  .
\end{align}
This implies, in particular,
\begin{align}
e\left(  \xi,c_{1}\right)   &  :=\tilde{\imath}_{2}-2c_{1}\tilde{\imath}%
_{3}+c_{1}^{2}\tilde{\imath}_{4}+2c_{1}\frac{\gamma}{\zeta}\tilde{\imath}%
_{2}\left(  \tilde{j}_{3}-2c_{1}\tilde{j}_{4}+c_{1}^{2}\tilde{j}_{5}\right) \\
&  \overset{\xi=0}{=}\frac{1}{3}-\frac{c_{1}}{5}-\frac{2\gamma}{3\zeta}\left(
1-\frac{\operatorname{arctanh}\sqrt{c_{1}}}{\sqrt{c_{1}}}\right)
\end{align}
for the combination of integrals which appears in the electric soft-mode
contribution (\ref{t2u}) to the vacuum energy density (\ref{e}).

We further examine the $c_{1}\rightarrow0$ limits%
\begin{align}
\tilde{\imath}_{n}\left(  \xi\right)   &  :=\tilde{\imath}_{n}\left(
\xi,c_{1}=0\right)  =\int_{0}^{1}\frac{\kappa^{2n}}{\kappa^{2}+\xi^{2}}%
d\kappa,\text{ \ \ \ \ \ \ \ \ \ \ \ \ }n\geq1,\\
\tilde{j}_{n}\left(  \xi\right)   &  :=\tilde{j}_{n}\left(  \xi,c_{1}%
=0\right)  =\int_{0}^{1}\frac{\kappa^{2n}}{\left(  \kappa^{2}+\xi^{2}\right)
^{2}}d\kappa,\text{ \ \ \ \ \ \ \ \ }n\geq1
\end{align}
which we distinguish from the general integrals (\ref{in}), (\ref{jn}) only by
their arguments. Both $\tilde{\imath}_{n}\left(  \xi\right)  $ and $\tilde
{j}_{n}\left(  \xi\right)  $ decay monotonically in $\xi\in\left[
0,\infty\right]  ,$ starting from the (finite) values%
\begin{align}
\tilde{\imath}_{n}\left(  0\right)   &  =\frac{1}{2n-1},\text{\ \ \ \ \ \ }%
n\geq1,\\
\tilde{j}_{n}\left(  0\right)   &  =\frac{1}{2n-3},\text{\ \ \ \ \ \ }n\geq2,
\end{align}
except for%
\begin{equation}
\tilde{j}_{1}\left(  \xi\right)  \overset{\xi\rightarrow0}{\longrightarrow
}\frac{\pi}{4}\frac{1}{\xi}-1+\frac{2}{3}\xi^{2}-\frac{3}{5}\xi^{4}+O\left(
\xi^{6}\right)  \label{j1lim}%
\end{equation}
which exhibits the already mentioned pole divergence at $\xi\rightarrow0$. The
$\tilde{\imath}_{n}\left(  \xi\right)  $, $\tilde{j}_{n}\left(  \xi\right)  $
further satisfy the recursion relations%
\begin{align}
\tilde{\imath}_{n}^{\prime}\left(  \xi\right)   &  \equiv\frac{d\tilde{\imath
}_{n}\left(  \xi\right)  }{d\xi}=-2\xi\tilde{j}_{n}\left(  \xi\right)
\leq0,\\
\tilde{j}_{n}^{\prime}\left(  \xi\right)   &  \equiv\frac{d\tilde{j}%
_{n}\left(  \xi\right)  }{d\xi}=-4\xi\tilde{k}_{n}\left(  \xi\right)  \leq0,
\end{align}
and
\begin{align}
\tilde{\imath}_{n}\left(  \xi\right)   &  =\frac{1}{2n-1}-\xi^{2}\tilde
{\imath}_{n-1}\left(  \xi\right)  ,\label{irecur}\\
\tilde{j}_{n}\left(  \xi\right)   &  =\frac{1}{2n-3}\left[  \frac{1}{1+\xi
^{2}}-\left(  2n-1\right)  \xi^{2}\tilde{j}_{n-1}\right]  \text{
\ \ \ \ \ \ \ \ \ }\left(  n>1\right)
\end{align}
which may be summed up into a finite power series and a transcendental
piece,\textbf{ }%
\begin{equation}
\tilde{\imath}_{n}\left(  \xi\right)  =\sum_{k=0}^{n-2}\frac{\left(
-1\right)  ^{k}\xi^{2k}}{2n-2k-1}+\left(  -1\right)  ^{n+1}\left(
1-\xi\arctan\frac{1}{\xi}\right)  \xi^{2n-2}.
\end{equation}

Explicit expressions for the first few $\tilde{\imath}_{n}\left(  \xi\right)
$ and $\tilde{j}_{n}\left(  \xi\right)  $, which appear in the main text, are
thus
\begin{align}
\tilde{\imath}_{1}\left(  \xi\right)   &  =1-\xi\arctan\frac{1}{\xi},\\
\tilde{\imath}_{2}\left(  \xi\right)   &  =\frac{1}{3}-\xi^{2}+\xi^{3}%
\arctan\frac{1}{\xi},\\
\tilde{\imath}_{3}\left(  \xi\right)   &  =\frac{1}{5}-\frac{1}{3}\xi^{2}%
+\xi^{4}-\xi^{5}\arctan\frac{1}{\xi},\\
\tilde{\imath}_{4}\left(  \xi\right)   &  =\frac{1}{7}-\frac{1}{5}\xi
^{2}+\frac{1}{3}\xi^{4}-\xi^{6}+\xi^{7}\arctan\frac{1}{\xi}%
\end{align}
as well as%
\begin{align}
\tilde{j}_{1}\left(  \xi\right)   &  =\frac{1}{2}\left(  -\frac{1}{1+\xi^{2}%
}+\frac{1}{\xi}\arctan\frac{1}{\xi}\right)  ,\\
\tilde{j}_{2}\left(  \xi\right)   &  =\frac{1}{2}\left(  3-\frac{1}{1+\xi^{2}%
}-3\xi\arctan\frac{1}{\xi}\right)  ,\\
\tilde{j}_{3}\left(  \xi\right)   &  =\frac{1}{2}\left(  \frac{5}{3}-5\xi
^{2}-\frac{1}{1+\xi^{2}}+5\xi^{3}\arctan\frac{1}{\xi}\right)  ,\\
\tilde{j}_{4}\left(  \xi\right)   &  =\frac{1}{2}\left(  \frac{7}{5}-\frac
{7}{3}\xi^{2}+7\xi^{4}-\frac{1}{1+\xi^{2}}-7\xi^{5}\arctan\frac{1}{\xi
}\right)  ,\\
\tilde{j}_{5}\left(  \xi\right)   &  =\frac{1}{2}\left(  \frac{9}{7}-\frac
{3}{5}\left(  3-5\xi^{2}+15\xi^{4}\right)  \xi^{2}-\frac{1}{1+\xi^{2}}%
+9\xi^{7}\arctan\frac{1}{\xi}\right)  .
\end{align}

Finally, we note that the $\tilde{\imath}_{n}\left(  \xi\right)  $, $\tilde
{j}_{n}\left(  \xi\right)  $ (and their analogs with additional powers of the
integrand's denominator) determine the coefficients of the expansion of the
integrals $\tilde{\imath}_{n}\left(  \xi;c_{1}\right)  $ and $\tilde{j}%
_{n}\left(  \xi;c_{1}\right)  $ in powers of $c_{1}$. For example, from
\begin{align}
\frac{1}{\kappa^{2}\left(  1-c_{1}\kappa^{2}\right)  +\xi^{2}}  &  =\frac
{1}{\kappa^{2}+\xi^{2}}+\frac{\kappa^{4}}{\left(  \kappa^{2}+\xi^{2}\right)
^{2}}c_{1}+\frac{\kappa^{8}}{\left(  \kappa^{2}+\xi^{2}\right)  ^{3}}c_{1}%
^{2}+O\left(  c_{1}^{3}\right)  ,\\
\frac{1}{\left(  \kappa^{2}\left(  1-c_{1}\kappa^{2}\right)  +\xi^{2}\right)
^{2}}  &  =\frac{1}{\left(  \kappa^{2}+\xi^{2}\right)  ^{2}}+\frac{2\kappa
^{4}}{\left(  \kappa^{2}+\xi^{2}\right)  ^{3}}c_{1}+\frac{3\kappa^{8}}{\left(
\kappa^{2}+\xi^{2}\right)  ^{4}}\allowbreak c_{1}^{2}+O\left(  c_{1}%
^{3}\right)
\end{align}
$\allowbreak$one has (with $\tilde{k}_{n}\left(  \xi\right)  :=\int_{0}%
^{1}\kappa^{2n}/\left(  \kappa^{2}+\xi^{2}\right)  ^{3}d\kappa$)
\begin{align}
&  \tilde{\imath}_{n}\left(  \xi;c_{1}\right)  =\tilde{\imath}_{n}\left(
\xi\right)  +\tilde{j}_{n+2}\left(  \xi\right)  c_{1}+\tilde{k}_{n+4}\left(
\xi\right)  c_{1}^{2}+O\left(  c_{1}^{3}\right)  ,\label{inexp}\\
&  \tilde{j}_{n}\left(  \xi;c_{1}\right)  =\tilde{j}_{n}\left(  \xi\right)
+2\tilde{k}_{n+2}\left(  \xi\right)  c_{1}+O\left(  c_{1}^{2}\right)  .
\label{jnexp}%
\end{align}

\newpage

\end{document}